%% file: noisy_optpol.tex
\documentclass[12pt]{article}
\input{noisy_optpol_header.tex}

\begin{document}

\title{Optimal Policy With Noisy Sufficient Statistics\thanks{Email: {\tt montiel.olea@gmail.com}, {\tt mikkelpm@uchicago.edu}, and {\tt ckwolf@mit.edu}. Mathematical proofs and computer code were generated in close collaboration with generative AI. We are grateful for comments from St\'{e}phane Bonhomme. Plagborg-M{\o}ller acknowledges that this material is based upon work supported by the NSF under Grant {\#}2238049 and by the Alfred P.\ Sloan Foundation. Wolf similarly thanks the Alfred P.\ Sloan Foundation.}}
\author{\begin{tabular}{ccccc}
Jos\'{e} Luis Montiel Olea && Mikkel Plagborg-M{\o}ller && Christian K.\ Wolf \\
{\small Cornell University} && {\small University of Chicago} && {\small MIT \& NBER}
\end{tabular}}
\date{\texorpdfstring{\bigskip}{ }October 29, 2026}
\maketitle

\vspace{-1.5em}
\begin{abstract}
We study how policy-makers should set $k$ policy instruments so as to best hit $m$ (perhaps dynamic) targets, in a linear-quadratic model where there is statistical uncertainty about the causal effects of the policies. Plugging the estimated effects into ``sufficient statistics'' optimal policy formulas, as commonly done in the literature, can result in infinite risk. In fact, maintaining the \emph{status quo} regardless of the data is uniquely minimax-optimal when $m>1$, and uniquely minimax-regret-optimal when $m,k>1$. Nevertheless, we recommend that policy-makers adopt a simple, dampened version of the popular plug-in rule, as in \citet{Fisher1962} and \citet{Brainard1967}. Not only does this \emph{cautious} rule outperform doing nothing over most of the parameter space, its worst-case risk is also close to the minimax value in empirically relevant models. Policy-makers should use causal effect estimates from an unbiased but dispersed estimator (such as a local projection) rather than from a moderately biased but precise estimator (such as a vector autoregression).
\end{abstract}

\clearpage

\section{Introduction}
\label{sec:intro}

Recent theoretical analyses of economic policy often derive formulas that express the optimal levels of policy instruments as functions of estimable ``sufficient statistics,'' where the latter are given by the causal effects of certain treatments or policy interventions. To bring those formulas to data, researchers typically proceed in two steps: first estimate the causal effects, and then plug them into the optimal policy formula. However, policy causal effects are often estimated with substantial noise, and good econometric properties of the causal effect estimators for \emph{point estimation} may not translate into good properties of the resulting down-stream \emph{policy actions}.\footnote{See \citet{Manski2021}, \citet{GuKoenker2023}, \citet{Andrews2026}, and \citet{Christensen2026} for recent discussions of this fundamental point in contexts different from ours.} Indeed, a classic literature building on the policy framework of \citet{Tinbergen1952} and \citet{Theil1958} argued that Bayes-optimal policy rules tend to be \emph{cautious}, dampening the sometimes erratic behavior of the plug-in rule \citep{Fisher1962,Brainard1967,Lindley1968,Zellner1971}. Subsequently, substantial progress was made by using statistical decision theory to analyze the properties of a variety of policy rules \citep{Zaman1981,ZamanStein1981,BergerBerlinerZaman1982,Berliner1983}.

Unfortunately, it is not straightforward to apply the results in this existing literature to contemporary problems in economic policy, for at least three reasons. First, earlier papers dealt primarily with static or backward-looking models that are potentially subject to the \citet{Lucas1976} critique. Optimal policy analysis over the past few decades, in particular in macroeconomics, has instead typically relied on rational expectations general equilibrium models. Second, the Bayes-optimal rules proposed in this literature raise the perennial question of how sensitive the recommended policy actions are to the choice of prior. Third, most theoretical progress on the statistical front was achieved for the case of a single policy target. In practice, policy-makers typically need to balance multiple targets---both because they care about many different outcome variables, and because their policies tend to affect the economy for many periods, leading to dynamic trade-offs.

In this paper, we bring the classical theory of optimal policy choice under statistical uncertainty into the modern era. We study a linear-quadratic control problem with $m \geq 1$ policy targets (that could arise, for example, from considering multiple horizons) and $k \geq 1$ policy instruments, where the causal effects of policy actions are measured with statistical noise. Exploiting recent advances in macroeconomic theory, we show that this framework nests a large class of optimal dynamic policy problems under rational expectations, which implies that our setting is robust to the \citeauthor{Lucas1976} critique. Our main theoretical result is that the cautious decision rule proposed in the earlier literature---a Bayes rule based on a diffuse prior---has surprisingly favorable \emph{worst-case} properties, in contrast to the plug-in rule. Hence, the cautious rule emerges as a robust choice for policy-making for a wide range of data-generating processes. Lastly, we argue that the down-stream policy problem informs the trade-off between bias and variance when estimating policy causal effects: bias should receive much larger weight than in the traditional mean squared error criterion.

The linear-quadratic decision problem we study is applicable to a range of settings across macroeconomics, public finance, and related fields. In macroeconomics, a vast theoretical literature studies linear-quadratic approximations to optimal dynamic policy design using fully specified, structural general equilibrium models \citep[e.g.,][]{Woodford2003, HansenSargent2013, Gali2015}. Building on \citet{Barnichon2023} and \citet{McKay2023}, we show that, for a wide range of structural models, the forward-looking optimal policy problem under commitment can be recast as a special case of our control problem, with the unknown parameters being the dynamic causal effects of \emph{shocks} to the policy instrument---i.e., the estimands of a vast empirical literature that traces back to \citet{Sims1980}. We show that the exact same logic also applies to problems in other areas of economics, where optimal policies depend on ``sufficient statistics'' estimated through natural experiments \citep{Chetty2009,Hendren2020}.

The plug-in approach to computing optimal policies, which treats the estimated policy causal effects as noiseless, is ubiquitous in applied work, such as in the recent optimal monetary policy exercises of \citet{Barnichon2023} and \citet{Caravello2026}. Unfortunately, we show that the plug-in policy rule has unappealing properties: it often generates counterproductive actions---actions worse than doing nothing---and can have unbounded worst-case \emph{risk} (expected loss). Intuitively, the problem is that the plug-in rule demands large policy actions if the estimated causal effects are small, making the actions highly sensitive to small amounts of noise.

An alternative policy rule is to simply do nothing and maintain the \emph{status quo}, regardless of the estimated policy effects. Strikingly, under mild conditions, we show that this maximally cautious decision rule that ignores the data is \emph{uniquely} minimax-optimal (and admissible) when $m>1$; that is, it uniquely delivers the best possible robust guarantee on risk. The intuitive reason is that, when there are multiple targets $m>1$, the policy causal effects could be so misaligned with the targets that any data-dependent policy rule only ends up adding noise, increasing the loss relative to doing nothing. Perhaps more surprisingly, if both $m,k > 1$, doing nothing is also uniquely minimax-regret-optimal. Our results amount to an extreme form of \citeauthor{Brainard1967}'s (\citeyear{Brainard1967}) attenuation principle: when causal effects are statistically uncertain, a policy-maker concerned about robustness will choose maximal caution. The obvious downside of the do-nothing rule is that it fails to exploit the information in the data, which is highly suboptimal at more favorable parameters.

Our main result is that the cautious rule---the Bayes-optimal rule based on a diffuse prior and Gaussian likelihood---is an attractive practical choice, as it exploits the estimated policy effects while maintaining worst-case risk often (very) close to the minimax value. We first prove that, differently from the plug-in rule, the risk of the cautious rule is bounded, \emph{uniformly} across causal effect parameters and possible data correlation structures. In fact, in the empirically relevant setting of a single policy instrument ($k = 1$), extensive analytical and numerical work indicates that the cautious rule incurs at worst 10.4\% excess risk relative to the minimax rule, for any $m \geq 2$.\footnote{This bound is sharp among bounds that are independent of the correlation structure of the statistical noise. While the bound is obtained from a combination of analytical dimension-reduction and numerical optimization, we can fully analytically prove a bound of 16.9\%.} At the same time, the upside of the cautious rule relative to doing nothing is unbounded: the rule adapts to the strength of the signal, converging to the oracle action as the signal starts to dominate the noise. In addition to its limited downside and unlimited upside, the cautious rule is admissible in low-dimensional problems with $mk \leq 4$ and approximately Bayes-optimal for sufficiently diffuse Gaussian priors.\footnote{\citet{Hurwicz1951} and \citet{HodgesLehmann1952} develop frameworks for weighing worst-case risk against best-case or average performance.}

How should the down-stream policy problem inform the choice of causal effect estimator? We argue that, while the worst-case risk of the cautious rule does not depend on the scale of the variance of the estimator, it depends sensitively on the potential bias of the estimator. Hence, estimators with low mean squared error may not be the optimal inputs to subsequent policy-making. Indeed, policy-makers who are seeking a robust guarantee on the performance of their policies prefer estimators with low bias (such as local projections, in the case of dynamic causal effect estimation) to estimators with low variance but potentially non-trivial bias (such as vector autoregressions with short lag lengths).

We illustrate our theoretical results with an application to optimal monetary policy, following \citet{Barnichon2023}. We first contrast the policy actions reported in their analysis, based on the plug-in rule, with our preferred cautious rule. Quantitatively, the statistical uncertainty about monetary policy causal effects is large enough that the cautious rule non-trivially dampens the plug-in rule, and yet small enough to often yield decisive policy action. To further investigate the properties of the various decision rules under consideration, we calibrate a simulation study to this empirical application. Here our theory is borne out: the cautious rule performs almost as well as the plug-in rule when signals are strong and instruments are well-aligned with targets, and attains much smaller losses in the tails.

\paragraph{Further related literature.}
Though Bayesian treatments of the linear-quadratic control problem have considered models with several targets $m$ \citep[Chapter 11]{Zellner1971}, existing results on the frequentist properties of policy rules are limited to the special case $m = 1$. Our analysis shows that this case is quite special, both qualitatively and quantitatively. To our knowledge, there are no parallels in the literature to our results on the unique minimax-optimality of doing nothing when $m>1$, the risk bounds for the cautious rule for general $m$, and the role of bias vs.\ variance. The poor performance of the plug-in rule has been pointed out in the $m=1$ case (\citealp{Brainard1967}; \citealp[p.\ 323]{Zellner1971}; \citealp[Section 5]{Zaman1981}); we provide a more general theoretical characterization of its properties.

Our approach differs from the robust control literature in macroeconomics \citep{Onatski2002,Giannoni2002,Hansen2008,Barlevy2009} by allowing policy-makers to observe a noisy signal about policy causal effects. That literature instead begins with \emph{a priori} restrictions either on deep structural parameters of an economic model or, more similarly to our analysis, directly on policy causal effects. We instead consider policy-makers who want to exploit the rich menu of dynamic causal effect estimates available for their policy instruments, keeping in mind that these estimates may be quite noisy.

Our paper is related to concurrent work by \citet{Kitagawa2026}, who examine a statistical decision problem similar to ours, also motivated by the recent advances of \citet{Barnichon2023} and \citet{McKay2023}. Their paper argues for Bayes or robust Bayes policy rules over the plug-in rule, but arrives at this conclusion through different criteria than ours. In particular, they do not derive worst-case risk properties of policy rules, which is the main technical contribution of our paper. Otherwise, the two papers are complementary: \citeauthor{Kitagawa2026} extend their framework to set-identified impulse responses and sequential decisions, while we discuss how the policy problem informs the bias-variance trade-off between causal effect estimators.

\paragraph{Outline.}
\cref{sec:framework} defines and motivates the decision-theoretic framework and candidate policy rules. \cref{sec:intuition} provides simple illustrations of our results, while \cref{sec:results} presents the general theory. \cref{sec:bias_var} discusses how policy-makers should trade off bias and variance in causal effect estimators. \cref{sec:sim} demonstrates the practical relevance of the theory through an empirical application and a calibrated simulation study. \cref{sec:concl} concludes. Proofs and some technical details are relegated to the appendix. 

\section{Framework}
\label{sec:framework}

We start by defining the statistical decision problem, and then give practical examples of economic applications that fit into our structure. Subsequently we define the various candidate policy rules whose properties will be analyzed in the rest of the paper.

\subsection{Statistical decision problem}
\label{subsec:problem}

A policy-maker chooses the values for $k$ policy instruments $a=(a_1,\dots,a_k)'$. The policies transmit linearly into $m$ outcome variables $\tilde{y} = \theta a$, where $\theta \in \mathbb{R}^{m \times k}$ is the matrix of policy causal effects. Though $\theta$ is unknown, the policy-maker observes a noisy signal $X$ that is informative about those causal effects. The goal of the policy-maker is to approximately hit some given, deterministic target values $y=(y_1,\dots,y_m)'$ for the $m$ outcomes $\tilde{y}=(\tilde{y}_1,\dots,\tilde{y}_m)'$. As in the voluminous literature on optimal control in macroeconomics, we adopt a quadratic loss function that penalizes positive and negative targeting errors $y-\tilde{y}$ equally; see \cref{subsec:econ_examples} for further discussion.\footnote{\citet[Chapter 11.4]{Zellner1971} discusses alternative loss functions.}

Specifically, we define the statistical decision problem---in the sense of \citet{Ferguson1967}---as follows. Versions of this problem with $m=1$ outcome have been called the ``control problem'' in the statistics literature \citep{ZamanStein1981}.

\begin{itemize}

\item We refer to the noisy signal $X$ observed by the policy-maker as the \emph{data}. The \emph{sample space} $\mathcal{X}$ of possible data realizations will be defined below when needed.

\item The \emph{action space} for the problem---that is, the menu of potential choices for the policy instruments that are available to the policy-maker---is $\mathbb{R}^k$.

\item The \emph{statistical model} $\{P_{\theta}\}_{\theta \in \Theta}$ describes how the probability distribution of the observed data $X$ depends on the unknown \emph{parameter} $\theta$, i.e., the policy causal effects. The \emph{parameter space} is denoted $\Theta \subset \mathbb{R}^{m \times k}$. Expectations under each distribution $P_{\theta}$ are written $E_\theta[\cdot]$, and similarly for variances.

In most of our analysis we will focus on the Gaussian leading case where $\mathcal{X} \equiv \mathbb{R}^{m \times k}$ and $X \in \mathcal{X}$ satisfies
\begin{equation} \label{eqn:X_gauss}
\ve(X) \sim N(\ve(\theta), \Sigma),
\end{equation}
for a known $mk \times mk$ symmetric positive definite $\Sigma$. If we view \eqref{eqn:X_gauss} as a ``limiting experiment,'' this model approximately captures settings where we have available a consistent and asymptotically normal estimator $X$ of the causal effects $\theta$. In practice, $\Sigma$ is replaced with a consistent estimator.

\item The \emph{loss function} provides a ranking of actions contingent on $\theta$. We work under the loss function $L(\theta,a) \equiv \nW{y-\theta a}^2$ that penalizes weighted squared deviations from the policy targets. Here $y \in \mathbb{R}^m \setminus \lbrace 0 \rbrace$ is a known target vector, $W$ is a known $m \times m$ symmetric positive definite matrix, and $\nW{x} \equiv \sqrt{x'Wx}$ is a weighted Euclidean norm.

\end{itemize}

A \emph{policy rule} (or decision rule) is a measurable map $a \colon \mathcal{X} \to \mathbb{R}^k$ from the sample space to the action space, so that $a(X)=(a_1(X),\dots,a_k(X))'$ is the realized policy choice given data $X$. We will mostly evaluate policy rules $a(\cdot)$ by their \emph{risk}, that is, expected loss: $R(\theta,a) \equiv E_\theta[L(\theta,a(X))]$, which depends on the true $\theta$ and may be infinite.

\paragraph{Extension: uncertain targets.}
As in \citet{Brainard1967}, our analysis extends immediately to settings where there is additional independent noise in the outcome variables $\tilde{y}$. This corresponds to a model where we replace $y$ with $y-u$, for an $m$-dimensional random vector $u$ with mean zero and finite variance. If $u$ is independent of the data $X$, then the risk of policy rule $a(X)$ equals
\[E_\theta\left[\nW{(y-u)-\theta a(X)}^2\right] = \tr\big\lbrace W\var(u)\big\rbrace + E_\theta\left[\nW{y-\theta a(X)}^2\right],\]
which is the same as in the earlier decision problem, up to a constant (provided $\var(u)$ does not depend on $\theta$). If instead $u$ and $X$ are dependent, our results may not go through; we focus on the independent case because it emerges naturally in the leading economic applications discussed next.

\subsection{Connection to the literature on optimal policy}
\label{subsec:econ_examples}

We now give two examples that showcase the relevance of our statistical decision problem for policy problems. The first one is taken from macroeconomics, and the second comes from public finance.

\begin{exm}[Optimal policy in structural business-cycle models]
\label{exm:macro}

Our first example nests linear-quadratic optimal policy analyses in a large family of linearized business-cycle models, see for example the New Keynesian textbook treatments of \citet{Woodford2003} and \citet{Gali2015}. The policy-maker's loss is
\begin{equation}
L = \mathbb{E}_0 \left [ \sum_{t=0}^{T} \beta^t \sum_{i=1}^n \omega_i \hat{y}_{it}^2 \right ], \label{eqn:loss_nk}
\end{equation}
where $\hat{y}_{1t},\dots,\hat{y}_{nt}$ are the $n$ outcome variables of interest (e.g., unemployment and inflation) in deviation from their respective targets, the $\omega_i$'s are the corresponding fixed (e.g., legislatively mandated) weights, $\beta \in (0,1)$ is the discount factor, $T$ is the maximal time horizon, and $\mathbb{E}_0$ is the agents' expectation operator at time 0.\footnote{$T$ is typically infinite in theoretical analyses. The mapping into our statistical decision problem requires it to be finite, though arbitrarily large.} A standard optimal policy problem with commitment is to choose the date-$0$ expected paths of a policy instrument (e.g., the federal funds rate, or government purchases) to minimize \eqref{eqn:loss_nk} subject to the constraints imposed by private-sector behavior, with the responses to future shocks then handled by solving the same problem at later dates.

To translate this optimal policy problem to our statistical decision problem, we leverage the recent work of \citet{Barnichon2023} and \citet{McKay2023}. Those papers establish that, for a large class of general equilibrium models and thus possible private-sector constraints, we can write the constraint set for the policy-maker's problem as
\begin{equation}
\mathbb{E}_0 \hat{y}_{it} = \mathbb{E}_0 \bar{y}_{it} + \sum_{j=1}^{k} \theta_{ijt} \hat{z}_j, \label{eqn:outcomes}
\end{equation}
where the first term, $\mathbb{E}_0 \bar{y}_{it}$, denotes expected values in the absence of further policy action (i.e., a ``baseline'' forecast under \emph{status quo} policy), the $\hat{z}_j$'s are deviations from the \emph{status quo} policy in the form of policy ``shocks'' \citep[as defined in][]{Sims1980,Ramey2016}, and $\theta_{ijt}$ is the causal effect of policy shock $j$ on outcome $i$ at horizon $t$. The optimal policy problem can then be re-expressed as choosing the ``shocks'' to minimize the loss \eqref{eqn:loss_nk}. Formally, equivalence of this re-cast problem with the original full-commitment problem requires a spanning condition: the $k$ policy shocks must allow the policy-maker to implement any desired path of her policy instrument(s). Short of that spanning condition (e.g., if $k \ll T$), solving this ``optimal shock'' problem instead means finding the optimal commitment policy within the more limited space of policy paths spanned by the $k$ available shocks in \eqref{eqn:outcomes}. Either way, baseline forecasts ($\mathbb{E}_0 \bar{y}_{it}$) together with policy shock effects ($\theta_{ijt}$) are ``sufficient statistics'' for \citeauthor{Lucas1976}-critique-robust policy evaluation.

Given this re-casting of the constraint set and optimal policy problem, the translation into our statistical decision problem is immediate. First, stack all the outcomes $\hat{y}_{it}$ over variables $i$ and time $t$ in the $m$-dimensional vector $\hat{y}$, where $m \equiv n(T+1)$. Defining analogously the $m$-dimensional vector $\bar{y}$ of baseline forecasts and $m \times k$ matrix $\theta$ of policy causal effects, we can by \eqref{eqn:outcomes} write the outcome vector $\hat{y}$ as
\begin{equation*}
\mathbb{E}_0 \left [ \hat{y} \right ] = \mathbb{E}_0 \left [ \bar{y} \right ] + \theta \hat{z},
\end{equation*}
where $\hat{z} \equiv (\hat{z}_1,\dots,\hat{z}_k)'$. We can now set $y \equiv \mathbb{E}_0 \left [ \bar{y} \right ]$, so the ``target'' is to undo any currently expected deviation of the outcomes from the (zero) bliss point, and $W \equiv \diag((\omega_1,\dots,\omega_n) \otimes (1,\beta,\dots,\beta^T))$. Finally, let the $k$-dimensional policy action be $a \equiv -\hat{z}$. With these substitutions, the policy loss \eqref{eqn:loss_nk} maps into our loss function up to an additive, policy-independent variance term. The noisy signal $X$ of the policy shock causal effects $\theta$ (i.e., impulse responses) is typically obtained from vector autoregressions or local projections, delivering asymptotically normally distributed estimators of $\theta$.\footnote{The $y$'s, corresponding here to forecasts of the policy-maker targets in the absence of further policy action, are plausibly uncertain. As discussed in \cref{subsec:problem}, this is nested in our set-up as long as $y$ is independent of past accumulated evidence on the effects of the policy instrument of interest.}

\end{exm}

Our empirical application in \cref{sec:sim} will be to optimal monetary policy, following \cref{exm:macro}. That application will feature large $m$, corresponding to two targets (unemployment and inflation) over many time horizons, and small $k$, corresponding to a low-dimensional estimable space of policy shock causal effects (transitory policy rate changes). Our upcoming theoretical analysis will thus pay particular attention to this practically relevant large-$m$, small-$k$ configuration.

\begin{exm}[Static corrective taxation in public finance]

Our second example is static and comes from public finance: the corrective taxation of externalities or internalities \citep[e.g.,][]{Diamond1973,Sandmo1975,Allcott2019}. Consider an economy with a single good consumed by $m$ groups of heterogeneous households, and let $y_i$ denote the \emph{status quo} consumption of group $i$ in excess of a target level, such as the efficient level or the level recommended by public health guidelines. This excess is positive if group $i$ over-consumes, and negative if it under-consumes. The policy-maker contemplates changing the tax on the good by some amount $a$. With $\theta_i$ denoting the resulting reduction in the consumption of group $i$ per unit of the tax, the remaining excess consumption after the tax change is $y_i - \theta_i a$. The policy-maker's loss is
\begin{equation}
L = \sum_{i=1}^m \omega_i \left ( y_i - \theta_{i}a \right )^2, \label{eqn:corrective}
\end{equation}
where the weights $\omega_i > 0$ are fixed and known, e.g., reflecting a public health mandate.\footnote{If the targets are the efficient consumption levels, marginal costs and the marginal externality or internality are constant (though possibly group-specific), and revenue and distributional considerations are set aside, then \eqref{eqn:corrective} is the second-order approximation of the welfare loss, i.e., the sum of the \citet{Harberger1964} triangles, with $\omega_i$ equal to half the absolute slope of group $i$'s inverse compensated demand curve. The optimal action under known $\theta$ (i.e., the oracle action of \cref{subsec:policy_rules}) then reduces to the uniform corrective tax of \citet{Diamond1973}.}

We can map this structure into our decision problem by setting $y\equiv (y_1,\dots,y_m)'$, $\theta \equiv (\theta_1,\dots,\theta_m)'$, and $W\equiv\diag(\omega_1,\dots,\omega_m)$, with $k=1$. Doing nothing, $a=0$, means retaining the current tax. While textbook analysis of corrective taxation often treats the consumption responses $\theta$ as known or estimated without noise, our policy-maker here observes only a noisy signal $X$ of $\theta$, obtained from empirical studies of household responses to tax changes \citep[e.g., from experimental or quasi-experimental estimates in the sufficient-statistics tradition, as in][]{Chetty2009,Taubinsky2018}. The excess consumption levels are taken as known (independent uncertainty about them can be accommodated as in \cref{subsec:problem}).

\end{exm}

\subsection{Candidate policy rules}
\label{subsec:policy_rules}

In our analysis we will use statistical decision theory to compare the properties of different policy rules. For reference, the infeasible \emph{oracle action} that requires knowledge of the policy causal effects $\theta$ is given by
\begin{equation} \label{eqn:oracle_action}
a_\theta \equiv  (\theta'W\theta)^\dagger \theta'Wy \in \argmin_{\alpha \in \mathbb{R}^k} \nW{y-\theta \alpha},
\end{equation}
where $M^\dagger$ denotes the generalized inverse of $M$ \citep[Equation 2.2]{Fisher1962}. Let $V(\theta) \equiv L(\theta,a_\theta)$ denote the oracle loss. We are interested in analyzing the properties of the following three feasible policy rules. 

\begin{enumerate}

\item {\it Plug-in rule.} A feasible \emph{plug-in rule} takes the oracle action and substitutes the unknown parameter $\theta$ with its maximum likelihood estimate. For the Gaussian model \eqref{eqn:X_gauss} with $m \geq k$, this corresponds to $a_P(X) \equiv (X'WX)^{-1} X'Wy$. The literature sometimes refers to this rule as the ``certainty-equivalent rule,'' since it exploits $E_\theta[X]=\theta$ but ignores the fact that $\var_\theta(\ve(X)) \neq 0$ \citep[p.\ 322]{Zellner1971}.

\item {\it Do-nothing rule.} A natural benchmark with which to compare any proposed policy rule is the rule that does nothing regardless of the data: $a(X)=0$ identically. We will write this rule simply as ``$0$'' and refer to it as the \emph{do-nothing} rule. Importantly, as discussed in \cref{subsec:econ_examples}, ``doing nothing'' is an application-specific concept, as it usually refers to maintaining the policy instruments at some \emph{status quo} values rather than literally forcing all instruments to the numerical value of 0.

\item {\it Cautious rule.} A Bayesian decision-maker would formulate a prior distribution $\theta \sim \pi$ and select the \emph{Bayes rule} $a_\pi(X)$ that minimizes the posterior expected loss. This yields the rule $a_\pi(X) = E[\theta'W\theta \mid X]^\dagger E[\theta \mid X]'Wy$, where the expectations are with respect to the posterior distribution of $\theta$ given $X$ \citep[Section 3]{Fisher1962}. If we consider the Gaussian model \eqref{eqn:X_gauss} and use the improper prior $\pi$ given by Lebesgue measure, then the (generalized) Bayes rule for this improper prior equals the \emph{cautious rule}
\begin{equation} \label{eqn:cautious}
a_C(X) \equiv (X'W X + C)^{-1}X'Wy,\quad \text{where} \quad C \equiv \big(\tr\lbrace W\Sigma^{ij}\rbrace\big)_{ij} \in \mathbb{R}^{k \times k},
\end{equation}
and $\Sigma^{ij} \equiv \cov(X_{\bullet i},X_{\bullet j})$ with $X_{\bullet i}$ the $i$-th column of $X$.\footnote{Equivalently, $C = E[Z'WZ]$, where $Z$ is an $m \times k$ random matrix with $E[Z]=0$ and $\var(\ve(Z))=\Sigma$. The cautious rule also obtains as the limit of Bayes rules for the sequence of proper priors $\ve(\theta) \sim N(0,\tau^2 I_{mk})$ as $\tau \to \infty$.} To our knowledge, this rule was first derived by \citet{Fisher1962}. It was then popularized within economics (for the case $m=1$) by \citet{Brainard1967}. In the statistics literature \citep[e.g.,][]{Zaman1981}, the rule is often credited to \citet{Zellner1971}.

\end{enumerate}

Despite being based on a diffuse prior, the cautious rule differs from the plug-in rule for any nonzero noise covariance matrix $\Sigma$. In the limit as sampling uncertainty vanishes, $\Sigma \to 0$, both the plug-in and cautious rules converge in probability to the oracle action, provided $\theta$ has full column rank (though their risk functions need not coincide, cf.\ \cref{prop:plugin_inf,prop:cautious_risk_best_case}).

\section{Illustrations}
\label{sec:intuition}

In this section we illustrate the main results of our paper, focusing on the case with $k=1$ policy instrument and Gaussian data. We compare the performance of the do-nothing, plug-in, and cautious rules in two illustrative toy models. First, we consider the simplest model with $m=1$ outcome variable, which has received a lot of attention in the earlier literature. While this example has a special structure relative to our more general problem, we think it is pedagogically useful as a starting point. We then consider a model with $m = 2$, which illustrates most of our main qualitative---and to a large extent quantitative---results.

\subsection{One outcome variable}
\label{subsec:intuition_m1}

We begin with the simplest possible model, with $m=1$ outcome variable and $k=1$ policy instrument. In practice, this could correspond to a single instrument (say, the interest rate) being set to stabilize some outcome of interest on average over a time window (e.g., inflation averaging 2\% over the next two years). Without loss of generality, let the target value be $y=1$ and the loss weight $W=1$, yielding the loss
\[L(\theta,a) = (1-a\theta)^2 = \theta^2\left(\tfrac{1}{\theta}-a\right)^2,\]
with oracle action $a_\theta = 1/\theta$, provided $\theta \neq 0$. Note that this loss function can be viewed as squared error loss for estimating the inverse causal effect $1/\theta$, but---importantly---the loss is weighted by $\theta^2$ so that estimation errors are penalized less near the origin. At $\theta=0$ the loss does not depend on the policy action at all, since the policy has no causal effect. To finish the description of the model, assume we observe Gaussian data
\[X \sim N(\theta,\sigma^2).\]
As mentioned above, this setting has been studied extensively in prior work; the relevant references will be provided here when suitable, and otherwise in \cref{sec:results}.

\begin{figure}[t]
\begin{tabular}{cc}
\textsc{Policy rules} & \textsc{Relative risk of cautious rule} \\
\includegraphics[width=0.475\linewidth]{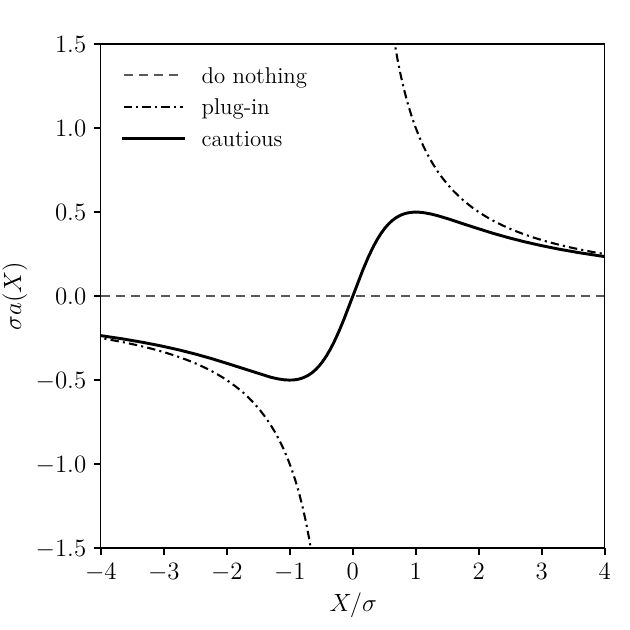} & \includegraphics[width=0.475\linewidth]{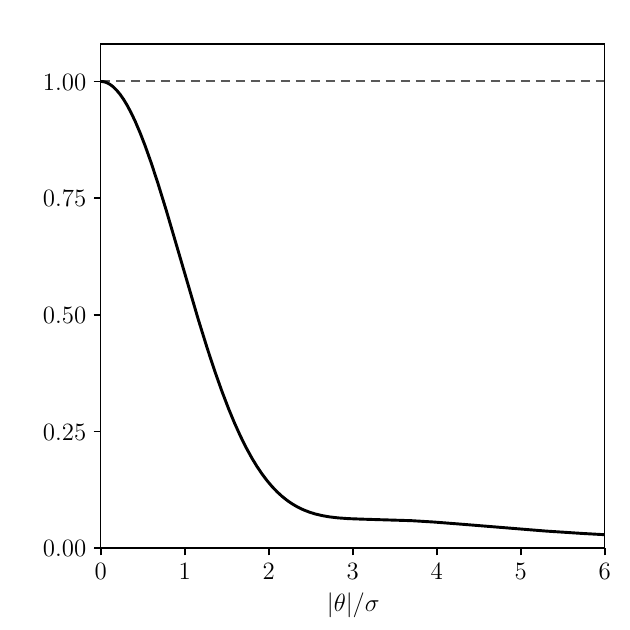}
\end{tabular}
\caption{Left panel: plot of three policy rules as functions of the data. Right panel: plot of the risk of the cautious rule relative to doing nothing, $R(\theta,a_C)/R(\theta,0)$. Gaussian model with $m=k=1$, $\Sigma=\sigma^2$, and $y=W=1$. Risk computed numerically via quadrature using the formula in \cref{app:prop:cautious_risk_spherical}.} \label{fig:rules_risk_m1}
\end{figure}

The candidate policy rules from \cref{subsec:policy_rules} lie on a spectrum. Consider first the simple plug-in rule, which here is $a_P(X)=1/X$. Note that this rule explodes to $\pm \infty$ near the origin $X \approx 0$: because the rule fails to take into account the uncertainty about the policy causal effect $\theta$, it overreacts to small changes in $X$ around the origin. As a result, the plug-in rule in this example does not have finite first or second moments, and thus \emph{infinite} risk: $R(\theta,a_P) = E_\theta[(1-\theta/X)^2] = \infty$ for all $\theta \neq 0$, and regardless of $\sigma>0$, reflecting the losses associated with very large actions taken even though the true $\theta$ is not commensurately close to zero.\footnote{Notice the contrast with point estimation problems with squared error loss. In such problems, the plug-in rule (maximum likelihood estimator) is uniquely minimax in one dimension, admissible for dimensions 1 and 2, and inadmissible in higher dimensions \citep{JamesStein1961}. In the control problem we consider, the plug-in rule is inadmissible even in one dimension, regardless of the amount of noise $\sigma^2$. See \citet[Sections 5--6]{Zaman1981} and \citet[Remark 2.4.1]{Berliner1983} for related points.} Moreover, the rule can also have a high probability of a \emph{counterproductive action}, relative to doing nothing: $P_\theta(L(\theta,a_P) > L(\theta,0)) = P_\theta(\theta(X-\theta/2)<0)$ equals 30.9\% when $|\theta|/\sigma=1$. The divergence of the plug-in action near $X \approx 0$ is visualized in the left panel of \cref{fig:rules_risk_m1}; the corresponding risk is infinite, and thus not shown in the right panel.

We next contrast the aggressive plug-in rule with the do-nothing rule, which simply sets $a(X)=0$. Appealingly, this ultra-conservative choice ensures that loss is uniformly controlled, since trivially $L(\theta,0) = 1$ for all $\theta$. Indeed, doing nothing is \emph{minimax-optimal}: no rule $a$ can have better worst-case risk than doing nothing, since $\sup_\theta R(\theta,a) \geq R(0,a) = 1$ \citep{Berliner1983}. Less appealingly, the exclusive focus on the worst case leads the do-nothing rule to ignore the signal $X$. This needlessly sacrifices performance outside of the worst case, as we shall see. The left panel of \cref{fig:rules_risk_m1} shows the do-nothing rule as the flat line, and the right panel correspondingly plots its constant risk of $1$.

Finally, the cautious rule dampens the aggressive plug-in rule, thus bringing the resulting actions closer to the ultra-conservative do-nothing rule. The amount of dampening depends on the estimated signal-to-noise ratio $X/\sigma$, as visualized in the left panel of \cref{fig:rules_risk_m1}:
\[a_C(X) = \frac{X}{X^2 + \sigma^2} = \frac{(X/\sigma)^{-2}}{1+(X/\sigma)^{-2}} \times 0 + \frac{1}{1+(X/\sigma)^{-2}} \times a_P(X).\]
It can be shown that this dampening of the plug-in rule necessarily reduces the probability of counterproductive action, i.e., $P_\theta(L(\theta,a_C)>L(\theta,0)) \leq P_\theta(L(\theta,a_P)>L(\theta,0))$ for all $\theta$ (see \cref{subsec:plugin}). Importantly, the dampening is particularly strong for small signals $X \approx 0$, which is precisely the region of the sample space where the plug-in rule behaves erratically. A key implication is that the cautious rule is bounded, $|a_C(X)| \leq 1/(2\sigma)$ for all $X$ (see \cref{lem:cautious_bound}(ii) in \cref{app:lemmas}), and attains finite risk. In fact, the right panel of \cref{fig:rules_risk_m1} suggests that the cautious rule here \emph{dominates} doing nothing in terms of risk, which is indeed true in this special case with $m=k=1$, as shown by \citet{Berliner1983}. The worst-case relative risk of 1 is attained at $\theta = 0$; at this point, the cautious rule achieves the same (minimax) risk as doing nothing. Hence, in settings with only one outcome variable, the cautious rule has a lot of desirable properties: it is minimax-optimal (though not uniquely), and it dominates the plug-in and do-nothing rules.

\subsection{Two outcome variables} 
\label{subsec:intuition_m2}

We now turn to a model with $m=2$ outcome variables, and still $k=1$ policy instrument.\footnote{\citet[pp.\ 14--15 and 21]{Fisher1962} also provides some graphical intuition for this case.} As we discussed in \cref{subsec:econ_examples}, the two outcomes could correspond to two different economic variables (e.g., unemployment and inflation), or alternatively to dynamic outcomes at two horizons of the same variable. We will see that there are some meaningful differences compared to the $m=1$ case. Importantly, as we will show later, the qualitative results for the toy model of this section hold very generally across choices of $(k,y,W,\Sigma)$ and $m \geq 2$; in fact, as long as $k=1$ (one policy instrument), the worst-case risk results are also \emph{quantitatively} insensitive to $(m,y,W,\Sigma)$.

Let the target values be $y=(1,1)'$ and loss weights $W=I_2$, yielding the loss function
\[L\big((\theta_1,\theta_2),a\big) = (1 - a\theta_1)^2 + (1-a\theta_2)^2.\]
The oracle action $a_\theta = (\theta_1^2+\theta_2^2)^{-1}(\theta_1+\theta_2)$ trades off the two targets, due to the availability of only a single policy instrument. Furthermore, we assume the signals about the two causal effects are two independent Gaussians:
\[X_1 \sim N(\theta_1,\sigma^2),\quad X_2 \sim N(\theta_2,\sigma^2),\quad X_1 \independent X_2.\]

The properties of the do-nothing and plug-in rules in this extended two-outcome model largely echo those of the above single-outcome case. The do-nothing rule is still minimax, for the same reason as before. The plug-in rule $a_P(X) = (X_1^2+X_2^2)^{-1}(X_1+X_2)$ again blows up to $\pm \infty$ near the origin $X_1 \approx X_2 \approx 0$, which leads to infinite risk: $R(\theta,a_P) = \infty$ for all $\theta \neq (0,0)$, regardless of $\sigma>0$ (see \cref{prop:plugin_inf}). And finally, the cautious rule
\[a_C(X) = \frac{X_1+X_2}{X_1^2+X_2^2 + 2\sigma^2} = \frac{X_1^2+X_2^2}{X_1^2+X_2^2 + 2\sigma^2} \times a_P(X)\]
still dampens the plug-in rule. As a result, the cautious rule continues to be bounded (see \cref{lem:cautious_bound}(ii) in \cref{app:lemmas}) and has a lower probability of counterproductive action than the plug-in rule (see \cref{subsec:plugin}).

\begin{figure}[t]
\centering
\includegraphics[width=0.75\linewidth]{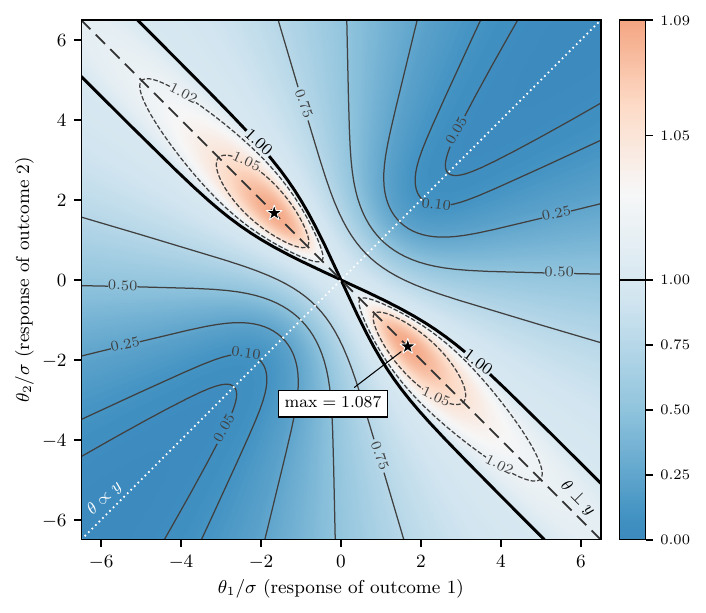}
\caption{Contour plot of the risk of the cautious rule relative to doing nothing, $R(\theta,a_C)/R(\theta,0)$. Gaussian model with $m=2$, $k=1$, $\Sigma=\sigma^2 I_2$, $y=(1,1)'$, and $W=I_2$. Risk computed numerically via quadrature using the formula in \cref{app:prop:cautious_risk_spherical}.}
\label{fig:risk_contour}
\end{figure}

A key difference compared to the single-outcome case is that the cautious rule no longer dominates doing nothing. To illustrate this, \cref{fig:risk_contour} shows a contour plot of the risk of the cautious rule relative to doing nothing, $R(\theta,a_C)/R(\theta,0)$. Unlike in the case of a single outcome, the cautious rule now has risk that \emph{exceeds} the minimax value $R(\theta,0) = 2$ in a narrow region around the ray marked ``$\theta \perp y$,'' where $\theta_2=-\theta_1$. Intuitively, on this ray, the policy instrument is useless, as the instrument moves the two targets in opposite directions with equal magnitude; hence, any improvement in hitting one target must come at the expense of missing the other. Indeed, the loss on this ray equals
\[L\big((\theta_1,-\theta_1),a\big) = (1 - a\theta_1)^2 + (1+ a\theta_1)^2 = 2 + 2a^2\theta_1^2,\]
and the oracle action is to do nothing ($a_\theta=0$). \emph{Any} data-dependent policy rule, including the cautious rule, will thus be outperformed by the do-nothing rule in at least a narrow slice around the ``$\theta \perp y$'' ray, since such a rule can only contribute unhelpful noise at these parameters. Indeed, doing nothing $a(X)=0$ is \emph{uniquely minimax-optimal} in this model (see \cref{prop:minimax} below).\footnote{\emph{A priori} restrictions on the parameter space could in principle reduce the minimax value. However, \cref{app:do_nothing} shows that for such restrictions to overturn the minimax-optimality of the do-nothing rule, the researcher would require quite detailed knowledge of causal effects. For example, \cref{fig:risk_contour} suggests that one would have to rule out any causal effects close to the problematic ``$\theta \perp y$'' ray.} Whereas \citet{Brainard1967} described the rule $a_C(X)$ as a more cautious variant of the plug-in rule $a_P(X)$, we have seen that even more caution is warranted if one adopts a worst-case perspective rather than a diffuse prior.

Crucially, however, this downside of the cautious rule relative to doing nothing is quite tightly limited, while its upside is unlimited. \cref{fig:risk_contour} shows that the maximum excess risk $R(\theta,a_C)/R(\theta,0)-1$ of the cautious rule equals a mere 8.7\% in the present model. On the other hand, the plot also shows that the risk of the cautious rule falls much below the do-nothing risk in a large part of the parameter space, especially along the ``$\theta \propto y$'' ray where the causal effects are perfectly aligned with the targets. Hence, the dampening of the plug-in rule is successful at limiting excessive sensitivity to the data for the adverse parameters on the ``$\theta \perp y$'' ray, and yet also sufficiently responsive to the data on the favorable ``$\theta \propto y$'' ray. A related advantage of the cautious rule is that it adapts to the strength of the signal: as $\sigma^2 \to 0$ for fixed $(\theta_1,\theta_2)$, its risk converges to the oracle loss (see \cref{prop:cautious_risk_best_case}).

\subsection{Summary}
\label{subsec:summary}

In the two toy models above, the cautious rule---a generalized Bayes rule that is familiar from the economic control literature---is an appealing practical choice: it has bounded worst-case risk often close to the minimax value, unlike the plug-in rule, and it substantially outperforms doing nothing over most of the parameter space. In the next section we argue that these properties hold generally.

\section{Main results}
\label{sec:results}

In this section we formalize the intuition from \cref{subsec:intuition_m1,subsec:intuition_m2,subsec:summary} in the general set-up of \cref{sec:framework}. First, we document some unfavorable properties of the plug-in rule. Second, we give conditions under which the do-nothing rule is uniquely minimax- and minimax-regret-optimal, as well as admissible. Third, we analyze the worst-case and best-case risk of the cautious rule.

Motivated by the applications in \cref{subsec:econ_examples} and \cref{sec:sim}, we focus on settings with $m>1$ outcomes, and we will pay particular attention to the case of a single policy instrument, $k = 1$. However, we will also comment on the $m=1$ case, which has been analyzed more thoroughly in the preceding literature.

\subsection{Properties of the plug-in rule}
\label{subsec:plugin}

The plug-in rule is a natural starting point that is common in applied work. We will see that the rule's ignorance of the statistical uncertainty in the data causes it to sometimes act too aggressively, resulting in undesirable decision-theoretic properties.

We begin with the undesirable risk properties: if $m$ is small relative to $k$, then plug-in rules have infinite risk; and even with large $m$, the risk is not uniformly bounded over $(\theta, \Sigma)$. Throughout we assume Gaussian data.

\begin{prop}[Plug-in rule has unbounded risk]
\label{prop:plugin_inf}

Consider the case of Gaussian data \eqref{eqn:X_gauss}. 

\begin{enumerate}[(i)]

\item If $k \leq m \leq k + 1$, then $R(\theta, a_P) = \infty$ for every $\theta \neq 0$. If instead $m \geq k + 2$, then $R(\theta, a_P) < \infty$ for every $\theta$.

\item For any $\varepsilon>0$, let $\mathcal{M}_\varepsilon$ denote the space of symmetric positive definite $mk \times mk$ matrices $\Sigma$ satisfying $\tr(\Sigma)\leq \varepsilon$, and let $R_\Sigma(\theta,a)$ denote the risk of rule $a$ when the covariance matrix in \eqref{eqn:X_gauss} equals $\Sigma$. Then, for any $m \geq k$,
\begin{equation*}
\sup_{\Sigma \in \mathcal{M}_\varepsilon} \sup_{\theta \in \mathbb{R}^{m \times k}} R_\Sigma(\theta, a_P) = \infty.
\end{equation*}

\end{enumerate}

\end{prop}

\begin{proof}
See \cref{app:prop:plugin_inf}.
\end{proof}

As already illustrated in \cref{sec:intuition}, the problem with plug-in rules is that they are unbounded. If the targets $y$ and data $X$ are aligned, and the magnitude of $X$ is small, then the plug-in rule suggests a very large action. Part (i) of \cref{prop:plugin_inf} shows that if observations near zero are frequent enough---as is the case when $m$ is sufficiently small relative to $k$---then the risk is infinite.\footnote{Both illustrative examples in \cref{sec:intuition} satisfy $k=1 \leq m \leq 2$.} For large $m$, the risk is finite. However, part (ii) of \cref{prop:plugin_inf} shows that, even in this case, there always exists a correlation structure $\Sigma$ for which the worst-case risk is arbitrarily large. Intuitively, if the noise structure is close to having reduced rank, then it is as if the target dimension $m$ is small, so risk diverges. Importantly, this is so even if we place an arbitrarily small bound $\varepsilon$ on the overall amount of noise, $\tr(\Sigma)$.

A second unappealing property of the plug-in rule is that, precisely because it prescribes aggressive actions, it will frequently yield a larger realized loss than that of doing nothing. Indeed, when $k=1$, the probability of counterproductive actions uniformly (weakly) exceeds that of the cautious rule:
\begin{equation*}
P_\theta\Big(L(\theta,a_C(X))>L(\theta,0)\Big) \leq P_\theta\Big(L(\theta,a_P(X))>L(\theta,0)\Big)\quad \text{for all } \theta \in \mathbb{R}^m.
\end{equation*}
This property is implied by a general result on the benefits of dampening: if $a(X)$ is any policy rule, then a dampening of that rule, defined as $\tilde{a}(X) \equiv \phi(X)a(X)$ with $\phi \colon \mathcal{X} \to [0,1]$ a measurable function, delivers $L(\theta,\tilde{a}(X)) \leq L(\theta,0)$ whenever $L(\theta,a(X)) \leq L(\theta,0)$.\footnote{This follows from the fact that the function $a \mapsto L(\theta,a)$ is convex and thus quasi-convex. Hence, $L(\theta,\phi a) \leq \max\lbrace L(\theta,0),L(\theta,a)\rbrace$ for $\phi \in [0,1]$.} The above conclusion then follows from the fact that the cautious rule is a scalar dampening of the plug-in rule when $k=1$: $a_C(X) = \lbrace X'WX/(X'WX+C) \rbrace \times a_P(X)$ with $C = \tr(W\Sigma)>0$.

Our simulations in \cref{sec:sim}, based on an empirical application to optimal monetary policy, will show that both of these concerns actually matter in practice: the plug-in rule tends to have high mean losses that are driven by adverse tail outcomes, and often prescribes counterproductive actions.

\subsection{Minimax-optimality and admissibility of doing nothing}
\label{subsec:minimax}

Since the plug-in rule has unbounded worst-case risk, one may ask which rule provides the best risk guarantee. We will now give weak conditions on the parameter space under which the do-nothing rule is uniquely minimax-optimal (for $m>1$) and uniquely minimax-regret-optimal (for $m,k>1$); it is also admissible in the former case. Though doing nothing is arguably too cautious, our results imply that the do-nothing rule is a relevant benchmark against which to measure other rules.

The minimax-optimality of the do-nothing rule is a consequence of the fact that, since it discards the data, it has constant loss $L(\theta,0)=\nW{y}^2$ regardless of $\theta$ and $\Sigma$. Define the set $\mathcal{S} \equiv \lbrace \theta \in \mathbb{R}^{m \times k} \colon \theta'Wy=0 \rbrace$ of policy effect matrices $\theta$ that are orthogonal to the targets $y$ (with respect to the $W$-weighted inner product).\footnote{$\mathcal{S}$ corresponds to the ``$\theta \perp y$'' ray in \cref{fig:risk_contour}.} For any $\theta \in \mathcal{S}$, the policy causal effects are so misaligned with the targets that any improvement along one target dimension must necessarily worsen performance along another dimension. Hence, for such unfavorable $\theta$, the oracle action \eqref{eqn:oracle_action} is to do nothing, $a_{\theta}=0$; any non-trivial rule can at best contribute noise, causing risk to weakly exceed the do-nothing value on $\mathcal{S}$, as illustrated in \cref{subsec:intuition_m2}.

In fact, when $m>1$ and weak conditions hold, doing nothing is \emph{uniquely} minimax-optimal as well as admissible. While there exist other decision problems where the minimax rule ignores the data---see \citet{ParmigianiTD}, \citet{Manski2004}, \citet{MontielOlea2026}, and the references therein---it is perhaps more surprising that the do-nothing rule is also \emph{minimax-regret-optimal} when $m,k>1$.

We now formally state these results. Some of the results require the data to be absolutely continuous. 

\begin{asn} \label{asn:density}
The distribution $P_\theta$ of $X$ has density $f_\theta(x)$ with respect to Lebesgue measure on a Euclidean space $\mathcal{X}$. $f_\theta(x)$ is everywhere positive and continuous in $\theta$ for every $x$.
\end{asn}

\begin{prop} [Doing nothing is minimax-optimal]
\label{prop:minimax}

Assume that $\Theta \cap \mathcal{S} \neq \emptyset$. Then:

\begin{enumerate}[(i)]

\item The minimax value of the decision problem equals
\[\inf_a\sup_{\theta\in\Theta}R(\theta,a)=\nW{y}^2,\]
attained by $a=0$.

\end{enumerate}

\noindent Now impose \cref{asn:density}, and assume that $m>1$ and $\Theta$ contains a neighborhood of the origin 0. Then:

\begin{enumerate}[(i)]
\setcounter{enumi}{1}

\item Doing nothing, $a=0$, is \emph{uniquely} minimax (up to a null set): any policy rule $\tilde{a}(X)$ that satisfies $\sup_{\theta\in\Theta}R(\theta,\tilde{a}) \leq \nW{y}^2$ must equal 0 almost everywhere.

\item Doing nothing, $a=0$, is admissible: no rule $\tilde{a}(X)$ satisfies $R(\theta,\tilde{a}) \leq R(\theta,0)$ for all $\theta \in \Theta$ with strict inequality for some $\theta$.

\item Let $r(\theta,a) \equiv R(\theta,a) - V(\theta)$ denote the \emph{regret} of a rule $a=a(X)$, where we recall that $V(\theta)=L(\theta, a_{\theta})$ is the oracle loss. If $k>1$, doing nothing, $a=0$, is uniquely minimax-regret-optimal (up to a null set):
\[\inf_a\sup_{\theta\in\Theta}r(\theta,a)=\nW{y}^2,\]
and any policy rule $\tilde{a}(X)$ satisfying $\sup_{\theta\in\Theta}r(\theta,\tilde{a})\leq\nW{y}^2$ must equal 0 almost everywhere.

\end{enumerate}

\end{prop}

\begin{proof}
(i) $R(\theta,0)=\nW{y}^2$ for all $\theta$. Conversely, \cref{lem:loss_decomp} in \cref{app:lemmas} implies that $R(\theta,a) \geq \nW{y}^2$ for every $\theta \in \mathcal{S}$ and $a$. Since $\Theta \cap \mathcal{S} \neq \emptyset$, $\sup_{\theta\in\Theta}R(\theta,a)\geq\nW{y}^2$.

(ii)--(iv) follow from \cref{lem:nesting}(iii) and \cref{prop:minimax_general}(i)--(iv) in \cref{app:do_nothing}.
\end{proof}

\subparagraph{Remarks:}\noindent

\begin{enumerate}

\item As discussed earlier, when $\theta \in \mathcal{S}$, non-trivial policy rules only contribute noise. This noise is costless only when $\theta=0$. For $m>1$ there exist parameters in $\Theta \cap \mathcal{S}$ other than the origin, which is why doing nothing is \emph{uniquely} minimax-optimal in this case.

\item Since the do-nothing rule is admissible when $m>1$, policy-makers inevitably face a trade-off when deciding to act: no non-trivial rule can \emph{uniformly} improve on inaction.

\item For ease of exposition, parts (ii)--(iv) impose the assumption that the parameter space $\Theta \subset \mathbb{R}^{m \times k}$ for causal effects contains a ball around the origin $0_{m \times k}$. Arguably, this is a weak restriction in many applications in economics, as we can rarely rule out small causal effects \emph{a priori}. In fact, \cref{app:do_nothing} shows that the conclusions of the proposition go through under even weaker conditions on the parameter space, essentially just requiring that the set $\Theta \cap \mathcal{S}$ is not too small, topologically speaking.

\item When $m=1$ and the data are Gaussian, as in \cref{subsec:intuition_m1}, the do-nothing rule is not uniquely minimax-optimal and it is dominated by the cautious rule \citep{Berliner1983}. See \cref{subsec:cautious_bounds} below.

\item When proving minimax-optimality in parts (i)--(ii), it suffices to consider $\theta \in \mathcal{S}$. However, for the minimax regret result in part (iv) we consider small perturbations of parameters $\theta \in \mathcal{S}$ with reduced column rank. Such parameters exist when $k>1$. Due to reduced rank, we are able to construct the perturbations such that they span the target values $y$, implying that the oracle loss $V(\theta)$ equals 0 at these perturbations. Thus, at these parameters the regret equals the risk, and the latter is lower-bounded by the minimax value attained on $\mathcal{S}$ in the limit as the perturbations become small. Said more simply, one can find parameters very close to the set $\mathcal{S}$ that are very favorable for the oracle but unfavorable for any data-dependent rule. At those parameters, large actions are required to achieve low regret, but those large actions would lead to large losses on the nearby, and statistically indistinguishable, misaligned $\theta$'s.

\end{enumerate}

The above analysis implies that worst-case risk or regret cannot be the sole criterion to select among non-trivial policy rules, except perhaps in special cases where detailed \emph{a priori} knowledge about the causal effects $\theta$ is available, allowing the researcher to severely restrict the parameter space.

\subsection{Bounds on the risk of the cautious rule}
\label{subsec:cautious_bounds}

Unlike the do-nothing and plug-in rules, the cautious rule is able to exploit information in the data without incurring potentially high risk. We begin by showing that the cautious rule has uniformly bounded excess risk relative to the do-nothing rule, with the bound very tight for the case with $k=1$ and Gaussian data. In other words, the potential downside of the cautious rule is limited, in sharp contrast with the plug-in rule. We then establish that the best-case risk is arbitrarily small when $m \geq k$, i.e., the cautious rule has unbounded upside relative to doing nothing; this follows from the more general result that the cautious rule adapts to settings with low noise relative to signal strength, unlike doing nothing.

\paragraph{The special case $m = 1$.}
In the case $m=1$, with Gaussian data \eqref{eqn:X_gauss} and unrestricted parameter space $\Theta = \mathbb{R}^{1 \times k}$, it follows from \citet[Theorem 2.3]{Berliner1983} that the cautious rule is minimax for any $k$.\footnote{The cautious rule corresponds to $a=b=1$ in \citeauthor{Berliner1983}'s notation. Note that \citeauthor{Berliner1983} normalizes $y=W=1$ and $\Sigma=I_k$ (in our notation), which is without loss of generality when $m=1$.} Thus, the cautious rule dominates the do-nothing rule in this case, since the risk function of the latter rule is constant and equal to the minimax value; in other words, there is actually no downside here.

That said, as already discussed in \cref{sec:intuition}, the case with $m=1$ target is quite special, since no non-trivial rule can be minimax for $m>1$ targets. It is also of somewhat limited practical relevance, as policy questions typically feature multiple targets---both different variables (e.g., unemployment and inflation), and different time horizons. We next characterize the trade-off in this more relevant case.

\paragraph{Limited downside.}
The worst-case risk of the cautious rule is bounded \emph{uniformly} in the unknown parameter $\theta$ and the correlation structure $\Sigma$, in contrast to the plug-in rule.

\begin{prop}[Limited downside of cautious rule] 
\label{prop:cautious_risk}
Let the cautious rule be defined by \eqref{eqn:cautious}, where $X \in \mathbb{R}^{m \times k}$ is a (not necessarily Gaussian) random matrix with $E_\theta[X]=\theta$ and $\var_\theta(\ve(X))=\Sigma$, and $\Sigma$ is positive definite. Then:
\begin{enumerate}[(i)]
\item For general $k \geq 1$,
\[\sup_{\theta \in \mathbb{R}^{m \times k}} \frac{R(\theta,a_C)}{R(\theta,0)} \leq \tfrac{1}{4}\left(1+\sqrt{1+k}\right)^2.\]
\item If $k=1$, and moreover $X-\theta$ and $\theta-X$ have the same distribution under $P_\theta$ for every $\theta \in \mathbb{R}^m$,
\[\sup_{\theta \in \mathbb{R}^m} \frac{R(\theta,a_C)}{R(\theta,0)}-1 \leq \tfrac{27}{160} = 0.16875.\]
\end{enumerate}
\end{prop}

\begin{proof}
See \cref{app:prop:cautious_risk_i} for part (i) and \cref{app:prop:cautious_risk_ii} for part (ii).
\end{proof}

Recall from \cref{subsec:minimax} that the do-nothing risk $R(\theta,0)$ equals the minimax value $\nW{y}^2$ for all $\theta$. The first part of the proposition states that the excess risk $R(\theta,a_C)/R(\theta,0)-1$ of the cautious rule relative to doing nothing is bounded by a finite constant that depends only on the number of instruments $k$, and not on the other parameters of the problem $(m,W,y,\Sigma)$. This general excess risk bound evaluates to 45.7\% for $k=1$ and 86.6\% for $k=2$---potentially non-trivial, but finite, in sharp contrast to the plug-in rule.

The second part of the proposition shows that the worst-case excess risk bound can be substantially tightened to 16.9\% when there is a single instrument, $k=1$, and the distribution of the noise $X-\theta$ is symmetric around zero (which includes the Gaussian case).

In the leading case of Gaussian data (corresponding to the limit experiment interpretation in \cref{subsec:problem}), extensive numerical and analytical results suggest that the bound can be tightened even further.

\begin{num} \label{num:cautious}

Assume Gaussian data \eqref{eqn:X_gauss} with $k=1$ and $m \geq 2$, and let $\mathcal{M}$ be the space of $m \times m$ symmetric positive definite matrices. Then 
\begin{equation}
\label{eqn:cautious_bound_numerical}
\sup_{\Sigma \in \mathcal{M},\theta \in \mathbb{R}^m} R_\Sigma(\theta,a_C)/R_\Sigma(\theta,0)-1 \approx 0.103859
\end{equation}
for any fixed $(W,y)$.\footnote{In $R_\Sigma(\theta,a_C)$, the rule $a_C$ is evaluated under the same $\Sigma$ as the expectation $E_\theta[\cdot]$.}

\end{num}

That is, for any $m \geq 2$, the \emph{sharp} uniform-in-$(\Sigma,\theta)$ bound on the excess risk of the cautious rule equals 10.4\% (up to rounding error). We support this claim in \cref{app:num_cautious} by establishing analytically that the supremum in \eqref{eqn:cautious_bound_numerical} is bounded above by the supremum of an explicit six-variable function over a bounded set (regardless of the dimension $m$). We compute the latter supremum numerically, using 1 million starting parameter values. At the numerical maximizer, we can prove that the analytical bound on the supremum in fact binds. Hence, provided that our six-variable numerical optimization has reached the global maximum, 10.4\% is the sharp bound.

While the results above show that the excess risk of the cautious rule is quite small regardless of the correlation structure $\Sigma$, the worst-case excess risk is even smaller for specific $\Sigma$. \cref{app:spherical} derives the exact worst-case risk for the case of spherical Gaussian noise and $k=1$: the worst-case excess risk equals 8.7\% for $m=2$, as in \cref{subsec:intuition_m2}, and quickly decays toward 0 for larger $m$.

\paragraph{Unlimited upside.}
Having established that the cautious rule has limited downside relative to doing nothing, we now show that the rule has unlimited upside.

\begin{prop}[Unlimited upside of cautious rule] 
\label{prop:cautious_risk_best_case}
Assume Gaussian data \eqref{eqn:X_gauss}. 
\begin{enumerate}[(i)]
\item Let $M$ be an $m \times k$ matrix of rank $k$ whose column space contains the target vector $y$ (and so implicitly $m \geq k$). Define the family of parameters $\theta_\delta \equiv \delta M$ for $\delta>0$. Then $R(\theta_\delta,a_C) \to 0$ as $\delta \to \infty$.
\item For any $\theta \in \mathbb{R}^{m \times k}$ with full column rank $k$, $R_\Sigma(\theta,a_C) \to V(\theta)$ as $\Sigma \to 0$.\end{enumerate}
\end{prop}

\begin{proof}
See \cref{app:prop:cautious_risk_best_case}. 
\end{proof}

The parameters $\theta_\delta$ in part (i) of \cref{prop:cautious_risk_best_case} correspond to the aligned ray ``$\theta \propto y$'' in \cref{subsec:intuition_m2}. As $\delta \to \infty$, the signal $\theta_\delta$ eventually dominates the noise in the data, and the cautious rule approaches the oracle. But since the target $y$ is contained in the column space of the parameters $\theta_\delta$, the oracle loss---and thus the limiting risk of the cautious rule---equals 0. By contrast, the do-nothing rule of course continues to incur risk equal to $\nW{y}^2$ even at this highly favorable parameter.

Part (ii) of \cref{prop:cautious_risk_best_case} shows that unlike the do-nothing rule, the cautious rule adapts to the noise in the data. Specifically, it achieves the oracle loss as the level of noise tends to zero. The unlimited upside of the cautious rule for fixed $\Sigma$ is the same fundamental phenomenon as the noise-adaptivity for fixed $\theta$: part (ii) implies part (i). This is because the risk of the cautious rule depends on the signal and noise levels only through their ratio, i.e., $R_{\Sigma}(\delta M,a_C) = R_{\Sigma/\delta^2}(M,a_C)$, see \cref{lem:normalization}(iii) in \cref{app:lemmas}.

\paragraph{Admissibility and average-risk-optimality.}
Since we are recommending the cautious rule, a natural question is whether it can be improved uniformly over the parameter space. \cref{app:admissibility} shows that the answer is negative in low-dimensional, Gaussian models: the cautious rule is admissible when $mk \leq 4$. In the classical control problem with $m=1$, it is known that $k \leq 4$ is both necessary and sufficient for admissibility of the cautious rule \citep[see][and references therein]{ZamanStein1981}.\footnote{In the $m=1$ case, characterizations of classes of admissible rules are given by \citet{Zaman1981} and \citet{BergerBerlinerZaman1982}. \citet[Theorem 2.4]{Berliner1983} shows that when $m=1$ and $k \geq 5$, the plug-in rule actually dominates the cautious rule.} It is an open question whether the cautious rule is dominated for $m>1$ and $k>4/m$.

An auxiliary result for our admissibility proof establishes that the cautious rule is approximately Bayes-optimal---``$\varepsilon$-Bayes'' in the terminology of \citet[Chapter 1.6, Definition 2]{Ferguson1967}---for sufficiently diffuse Gaussian priors, regardless of $m,k$ (see \cref{lem:gauss} in \cref{app:prop:adm}). We can interpret this result as saying that not only is the cautious rule nearly optimal with respect to \emph{worst-case} risk when $k=1$ (as argued above), it is also nearly optimal with respect to \emph{average} risk, if we average using a diffuse Gaussian prior. Loosely speaking, the cautious rule is therefore nearly optimal in a \citet{HodgesLehmann1952} sense (though other rules may exist with similar properties).

\section{Bias vs.\ variance}
\label{sec:bias_var}

Which causal effect estimates $X$ should policy-makers rely on? In the familiar point estimation problem with squared error loss, variance and squared bias are weighted equally when ranking estimators. We now argue that in the present policy problem, more weight should be placed on bias than on variance, provided that the goal is to ensure good performance across a wide range of settings (i.e., limit the worst-case risk). Subsequently, we connect this discussion to the choice in macroeconometrics between two popular estimators of dynamic policy effects: local projections (LPs) and vector autoregressions (VARs).

\paragraph{The role of variance.}
In our policy decision problem, the overall amount of estimation noise---i.e., the \emph{scale} of $\Sigma$---does not affect the worst-case risk of any of the candidate policy rules in \cref{subsec:policy_rules}. For all these rules $a(\cdot)$, only the signal-to-noise ratio matters for their risk: for any $\delta>0$, $R_{\delta^2\Sigma}(\theta,a) = R_{\Sigma}(\theta/\delta,a)$.\footnote{This is immediate for the do-nothing and plug-in rules, and is implied by \cref{lem:normalization} in \cref{app:lemmas} for the cautious rule. Recall the convention that the formula for the cautious rule $a_C$ is evaluated at the true $\Sigma$.} Hence, $\sup_{\theta \in \Theta} R_{\delta^2 \Sigma}(\theta,a)$ does not depend on $\delta$ if $\Theta$ is a cone. Of course, the risk does depend on the scale of $\Sigma$ at a \emph{fixed} parameter $\theta$ (recall \cref{prop:cautious_risk_best_case}(ii)), but if $\Sigma$ is scaled up, then the \emph{worst-case} parameter $\theta$ just moves away from the origin, and the resulting worst-case risk remains unchanged.

While the scale of $\Sigma$ does not affect worst-case risk, its \emph{orientation} relative to the target vector $y$ could matter, but not by much in the case of the cautious rule when $k=1$. This is a direct consequence of the analysis in \cref{subsec:minimax,subsec:cautious_bounds}: the cautious rule has worst-case risk near the minimax value, and the latter does not depend on $\Sigma$ at all. Hence, the variance of the causal effect estimator $X$ is not of crucial importance to policy-makers who adopt the cautious rule and are primarily concerned about worst-case performance.

\paragraph{The role of bias.}
In contrast to the variance, any potential \emph{bias} in the estimator $X$ of the policy causal effects $\theta$ can have much more serious consequences for down-stream policies. This is seen most starkly by revisiting the simplest model with $m=k=1$ from \cref{subsec:intuition_m1}. We saw that in the absence of bias, the cautious rule dominates doing nothing. Yet the following result shows that even an arbitrarily small amount of potential bias can render the do-nothing rule uniquely minimax-optimal.
\begin{prop}[Biased signal]
\label{prop:risk_bias_m1}
Consider a biased Gaussian signal $X \sim N(\theta+b,\sigma^2)$ with $m = k = 1$. Normalize $y=W=1$ without loss of generality. Let $R(\theta,b,a) \equiv E_{\theta,b}[(y-\theta a(X))^2]$ denote the risk of a rule $a(\cdot)$, and let $r(\theta,b,a)\equiv R(\theta,b,a)-V(\theta)$ denote the regret. Assume that the parameter space $\Theta$ for $\theta$ contains an interval around the origin, and let the parameter space for bias $b$ be the interval $[-K,K]$ for some $K>0$. Then doing nothing $a=0$ is \emph{uniquely} minimax-optimal and minimax-regret-optimal:
\begin{equation*}
\inf_a \sup_{\theta \in \Theta, |b|\leq K} R(\theta,b,a) = \inf_a \sup_{\theta \in \Theta, |b|\leq K} r(\theta,b,a) = 1,
\end{equation*}
and any rule attaining either infimum equals $0$ almost everywhere.
\end{prop}

\begin{proof}
See \cref{app:prop:risk_bias_m1}.
\end{proof}

The proposition reveals that a potentially biased signal is entirely worthless in the minimax sense: ignoring it attains the no-information benchmark, while using it non-trivially will lead to strictly larger worst-case risk. The logic of its proof is the same logic that made doing nothing uniquely minimax for $m > 1$ in \cref{prop:minimax}: there exist parameter configurations at which acting is useless but costly. The only difference is that now it is the bias, and not the geometry of multiple targets, that renders the signal useless.

For $m > 1$, doing nothing is uniquely minimax already without bias (\cref{prop:minimax}), and it trivially remains so with bias. The relevant question for our recommended cautious rule is thus how quickly its risk increases as a function of the bias. We show in \cref{app:cautious_bias} that, when $k=1$, the worst-case risk of the cautious rule is of first order in the amount of potential bias. Intuitively, at the least favorable parameter $\theta$, the policy instrument is nearly useless, but even a small bias can make the signal appear to be aligned with the targets, thus inducing the policy-maker to act too aggressively. It follows that an estimator that is preferred on mean-squared-error grounds can be inferior when used as an input to policy, unless the potential for bias is extremely limited.

\paragraph{Implications for the choice between LPs and VARs.}
The preceding discussion has immediate implications for the question of how to choose between LP and VAR impulse response estimators when the ultimate goal is to inform dynamic policy-making, as in \cref{exm:macro}. \citet{MPQW2026} prove that in realistic data-generating processes (DGPs), LPs and VARs have opposite (asymptotic) bias-variance properties: LPs have low bias but high variance, while VARs (with conventional short lag lengths) have low variance but potentially non-trivial bias due to small amounts of dynamic misspecification. For squared error estimation loss, \citet{Li2024} argue through simulations with empirically calibrated DGPs that the variance gain of VARs is so large that they tend to be preferred over LPs despite the potentially higher bias. In the policy decision problem, however, a policy-maker who is concerned about worst-case performance will typically prefer to observe LP estimates rather than VAR estimates, since in this problem LP's bias advantage matters much more than VAR's variance advantage. For VAR estimates to deliver similar worst-case guarantees to LP estimates, one would need to condition on many more lags than is common in the literature \citep[cf.][]{MPQW2026}.

\section{Application and simulations}
\label{sec:sim}

We demonstrate the practical relevance of our results with an application to optimal monetary policy, building on \citet{Barnichon2023}. We first map their analysis into our statistical decision problem. Then we show that our preferred cautious rule gives substantively different policy recommendations than their plug-in approach. Finally, we calibrate a simulation study to the application to illustrate the virtues of the cautious rule.

\subsection{Optimal monetary policy}
\label{subsec:bm_replication}

We provide a brief sketch of the setting for the application, which closely follows the analysis of \citet{Barnichon2023}. Further details are provided in \cref{app:simulations}.

\paragraph{Decision problem.}
We consider the problem of a monetary policy-maker, mapped into our statistical decision problem using the arguments in \cref{subsec:econ_examples}. The policy-maker has two objectives, inflation and unemployment (both in deviations from some target, as specified below) over the next five years. The vector $y$ stacks forecasts for those two variables up to 19 quarters out, yielding a target dimension of $m = 2 \times 20 = 40$. The baseline forecasts are constructed under the current expected policy stance. We furthermore assume equal weights and no discounting, so $W = I$.

The policy-maker can impact the evolution of her objectives by moving her policy instruments, i.e., current and expected future interest rates. We restrict attention to a single instrument ($k = 1$), corresponding to a one-off conventional monetary policy ``shock'' to the policy rate today. The policy-maker's problem is to choose the scalar policy action to minimize her expected loss given the forecasts $y$ and given a statistical signal about the dynamic causal effects $\theta$ of the policy, both to be discussed next.

\paragraph{Measurement.}
Our measurement of both baseline forecasts $y$ and policy causal effects $\theta$ closely follows \citet{Barnichon2023}. Beginning with forecasts, we follow those authors and use the Summary of Economic Projections as well as the Monetary Policy Report to construct point forecasts, with the target values constructed as (real-time) long-run unemployment and inflation projections. This gives forecasts at each date ranging from June 1980 to December 2022, delivering 90 forecast vintages overall. We compute optimal one-off policy shocks for each date in this forecast sample, with the decision at a given date being completely independent of decisions at other dates. This approach is in keeping with the analysis of \citeauthor{Barnichon2023}, and it should be interpreted as delivering a sequence of optimal one-off policy decisions for a range of practically relevant policy target vectors $y$, and \emph{not} as a sequence of dynamic optimal policy decisions.\footnote{For the full dynamic optimal policy problem, the effect of lagged decisions on future forecasts should also be taken into account. Doing so is conceptually straightforward \citep[see][]{Caravello2026}, but for maximal comparability we here stay close to the analysis of \citet{Barnichon2023}.}

Turning to estimation of policy causal effects, we restrict attention to a single, conventional monetary policy shock, identified following \citet{Eberly2020}. Alternative shock measures, e.g., ``slope'' shocks to the yield curve, are only estimated over much shorter samples and thus excessively noisy. We estimate the causal effects of the rate shock using LPs, delivering point estimates and estimated asymptotic covariance matrices, $(X, \Sigma)$. Following \citeauthor{Barnichon2023}, we use the full available estimation sample, instead of only using data up to the forecast date used in defining each target vector $y$; the resulting estimated monetary policy shock impulse response functions have the typical textbook shapes and are displayed in \cref{app:simulations}. As expected, our LP impulse response estimates are somewhat less precise than those from the Bayesian VAR of \citeauthor{Barnichon2023}.\footnote{As detailed in \cref{app:simulations}, our LPs control contemporaneously for unemployment and inflation. This aligns with the recursive variable ordering in the replication code of \citeauthor{Barnichon2023}.}

\begin{figure}[tp!]
\centering \vspace{0.5\baselineskip}
\includegraphics[width=0.8\textwidth]{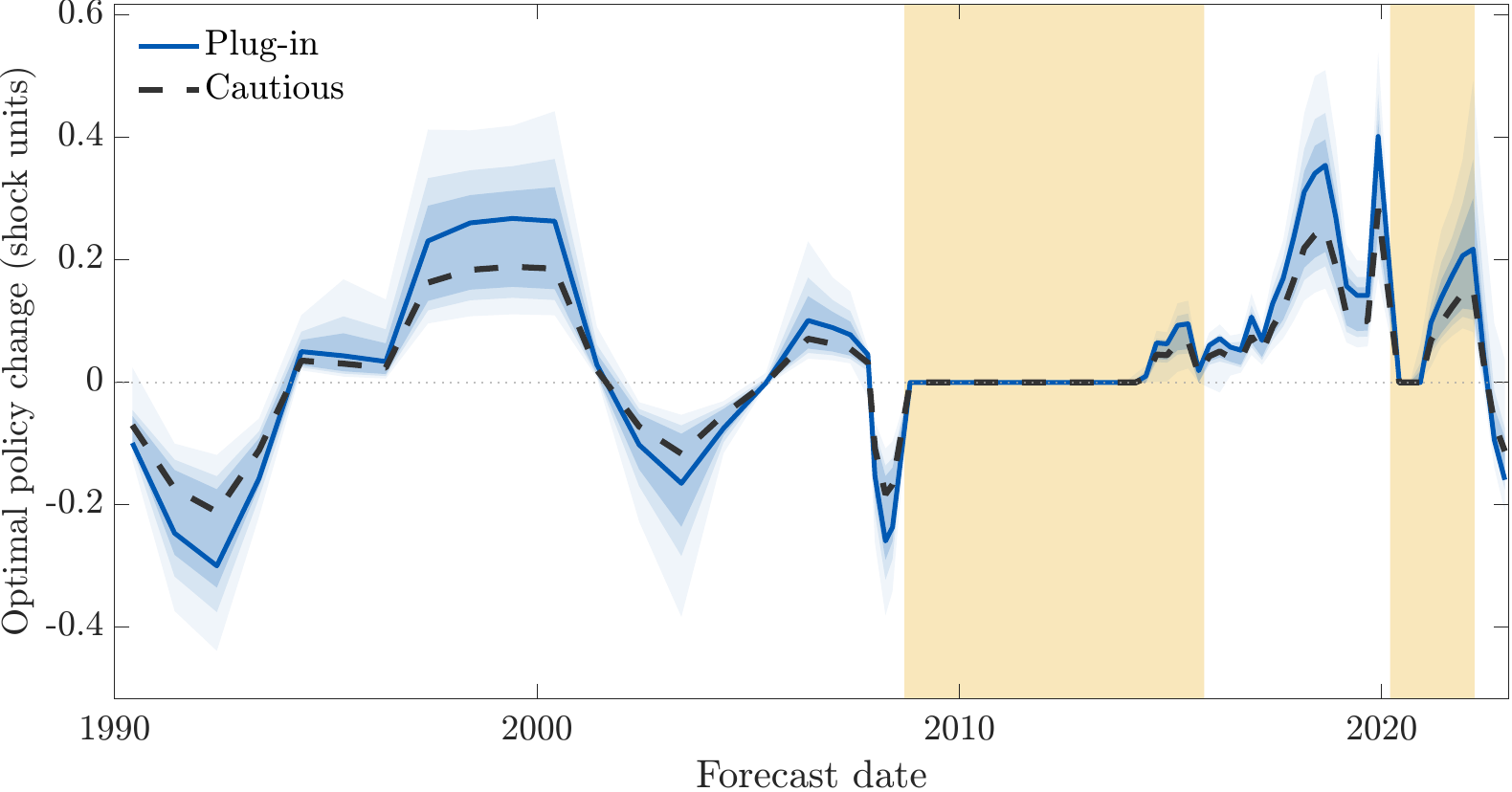}
\caption{Plug-in (colored solid) vs.\ cautious (black dashed) policy decisions based on LP estimates, with the blue shaded areas indicating 60, 75, and 90 percent equal-tailed posterior credible bands for the oracle action. The yellow areas indicate a binding zero lower bound on the federal funds rate. Forecast dates as in \citet{Barnichon2023}.}
\label{fig:opp_cautious}
\end{figure}

\paragraph{Dampening of the cautious rule.}
\cref{fig:opp_cautious} compares the policy actions implied by the plug-in and cautious rules, respectively. The horizontal axis indicates the forecast date used to construct the target vector $y$; hence, each point on the horizontal axis corresponds to a separate policy problem with different $y$ but the same $X$. Compared to the plug-in actions (blue line), the cautious rule (black dashed line) implies non-trivial dampening, but this dampening is not excessive---$a_C(X)$ is away from zero on most dates. The uncertainty bands (light blue areas) are posterior credible bands for the oracle action $a_\theta$ under a flat prior, as in \citet{Barnichon2023}. These bands are often wide, which is why the cautious rule dampens the plug-in rule. A quantitative measure of the amount of dampening is the ratio of the plug-in to cautious rule denominators: $(X'WX)/(X'WX + C)$. At the least-squares point estimates $X$, this ratio equals $0.706$, corresponding to the observed non-trivial, but not excessive, dampening.

\subsection{Calibrated simulation study}
\label{subsec:sim_cautious}

Whereas the preceding analysis revealed that the plug-in and cautious decisions differ materially in an application, it is necessarily silent on the relative statistical performance of the two rules. For that we now turn to a calibrated simulation study.

\paragraph{Set-up.}
We consider a simple simulation DGP: a reduced-form VAR estimated on the history of unemployment, inflation, the federal funds rate, and the measured monetary policy shock of \citeauthor{Eberly2020}. We then use that estimated VAR (at the least-squares point estimates) to define the ground truth monetary policy causal effects $\theta$, simulate 10,000 random samples from this DGP (assuming normal errors, and with the same sample length as in the original empirical exercise), and estimate the monetary shock local projections on those samples. Finally, for each artificial sample, we construct the plug-in and cautious decisions and evaluate the policy losses using the population ground truth. We evaluate the loss for each target vector $y$ (fixed at the values observed in the real data) at each forecast date. We then compute the difference between plug-in and cautious rule losses, normalized by the do-nothing loss.\footnote{Note that we depart from \citeauthor{Barnichon2023} and consider their full sample (from 1980 onwards). We do so since, for the purpose of illustrating the behavior of the different decision rules, having a large range of forecasts $y$ is useful. We also do not impose any truncation related to the zero lower bound, as such truncation, while clearly relevant in practice, would obscure the desired comparison.} Supplementary implementation details are reported in \cref{app:simulations}.

\paragraph{Results.}
\begin{figure}[tp!]
\centering \vspace{0.5\baselineskip}
\includegraphics[width=\textwidth]{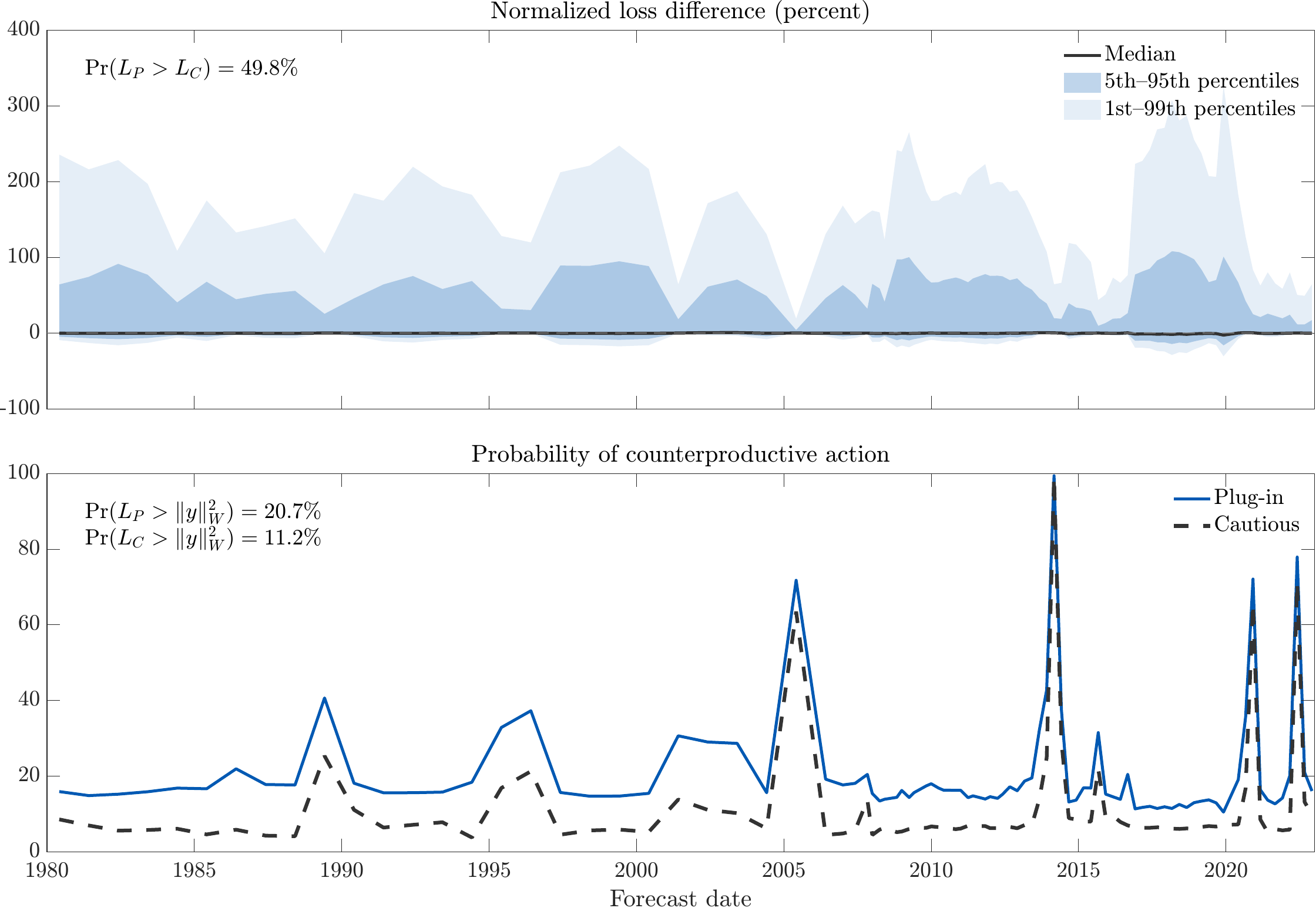}
\caption{Relative performance of the plug-in and the cautious rules, based on LPs estimated on artificial samples from the ground truth VAR(5). Top panel: normalized (by $\nW{y}^2$) loss difference of plug-in vs.\ cautious, median (solid black) and percentiles (shaded blue). Bottom panel: probability of an action that leads to a loss in excess of do-nothing, across samples.}
\label{fig:cautious_loss_var_simul}
\end{figure}

\cref{fig:cautious_loss_var_simul}, which displays the main results from our simulation exercise, bears out, and provides a further quantitative assessment of, our theoretical results. The top panel reports medians as well as percentile ranges of the normalized loss difference. The median loss difference of the plug-in relative to the cautious rule is essentially zero, with the cautious loss smaller 49.8\% of the time. More importantly, the figure clearly reveals the insurance value of the cautious rule, visible in the distributional percentiles: the distribution is heavily skewed, with a very long right tail in the relative loss distribution. In fact, at the 99th percentile, the plug-in rule can have excess losses that are \emph{multiples} of the do-nothing loss, consistent with our theoretical analysis in \cref{subsec:plugin}: in those cases, policy causal effects estimates are relatively small, and so the plug-in rule suggests very large actions. This asymmetry, and the associated insurance value, is also reflected in mean losses, with the pooled mean normalized loss difference equal to 10.2\% of the do-nothing loss in favor of the cautious rule. Finally, the bottom panel provides a different perspective on the same message, showing the probability of an action (for each date, across all the simulations) that leads to a loss larger than that of do-nothing. This probability is always larger for the plug-in rule, often by a factor of two, consistent with the discussion at the end of \cref{subsec:plugin}.

To further interpret the magnitudes of the cautious rule's insurance value and mean loss reduction, we use the central bank's objective function to re-express the loss reduction in terms of equivalent inflation or unemployment gaps. Specifically, we solve for the size of a persistent inflation or unemployment gap that would induce the same loss as the plug-in rule's increase in expected loss (averaged across all dates) over the cautious rule. This calculation shows that the cautious rule's loss reduction is equivalent to the loss from suffering an inflation or unemployment gap of 0.47 percentage points that persists for five years, an economically meaningful amount.

\section{Conclusion}
\label{sec:concl}

In this paper we revisited the classic problem of a policy-maker setting her $k$ instruments to best hit $m$ targets, in an environment where the causal effects of the policies are measured with noise. We argued that this framework can be used to study modern policy problems in various areas of economics, where optimal policies are expressed as functions of estimable ``sufficient statistics.'' The plug-in rule that ignores estimation uncertainty has unbounded worst-case risk and frequently prescribes counterproductive actions, while the (typically uniquely) minimax-optimal response of doing nothing sacrifices performance in large parts of the parameter space. A cautious rule---dampening the plug-in action in line with the estimation noise \citep{Fisher1962,Brainard1967,Lindley1968,Zellner1971}---strikes an appealing balance: its downside relative to doing nothing is uniformly bounded and small in empirically relevant settings, while its upside is unlimited.

We therefore recommend that policy-makers implement the cautious rule and avoid the plug-in rule. Ideally, the policy effect estimates should be obtained from an estimator with low bias (such as a local projection in the case of impulse response estimation). Estimators with lower dispersion but non-negligible potential bias (such as vector autoregressions with short lag lengths) will tend to yield higher risk in the worst case.

\clearpage

\appendix
\begin{appendices}
\crefalias{section}{appendix}
\crefalias{subsection}{subappendix}
\numberwithin{equation}{section}
\numberwithin{figure}{section}
\numberwithin{lem}{section}
\numberwithin{prop}{section}
\numberwithin{cor}{section}
\numberwithin{asn}{section}

\section{Additional results}

\subsection{Properties of the do-nothing rule}
\label{app:do_nothing}

Here we show that parts (ii)--(iv) of \cref{prop:minimax} in \cref{subsec:minimax} hold under weaker conditions on the parameter space $\Theta$. We consider the following conditions, only one of which is imposed at a time.
\begin{asn} \label{asn:parameter_space}
\begin{enumerate}[(a)]
\item \label{itm:parameter_space_i} $\Theta \cap \mathcal{S} \neq \emptyset$.
\item \label{itm:parameter_space_ii} There exists a finite number $J$ of parameters $\theta_1,\dots,\theta_J \in \Theta \cap \mathcal{S}$ such that the stacked $Jm \times k$ matrix $(\theta_1',\dots,\theta_J')'$ has full column rank $k$.
\item \label{itm:parameter_space_iii} $\Theta$ contains a neighborhood, open with respect to $\mathcal{S}$, of some $\theta_0 \in \mathcal{S} \setminus \lbrace 0 \rbrace$.
\item \label{itm:parameter_space_iv} $\Theta$ contains a neighborhood, open with respect to $\mathbb{R}^{m \times k}$, of some $\theta_0 \in \mathcal{S}$ with $0 < \rk(\theta_0) < k$.
\end{enumerate}
\end{asn}
Note that conditions (\ref{itm:parameter_space_i}) and (\ref{itm:parameter_space_ii}) allow $\Theta$ to be a finite set. Recall that condition (\ref{itm:parameter_space_i}) is imposed in \cref{prop:minimax}(i). It turns out that the four conditions above are nested:

\begin{lem} \label{lem:nesting}

\begin{enumerate}[(i)]

\item In \cref{asn:parameter_space}, we have (\ref{itm:parameter_space_iv}) $\Rightarrow$ (\ref{itm:parameter_space_iii}) $\Rightarrow$ (\ref{itm:parameter_space_ii}) $\Rightarrow$ (\ref{itm:parameter_space_i}).

\item \cref{asn:parameter_space}(\ref{itm:parameter_space_ii}) and (\ref{itm:parameter_space_iii}) can only hold if $m \geq 2$, while \cref{asn:parameter_space}(\ref{itm:parameter_space_iv}) can only hold if $m,k \geq 2$.

\item If $\Theta$ contains a neighborhood of 0 and $m \geq 2$, then \cref{asn:parameter_space}(\ref{itm:parameter_space_ii}) holds. If moreover $k \geq 2$, then \cref{asn:parameter_space}(\ref{itm:parameter_space_iv}) holds.

\end{enumerate}

\end{lem}
\begin{proof}
See \cref{app:lem:nesting}.
\end{proof}

We now state our generalization of \cref{prop:minimax}(ii)--(iv).

\begin{prop} \label{prop:minimax_general}

Impose \cref{asn:density}.

\begin{enumerate}[(i)]

\item If \cref{asn:parameter_space}(\ref{itm:parameter_space_ii}) holds, then $a=0$ is uniquely minimax up to a null set.

\item If \cref{asn:parameter_space}(\ref{itm:parameter_space_ii}) holds, then $a=0$ is admissible.

\item If \cref{asn:parameter_space}(\ref{itm:parameter_space_iv}) holds, then $\inf_a\sup_{\theta\in\Theta}r(\theta,a)=\nW{y}^2$, attained by $a=0$.

\item If \cref{asn:parameter_space}(\ref{itm:parameter_space_iv}) holds, then $a=0$ is uniquely minimax regret up to a null set.

\end{enumerate}

\end{prop}

\begin{proof}
See \cref{app:prop:minimax_general}.
\end{proof}

\subsection{Support for \texorpdfstring{\cref{num:cautious}}{Numerical result \ref{num:cautious}}}
\label{app:num_cautious}

Here we provide analytical and numerical support for \cref{num:cautious} in \cref{subsec:cautious_bounds}. Consider the case of Gaussian data \eqref{eqn:X_gauss} with $k=1$. Throughout this section, normalize $W = I_m$, $\tr(\Sigma) = 1$, and $\|y\| = 1$ without loss of generality (see \cref{lem:normalization} in \cref{app:lemmas}). Thus, $a_C(X) = X'y/(1+\|X\|^2)$.

We first show analytically that the worst-case cautious risk can be upper-bounded by a six-variable function. Write $x^\pm \equiv \max(\pm x,0)$.

\begin{prop} \label{prop:reduction}
Assume $m \geq 2$ and the Gaussian model \eqref{eqn:X_gauss}, except that $\Sigma$ need only be positive semidefinite with $y'\Sigma y>0$. Put $\tau_1 \equiv \theta'y$ and $\tau_2 \equiv \|\theta-\tau_1y\|$, and let $\vartheta$ be a unit-length vector that is orthogonal to $y$ and satisfies $\theta = \tau_1y+\tau_2\vartheta$. Define
\[\sigma_1^2 \equiv y'\Sigma y, \quad \sigma_2^2 \equiv \vartheta'\Sigma\vartheta, \quad \rho \equiv \frac{y'\Sigma\vartheta}{\sigma_1\sigma_2}, \quad \sigma_\perp^2 \equiv \frac{\|\Sigma y\|^2}{\sigma_1^2}-\sigma_1^2-\rho^2\sigma_2^2,\]
with $\rho \equiv 0$ if $\sigma_2 = 0$. Let $(\check{X}_1,\check{X}_2)$ be bivariate normal with means $(\tau_1,\tau_2)$, standard deviations $(\sigma_1,\sigma_2)$, and correlation $\rho$, let $\zeta \equiv (\check{X}_1-\tau_1)/\sigma_1 \sim N(0,1)$, and set
\[\check{X}_{2\mid 1} \equiv \tau_2+\rho\sigma_2\zeta, \quad \underline{S} \equiv 1+\check{X}_1^2+\check{X}_2^2,\]
\[\bar{S} \equiv 1+\check{X}_1^2+\check{X}_{2\mid 1}^2+\sigma_\perp^2 \zeta^2+\left(1-\sigma_1^2-\rho^2\sigma_2^2-\sigma_\perp^2\right),\]
\begin{equation} \label{eqn:Psi}
\Psi \equiv \left(\tau_1^2+\tau_2^2\right)E\left[\frac{\check{X}_1^2}{\underline{S}^2}\right]-2E\left[\frac{(\tau_1\check{X}_1)^+}{\bar{S}}\right]+2E\left[\frac{(\tau_1\check{X}_1)^-}{\underline{S}}\right].
\end{equation}
Then:
\begin{enumerate}[(i)]
\item $R(\theta,a_C)-1 \leq \Psi$. Moreover, $\sigma_1>0$, $\sigma_2,\sigma_\perp \geq 0$, $|\rho| \leq 1$, and $\sigma_1^2+\sigma_2^2+\sigma_\perp^2 \leq 1$.
\item If $|\rho| = 1$ and $\sigma_1^2+\sigma_2^2 = 1$, then $R(\theta,a_C)-1 = \Psi$ and $\Sigma = vv'$ with $v \equiv \sigma_1y+\rho\sigma_2\vartheta$.
\end{enumerate}
\end{prop}

\noindent Since the law of $(\check{X}_1,\check{X}_2)$ is determined by $(\tau_1,\tau_2,\sigma_1,\sigma_2,\rho)$, the quantity $\Psi$ is a function of the six scalars $(\tau_1,\tau_2,\sigma_1,\sigma_2,\rho,\sigma_\perp)$ alone, whatever the dimension $m$.

\begin{proof}
See \cref{app:prop:reduction}.
\end{proof}

\begin{lem} \label{lem:compact}
If $\|\theta\|^2 \geq 48$ then $R(\theta,a_C)-1 \leq 0.098$.
\end{lem}

\begin{proof}
See \cref{app:lem:compact}.
\end{proof}

By \cref{prop:reduction}(i) and \cref{lem:compact}, $\sup_{\theta,\Sigma} R_\Sigma(\theta,a_C)-1 \leq \max\lbrace 0.098,\sup \Psi \rbrace$, the supremum being over the \emph{bounded} set
\[\tau_1 \in \mathbb{R}, \quad \tau_2 \geq 0, \quad \tau_1^2+\tau_2^2 \leq 48, \quad |\rho| \leq 1, \quad \sigma_1>0,\quad \sigma_2,\sigma_\perp \geq 0, \quad \sigma_1^2+\sigma_2^2+\sigma_\perp^2 \leq 1 .\]
Numerical Nelder-Mead optimization with 1 million starting points (from a scrambled Sobol sequence), combined with Gauss-Hermite quadrature for the expectations, gives $\sup \Psi = 0.103859$,\footnote{47\% of the starting points yield this optimized value (up to small numerical error).} attained at
\[\tau_1 = 0.101485, \quad \tau_2 = 1.739516, \quad \sigma_1 = 0.679975, \quad \sigma_2 = 0.733236, \quad \rho = 1, \quad \sigma_\perp = 0 .\]
At this numerically determined global maximizer, we have $|\rho| = 1$ and $\sigma_1^2+\sigma_2^2 = 1$. Hence, by \cref{prop:reduction}(ii), $R(\theta,a_C)-1 = \Psi$ at the numerical maximizer, justifying the (numerically approximate) equality in \cref{num:cautious}.\footnote{The least favorable $\Sigma$ in \cref{prop:reduction}(ii) has rank 1, but it can be approximated arbitrarily well by a sequence of positive definite matrices.} At the least favorable configuration described above, we have $\|\theta\| = \sqrt{\tau_1^2+\tau_2^2} \approx 1.74$, and the vectors $\theta$ and $y$ form an angle of $\arccos\left(\tau_1/\|\theta\|\right) \approx 86.7^\circ$.

\subsection{Spherical noise}
\label{app:spherical}

Here we provide an exact characterization of the worst-case excess risk of the cautious rule in the case where $k=1$ and $\Sigma$ is spherical, i.e., the coordinates of $X$ are independent with common variance.

\begin{prop} \label{prop:cautious_risk_spherical}

Assume Gaussian data \eqref{eqn:X_gauss} with $k=1$, $m \geq 2$, $W=I_m$, $\Sigma=\sigma^2 I_m$, and $\sigma>0$. Then
\[\sup_{\theta \in \mathbb{R}^m} \frac{R(\theta,a_C)}{R(\theta,0)}-1 = \sup_{\tau>0} \left(\tau^2 \int_0^\infty \nu (1+2\nu)^{-m/2-1} \exp\left\lbrace -m\nu - \frac{\nu\tau^2}{1+2\nu} \right\rbrace d\nu\right).\]

\end{prop}

\begin{proof}
See \cref{app:prop:cautious_risk_spherical}.
\end{proof}

\begin{figure}[t]
\centering
\includegraphics[width=0.9\linewidth]{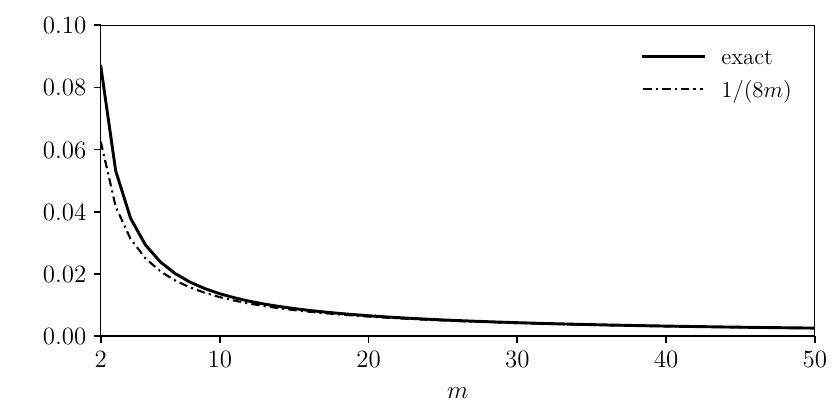}
\caption{Worst-case excess risk of the cautious rule in the spherical Gaussian case, as a function of $m$, computed as in \cref{prop:cautious_risk_spherical} using numerical quadrature and optimization. For $m = 1$ the excess risk is zero (not shown).}
\label{fig:spherical_wc}
\end{figure}

The proof shows that for spherical covariance structures, the least favorable parameter vector $\theta$ is orthogonal to $y$. \cref{fig:spherical_wc} plots the right-hand side of the above proposition for target dimensions $m=2,3,\dots$. The largest excess risk occurs for $m=2$ and equals $8.7\%$ (confirming \cref{fig:risk_contour} in \cref{subsec:intuition_m2}); the excess risk then decays to 0 as $m \to \infty$. As indicated by the figure, the excess risk asymptotes to $1/(8m)$ as $m \to \infty$.\footnote{The asymptote is suggested by the following calculation. Write the integral appearing in \cref{prop:cautious_risk_spherical} as $\int_0^\infty q(\nu) e^{-mp(\nu)}\,d\nu$ with $q(\nu) \equiv \nu/(1+2\nu)$, $p(\nu) \equiv \nu+\frac{1}{2}\log(1+2\nu) + \tilde{\tau}^2\nu/(1+2\nu)$, and $\tilde{\tau}^2 \equiv \tau^2/m$. Taylor expansions give $p(\nu) = p_0\nu + O(\nu^2)$ and $q(\nu) = q_0\nu + O(\nu^2)$, with $p_0 \equiv 2+\tilde{\tau}^2$ and $q_0\equiv 1$. As $m \to \infty$ with $\tilde{\tau}^2$ held fixed, Laplace's method \citep[{\S}2.3(iii)]{DLMF} then implies that the integral equals $\Gamma(2)q_0/(mp_0)^2 + O(m^{-3}) = [m(2+\tilde{\tau}^2)]^{-2} + O(m^{-3})$, where $\Gamma(\cdot)$ is the Gamma function. Hence, under additional regularity conditions the supremum in \cref{prop:cautious_risk_spherical} is approximated by $\sup_{\tau^2 \geq 0}\tau^2[m(2+\tau^2/m)]^{-2} = 1/(8m)$ (attained at $\tau^2=2m$).} 

\subsection{Admissibility of the cautious rule} \label{app:admissibility}

\begin{prop} \label{prop:adm}

Consider the case of Gaussian data \eqref{eqn:X_gauss}, and let $mk \leq 4$. Then the cautious rule is admissible: there is no policy rule $a(X)$ satisfying $R(\theta,a) \leq R(\theta,a_C)$ for all $\theta \in \mathbb{R}^{m \times k}$ with strict inequality for some $\theta$.

\end{prop}

\begin{proof}
See \cref{app:prop:adm}.
\end{proof}

Extending the $m=1$ arguments of \citet{ZamanStein1981}, our proof applies Blyth's method \citep[Theorem 5.7.13]{LehmannCasella1998} to construct a sequence of priors which approximately rationalize the cautious rule without letting prior mass escape to infinity too quickly. For $mk \leq 3$, Gaussian priors suffice, while for $mk=4$ a more involved scale mixture of Gaussian kernels is required.\footnote{One subtlety in our proof is that, unlike in point estimation problems, it is not immediate that a potential dominating rule must have a continuous risk function. \citet{Zaman1981} and \citet{ZamanStein1981} assert this continuity without a detailed argument. Our proof shows that any dominating rule must have finite second moment at every $\theta$ (which would be obvious in the point estimation problem with squared error loss, but not here), allowing us to appeal to known continuity results for exponential families.}

\subsection{The cautious rule with a biased signal}
\label{app:cautious_bias}

We show that allowing room for bias strictly increases the worst-case risk of the cautious rule, with that increase being of first order in the potential magnitude of the bias.

For the purposes of this subsection, replace the unbiased Gaussian signal \eqref{eqn:X_gauss} by
\begin{equation} \label{eqn:X_gauss_bias}
\ve(X) \sim N\big(\ve(\theta) + \ve(b), \Sigma\big),
\end{equation}
where the bias $b \in \mathbb{R}^{m \times k}$ is constrained to lie in the set
\begin{equation} \label{eqn:bias_set}
\mathcal{B}_K \equiv \big\lbrace b \in \mathbb{R}^{m \times k} \colon \ve(b)\ve(b)' \preceq K^2 \Delta \big\rbrace = \big\lbrace b \colon \ve(b) = K\Delta^{1/2}v, \; \|v\| \leq 1 \big\rbrace,
\end{equation}
for a scalar $K \geq 0$ and a symmetric positive semidefinite $mk \times mk$ matrix $\Delta \neq 0$. Here the matrix $\Delta$ determines the directions in which the signal can be biased, and $K$ scales the magnitude of the bias. The risk of a policy rule $a$ now depends on both $\theta$ and $b$, and we write it as $R(\theta,b,a) \equiv E_{\theta,b}[L(\theta,a(X))]$. See \cref{app:notation} for definitions of our notation.

\begin{prop}
\label{prop:bias_costly}

Consider the biased Gaussian model \eqref{eqn:X_gauss_bias}--\eqref{eqn:bias_set} with $k = 1$, $m \geq 2$, $\Theta = \mathbb{R}^m$, $\Delta \succ 0$, and $K > 0$, normalized so that $W = I_m$ and $\|y\| = 1$. Let $\gamma \equiv \sup_{\theta \in \Theta} R(\theta,0,a_C) - 1$ denote the worst-case excess risk of the cautious rule in the unbiased model, let $\theta^*$ attain this supremum, and write $E_{\theta^*}$ for expectations under $X \sim N(\theta^*, \Sigma)$. Then
\begin{equation*}
\sup_{\theta \in \Theta} \sup_{b \in \mathcal{B}_K} R(\theta,b,a_C) \; \geq \; 1 + \gamma + \max \left \{ K^2 \lambda_{\max}(\Delta) \, E_{\theta^*}[a_C(X)^2], \; 2 K \|\omega^*\| \right \} \; > \; 1 + \gamma,
\end{equation*}
where $\omega^* \equiv \Delta^{1/2} \big( y \, E_{\theta^*}[a_C(X)] - \theta^* E_{\theta^*}[a_C(X)^2] \big) \neq 0$. In particular, the bias contribution to the worst-case risk is of first order in $K$.
\end{prop}

\begin{proof}
See \cref{app:prop:bias_costly}.
\end{proof}

\clearpage

\section{Application and simulation details}
\label{app:simulations}

\subsection{Replication of \texorpdfstring{\citet{Barnichon2023}}{Barnichon and Mesters (2023)}}
\label{app:subsec:bm}

Supplementing the discussion in \cref{subsec:bm_replication}, we provide further details on our replication and extension of the optimal policy analysis in \citet{Barnichon2023}.

\paragraph{Forecasts.}
Our construction of unemployment and inflation point forecasts, together with the target values for those variables, closely follows the replication code of \citeauthor{Barnichon2023}. Using the notation of our decision problem, this delivers target vectors $y$ for 90 dates, ranging from June 1980 to December 2022. Specifically, for each forecast date, we stack unemployment and inflation forecast gaps relative to date-specific long-run projections, over horizons $t=0,\ldots,19$. Differently from those authors, we restrict attention to the point estimates of those forecasts, and do not take into account further forecasting uncertainty.\footnote{As discussed in \cref{sec:framework}, incorporating uncertainty is straightforward as long as we assume independence between that forecast uncertainty and uncertainty about monetary policy causal effects.}

\paragraph{Policy shock IRFs.}
Our data construction for the policy shock analysis also follows \citeauthor{Barnichon2023}: on a sample from 1991:Q1 to 2019:Q2 we follow their construction to get time series for unemployment, inflation (both de-trended), and the federal funds rate; furthermore, we take their series for the \citet{Eberly2020} federal funds rate shock. Differently from \citeauthor{Barnichon2023}, however, and in keeping with our results in \cref{sec:bias_var}, we estimate shock propagation using local projections. Specifically, in line with the identifying assumptions in their replication code, we consider local projections that control for the contemporaneous values of unemployment and inflation.\footnote{The published paper states that shocks are ordered first (p.\ 2832). The replication code, however, order the shocks after unemployment and inflation; for maximal comparability we follow that choice.} We include a constant, five lags of all four variables, and construct conventional homoskedastic standard errors.

\begin{figure}[tp!]
\centering \vspace{0.5\baselineskip}
\includegraphics[width=\textwidth]{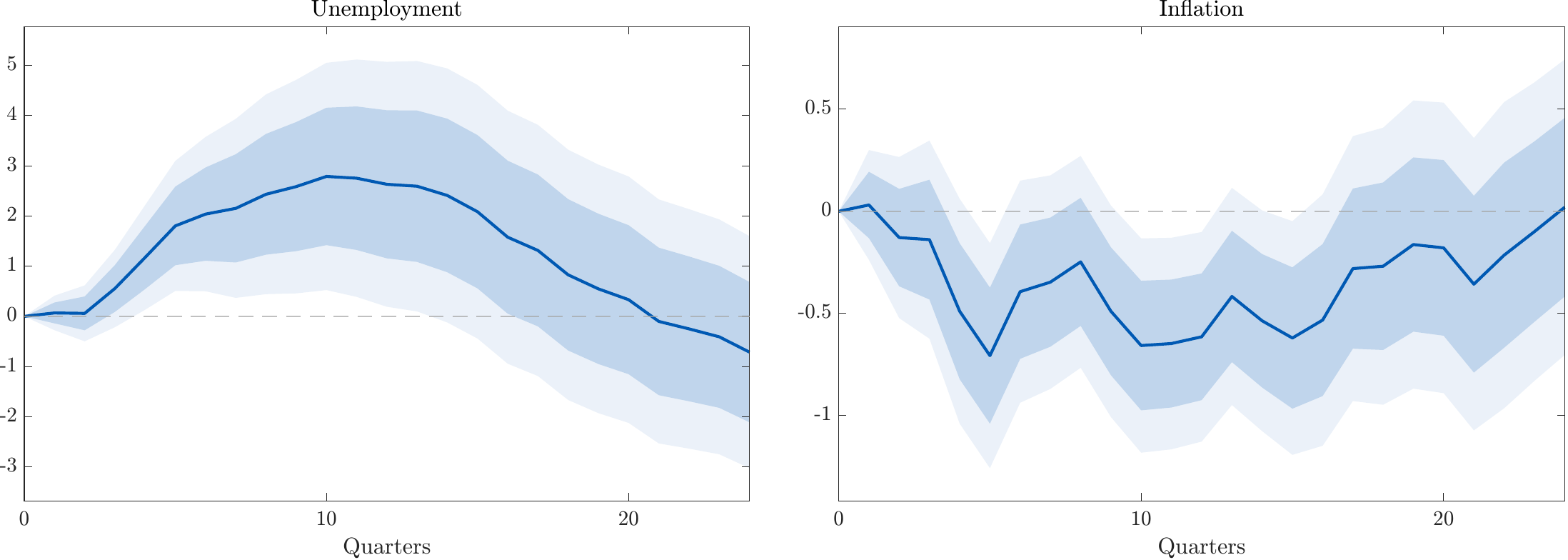}
\caption{LP-estimated impulse responses of unemployment and inflation to the federal funds rate shock of \citet{Eberly2020}. Solid: point estimates. Shaded: 68 per cent and 90 per cent bands.}
\label{fig:ffr_shock_irfs}
\end{figure}

Results are displayed in \cref{fig:ffr_shock_irfs}, which shows the estimated dynamic causal effects of the federal funds rate innovation on unemployment and inflation. The results are very similar to those reported by \citeauthor{Barnichon2023} (see their Figure S1, top panel), with somewhat wider uncertainty bands, as expected.

\paragraph{Optimal policy.}
\cref{fig:opp_cautious} is constructed as follows. The plug-in and cautious rule actions are computed by inserting the LP estimate $X$ and covariance matrix $\Sigma$ into the formulas in \cref{subsec:policy_rules}. Note that the plug-in policy rule is equivalent with the ``optimal policy perturbation'' (OPP) reported by \citeauthor{Barnichon2023}; consistent with that, our plug-in line is close to the contemporaneous federal funds rate OPP reported in their Figure 1.\footnote{Differently from us, \citeauthor{Barnichon2023} report a \emph{joint} OPP with respect to two policy shocks; for reasons described in the main text, we only consider a single shock.} The shaded posterior bands are percentile ranges for the oracle action $a_\theta$ evaluated at 20,000 draws from the flat-prior posterior distribution $\theta \sim N(X,\Sigma)$. Following \citeauthor{Barnichon2023}, we constrain the actions to be nonnegative during the historical zero lower bound periods, indicated by the yellow areas in \cref{fig:opp_cautious}.

\subsection{Simulation study}

This section provides further implementation details for our simulation study in \cref{subsec:sim_cautious}.

We begin with the specification of the DGP. On the same sample as that used above for LP policy causal effect estimation, we estimate a reduced-form VAR in unemployment, inflation, the federal funds rate, and the \citet{Eberly2020} federal funds rate shock. The VAR is estimated with a constant and five lags. We then use this reduced-form VAR for two purposes. First, we simulate $S \equiv \text{10,000}$ artificial samples from the reduced-form VAR, assuming normal errors, with each sample of the same length as the original empirical sample, and with $500$ periods for burn-in. Second, we define the VAR-implied impulse responses of unemployment and inflation to the federal funds rate shock---obtained from a recursive ordering, with the federal funds rate shock ordered third, after unemployment and inflation---as the population ground truth policy causal effects.

We next turn to the computations of the decisions and the associated losses. For each of the 10,000 samples, we first follow the LP estimation strategy described in \cref{app:subsec:bm} to get an estimate of policy causal effects $X$ and associated covariance matrix. Using these estimates, we construct, for each target vector $y$ (i.e., on each forecast date), the plug-in and cautious rule decisions, and then---using the VAR-implied ground truth causal effects $\theta$---recover the associated losses $L(\theta, a_P(X))$ and $L(\theta, a_C(X))$. For the top panel of \cref{fig:cautious_loss_var_simul}, we report the normalized loss difference
\begin{equation*}
\frac{L(\theta,a_P(X_s);y_d) - L(\theta,a_C(X_s);y_d)}{\nW{y_d}^2},
\end{equation*}
where $s$ indexes the Monte Carlo sample and $d$ indexes the forecast date. At each date $d=1,\dots,D\equiv 90$, $y_d$ is fixed at the value observed in the real data. We then plot percentiles of date-specific distributions across simulations. For the bottom panel, we report the share of samples for which the loss exceeds the do-nothing benchmark of $\nW{y_d}^2$. The probability annotations in both panels pool across all simulation-date observations. Finally, to convert loss differences into economic units, we compute the un-normalized average loss difference,
\begin{equation*}
\overline{\Delta L} \equiv \frac{1}{SD} \sum_{s=1}^{S}\sum_{d=1}^{D} \left[ L(\theta,a_P(X_s);y_d) - L(\theta,a_C(X_s);y_d) \right]
\end{equation*}
and then report
\begin{equation*}
\sqrt{\frac{\overline{\Delta L}}{H}} = \sqrt{\frac{4.35068}{20}} \approx 0.47 \quad \text{percentage points},
\end{equation*}
where $H \equiv 20$ is the total number of quarterly horizons in the policy-maker's objective.

\clearpage

\section{Proofs}

\subsection{Notation}
\label{app:notation}

We define the following additional notation. $\|\cdot\|$ and $\nop{\cdot}$ denote the Frobenius and operator norms, respectively. $\lambda_{\min}(\cdot)$ and $\lambda_{\max}(\cdot)$ denote the smallest and largest eigenvalues, respectively. $\col(\theta)$ and $\ker(\theta)$ denote the column space and kernel of $\theta$, respectively. For a symmetric positive definite matrix $M$, $M^{1/2}$ denotes its symmetric positive definite square root. $M_1 \preceq M_2$ denotes the Loewner order, i.e., $M_2 - M_1$ is positive semidefinite, while $M_1 \prec M_2$ means $M_2-M_1$ is positive definite. $|M|$ denotes the determinant of a square matrix $M$. For brevity, we henceforth drop the subscript $\theta$ from the expectation $E_\theta[\cdot]$.

\subsection{Auxiliary lemmas}
\label{app:lemmas}

\begin{lem} \label{lem:loss_decomp}
For all $\theta\in\mathbb{R}^{m\times k}$ and $\alpha\in\mathbb{R}^k$,
\[\nW{y-\theta\alpha}^2=V(\theta)+\nW{\theta\alpha-\theta a_\theta}^2, \quad\text{so}\quad R(\theta,a)=V(\theta)+E\left[\nW{\theta a(X)-\theta a_\theta}^2\right].\]
If $\theta \in \mathcal{S}$, then $\theta a_\theta=0$, $V(\theta)=\nW{y}^2$, and $R(\theta,a)=\nW{y}^2+E[\nW{\theta a(X)}^2]$.
\end{lem}
\begin{proof}
Write $y-\theta\alpha=(y-\theta a_\theta)+(\theta a_\theta-\theta\alpha)$, where $\theta a_\theta$ is the orthogonal projection of $y$ onto $\col(\theta)$ with respect to the inner product $\langle u,v\rangle_W \equiv u'Wv$. The first term is $\langle\cdot,\cdot\rangle_W$-orthogonal to $\col(\theta)$ and the second lies in $\col(\theta)$, so the cross term vanishes. The rest is obvious.
\end{proof}

\begin{lem} \label{lem:normalization}
Let $a\in\lbrace a_P,a_C\rbrace$, and write $Z\equiv X-\theta$. The following normalizations may be imposed without loss of generality when analyzing the risk $R(\theta,a)$ relative to $R(\theta,0)=\nW{y}^2$:
\begin{enumerate}[(i)]
\item $\nW{y}=1$;
\item $W=I_m$;
\item if $a=a_C$: for any invertible $H\in\mathbb{R}^{k\times k}$, replacing $(\theta,Z)$ by $(\theta H',ZH')$, so that $X$ is replaced by $XH'$ and $\Sigma$ by $(H\otimes I_m)\Sigma(H\otimes I_m)'$, replaces $C$ by $HCH'$ and leaves the loss $\nW{y-\theta a_C(X)}^2$ unchanged. In particular, we may take $C=I_k$ (with $H=C^{-1/2}$).
\end{enumerate}
\end{lem}
\begin{proof}
(i) Immediate: $a$ is linear in $y$ and the loss is quadratic in $y$, so $R(\theta,a)$ is proportional to $\nW{y}^2$.

(ii) This is the change of variables $\bar{y}\equiv W^{1/2}y$, $\bar{\theta}\equiv W^{1/2}\theta$, $\bar{X}\equiv W^{1/2}X$, and $\bar{\Sigma}\equiv(I_k\otimes W^{1/2})\Sigma(I_k\otimes W^{1/2})$: the problem $(y,\theta,W,\Sigma)$ is the problem $(\bar{y},\bar{\theta},I_m,\bar{\Sigma})$, and $C$ is unchanged.

(iii) The new $C$ matrix equals $E[(ZH')'W(ZH')]=HCH'$. The rule is equivariant,
\begin{align*}
a_C(XH')&=\left(HX'WXH'+HCH'\right)^{-1}HX'Wy\\
&=(H')^{-1}(X'WX+C)^{-1}X'Wy=(H')^{-1}a_C(X),
\end{align*}
so the fitted outcome $(\theta H')a_C(XH')=\theta H'(H')^{-1}a_C(X)=\theta a_C(X)$ is unchanged, and hence so is the loss.
\end{proof}

\begin{lem} \label{lem:cautious_bound}
Let $\theta,X \in \mathbb{R}^{m \times k}$, and write $Z \equiv X-\theta$, $z \equiv \nop{W^{1/2}ZC^{-1/2}}$, and $\xi \equiv \Sigma^{-1/2}\ve(Z)$. Then:
\begin{enumerate}[(i)]
\item $\nW{Xa_C(X)} \leq \nW{y}$ and $\nW{y-Xa_C(X)} \leq \nW{y}$.
\item $\|a_C(X)\| \leq \dfrac{\nW{y}}{2\sqrt{\lambda_{\min}(C)}}$.
\item $\nW{y-\theta a_C(X)} \leq \tfrac12\left(1+\sqrt{1+z^2}\right)\nW{y}$.
\item $z^2 \leq k\|\xi\|^2$, so that $L(\theta,a_C(X)) \leq \tfrac14\left(1+\sqrt{1+k\|\xi\|^2}\right)^2\nW{y}^2$.
\end{enumerate}
If, moreover, $X$ is random with $E[X]=\theta$ and $\var(\ve(X))=\Sigma$, then $E[z^2] \leq k$ and $E\|\xi\|^2 = mk$.
\end{lem}

\begin{proof}
Write $a \equiv a_C(X)$ and
\[\ell \equiv \nW{y-Xa}, \quad \psi \equiv \|C^{1/2}a\|.\]

(i) The normal equation $(X'WX+C)a = X'Wy$ gives $\nW{Xa}^2+\psi^2 = a'(X'WX+C)a = y'WXa$, so that
\[\ell^2+\psi^2 = \left(\nW{y}^2-2y'WXa+\nW{Xa}^2\right)+\psi^2 = \nW{y}^2-y'WXa = y'W(y-Xa) \leq \nW{y}\,\ell,\]
where the inequality uses Cauchy-Schwarz. Equivalently, the pair $(\ell,\psi)$ lies in the disk of radius $\frac12\nW{y}$ centered at $(\frac12\nW{y},0)$:
\begin{equation} \label{eqn:disk}
\left(\ell-\tfrac12\nW{y}\right)^2+\psi^2 \leq \tfrac14\nW{y}^2.
\end{equation}
In particular, $\ell \leq \nW{y}$. Moreover, $\nW{Xa}^2 \leq \nW{Xa}^2+\psi^2 = y'WXa \leq \nW{y}\nW{Xa}$ by Cauchy-Schwarz, so $\nW{Xa} \leq \nW{y}$.

(ii) By \eqref{eqn:disk}, $\psi \leq \frac12\nW{y}$, so $\|a\| \leq \nop{C^{-1/2}}\psi \leq \nW{y}/(2\sqrt{\lambda_{\min}(C)})$.

(iii) Since $y-\theta a = (y-Xa)+Za$ and $\nW{Za} = \|W^{1/2}ZC^{-1/2}C^{1/2}a\| \leq z\psi$,
\begin{align*}
\nW{y-\theta a} &\leq \ell+z\psi \quad (\text{triangle inequality}) \\
&= \tfrac12\nW{y}+\left[\left(\ell-\tfrac12\nW{y}\right)\times 1+\psi\times z\right] \\
&\leq \tfrac12\nW{y}+\sqrt{\left(\ell-\tfrac12\nW{y}\right)^2+\psi^2} \times \sqrt{1+z^2} \quad (\text{Cauchy-Schwarz in } \mathbb{R}^2) \\
&\leq \tfrac12\left(1+\sqrt{1+z^2}\right)\nW{y} \quad (\text{by \eqref{eqn:disk}}).
\end{align*}

(iv) Since $\ve(W^{1/2}ZC^{-1/2})=(C^{-1/2}\otimes W^{1/2})\ve(Z)$ and $\ve(Z)=\Sigma^{1/2}\xi$,
\[z^2 \leq \|W^{1/2}ZC^{-1/2}\|^2 = \ve(Z)'(C^{-1}\otimes W)\ve(Z) = \xi'A\xi, \quad A \equiv \Sigma^{1/2}(C^{-1}\otimes W)\Sigma^{1/2}.\]
The largest eigenvalue of the positive semidefinite matrix $A$ is at most its trace. Moreover, note that 
\[\tr(A) = \tr\left((C^{-1}\otimes W)\Sigma\right) = \sum_{i,j}(C^{-1})_{ij}\tr(W\Sigma^{ji}) = \tr(C^{-1}C) = k.\] Hence $z^2 \leq \xi'A\xi \leq k\|\xi\|^2$, and squaring part (iii) gives part (iv). Finally, if $E[X]=\theta$ and $\var(\ve(X))=\Sigma$, then $E[\xi\xi']=I_{mk}$, so $E\|\xi\|^2 = mk$ and $E[z^2] \leq E[\xi'A\xi] = \tr(A) = k$.
\end{proof}

\subsection{Proof of \texorpdfstring{\cref{prop:plugin_inf}}{Proposition \ref{prop:plugin_inf}}}
\label{app:prop:plugin_inf}

Following \cref{lem:normalization}(i)--(ii) in \cref{app:lemmas}, we will assume throughout that $\nW{y}=1$ and $W=I_m$.

(i) Fix an instrument $j$, write its column of $X$ as $X_{\bullet j}$, and collect the other $k-1$ columns in $X_{\bullet,-j}$, where $X_{\bullet,-j}$ almost surely has full column rank. Let $U$ be an $m\times n$ matrix whose columns form an orthonormal basis for the orthogonal complement of the columns of $X_{\bullet,-j}$, with $n\equiv m-k+1$, and set
\[p\equiv U'X_{\bullet j}, \quad q\equiv U'y,\]
so that $p$ and $q$ are the residualized effect of instrument $j$ and the residualized target. For any fixed value $b$ of the $j$-th action, optimizing over the other action coefficients leaves the objective
\[\min_{c\in\mathbb{R}^{k-1}}\|y-X_{\bullet j}b-X_{\bullet,-j}c\|^2=\|U'(y-X_{\bullet j}b)\|^2=\|q-pb\|^2.\]
Consequently, we can write the plug-in action for instrument $j$ as
\[a_{P,j}(X)=\frac{p'q}{\|p\|^2}.\]
Conditional on $X_{\bullet,-j}$, this is exactly the formula from the univariate case of \cref{lem:plugin_k1}, now with dimension $n$ and target $q$. Note for future reference that $q\neq0$ almost surely and $\|q\|\leq\|y\|=1$.

We now proceed as in the univariate case in \cref{lem:plugin_k1} below. The conditional distribution of $X_{\bullet j}$ given $X_{\bullet,-j}$ is Gaussian, with a positive definite covariance $\Omega_j$ independent of the realized value of $X_{\bullet,-j}$. Hence $p$ is conditionally Gaussian with covariance $U'\Omega_jU$, and
\[\lambda_{\min}(U'\Omega_jU)\geq\lambda_{\min}(\Omega_j)>0.\]
Its conditional density is continuous and strictly positive at zero, and is bounded above by
\[K_j\equiv (2\pi\lambda_{\min}(\Omega_j))^{-n/2},\]
a constant that does not depend on $X_{\bullet,-j}$. We can now, as in the univariate analysis, distinguish two cases. First, if $k\leq m\leq k+1$, then $n=1$ or $n=2$, and so the divergent-integral argument of the proof of \cref{lem:plugin_k1} applies conditional on $X_{\bullet,-j}$, giving $E[a_{P,j}(X)^2\mid X_{\bullet,-j}]=\infty$ almost surely. It follows that $E\|a_P(X)\|^2=\infty$. Second, if $m\geq k+2$, then $n\geq3$. Since $\|q\|\leq1$, the exact same upper bound argument as in the proof of \cref{lem:plugin_k1} gives that $E[a_{P,j}(X)^2\mid X_{\bullet,-j}]$ is uniformly upper-bounded. We can then take expectations and sum over $j$'s to get that $E\|a_P(X)\|^2<\infty$.

Finally, to go from here to risk, note that the same loss inequalities as in the proof of \cref{lem:plugin_k1} now take the form
\[\frac12\|\theta a\|^2-1\leq\|y-\theta a\|^2\leq2+2\|\theta a\|^2.\]
For every $\theta$, the upper bound $\|\theta a\|\leq\nop{\theta}\|a\|$ implies finite risk when $m\geq k+2$. If instead $k\leq m\leq k+1$, risk is infinite for any $\theta \neq 0$. To see this, let $h'$ be a nonzero row of $\theta$ and choose an orthogonal matrix $Q$ whose first column is $h/\|h\|$. The transformed data $\bar{X}\equiv XQ$ are Gaussian with mean $\theta Q$ and positive definite covariance $(Q'\otimes I_m)\Sigma(Q\otimes I_m)$; now write
\[a_P(\bar{X})=\big(Q'X'XQ\big)^{-1}Q'X'y=Q'a_P(X).\]
Applying the coordinate-wise infinite-second-moment result above to $\bar{X}$ therefore gives
\[E\|\theta a_P(X)\|^2\geq E\big[(h'a_P(X))^2\big]=\|h\|^2 E\big[a_{P,1}(\bar{X})^2\big]=\infty,\]
so the lower loss bound implies infinite risk.

(ii) If $k\leq m\leq k+1$, part (i) already gives infinite risk for every $\theta\neq0$ and every positive definite $\Sigma$. It thus remains to consider $m\geq k+2$. Let $e_j$ denote the $j$-th standard unit vector in $\mathbb{R}^m$. Take $y=e_1$ (after an orthogonal change of coordinates) and fix the full-column-rank true effect matrix
\[\theta_0\equiv [e_1,e_2,\ldots,e_k].\]
Now let $\xi=(\xi_{ij})$ have independent standard normal entries, and for $\eta>0$ define
\begin{align*}
(X_\eta)_{\bullet 1}&\equiv (1+\xi_{11})e_1+\sqrt{\eta}\sum_{i=2}^m\xi_{i1}e_i,\\
(X_\eta)_{\bullet j}&\equiv e_j+\sqrt{\eta}\sum_{i=1}^m\xi_{ij}e_i, \quad j=2,\ldots,k,
\end{align*}
and so, vectorized,
\[\ve(X_\eta)\sim N(\ve(\theta_0),\Sigma_\eta), \quad \Sigma_\eta \equiv \diag(1,\eta I_{mk-1})\succ0.\]
Writing $G\equiv 1+\xi_{11}$, we have, as $\eta\to 0$,
\[X_\eta\longrightarrow X_0=[Ge_1,e_2,\ldots,e_k] \quad \text{almost surely}.\]
Since $G\neq0$ almost surely, $X_0$ has full column rank almost surely. Continuity of the plug-in rule at such matrices gives
\[a_P(X_\eta)\longrightarrow\left(\frac{1}{G},0,\ldots,0\right)', \quad L(\theta_0,a_P(X_\eta))\longrightarrow\left(1-\frac{1}{G}\right)^2 \quad \text{almost surely}.\]
By \cref{lem:plugin_k1}, the limiting loss has infinite expectation. Fatou's lemma therefore yields
\[\liminf_{\eta\downarrow0}R_{\Sigma_\eta}(\theta_0,a_P)\geq E\left[\left(1-\frac{1}{G}\right)^2\right]=\infty.\]
Every $\Sigma_\eta$ with $\eta>0$ is positive definite, and every risk along this sequence is finite. This proves the stated supremum is infinite, with $\theta$ held fixed at $\theta_0$.

Note that this argument is all about the covariance \emph{shape}---the key is that uncertainty for the first entry is much larger than uncertainty for the other entries, essentially returning us to the univariate case. Formally, replacing $\Sigma_\eta$ by $\bar{\varepsilon}\Sigma_\eta/\tr(\Sigma_\eta)$, where $\bar{\varepsilon} \equiv \varepsilon\lambda_{\min}(W)$, normalizes the trace to $\bar{\varepsilon}$; by \cref{lem:normalization}(ii) in \cref{app:lemmas}, the corresponding covariance matrix in the original coordinates then has trace at most $\bar{\varepsilon}/\lambda_{\min}(W)=\varepsilon$ and so lies in $\mathcal{M}_\varepsilon$. This changes the limiting distribution only in that $G$ becomes $1+\sqrt{\bar{\varepsilon}}\,\xi_{11}\sim N(1,\bar{\varepsilon})$, whose density is still continuous and strictly positive at zero, so the exact same proof applies. \qed

\begin{lem} \label{lem:plugin_k1}
Assume Gaussian data \eqref{eqn:X_gauss} with $\nW{y}=1$ and $W=I_m$. Suppose that $k=1$.
\begin{enumerate}[(i)]
\item If $m\leq2$, then $E\|a_P(X)\|^2=\infty$. Consequently, $R_\Sigma(\theta,a_P)=\infty$ for every $\theta\neq0$.
\item If $m>2$, then $E\|a_P(X)\|^2<\infty$ and $R_\Sigma(\theta,a_P)<\infty$ for every $\theta$.
\end{enumerate}
\end{lem}
\begin{proof}
Note that the Gaussian density $f_\theta$ of $X\in\mathbb{R}^m$ is bounded above by a finite constant $K$, and is continuous and strictly positive at zero. By Cauchy-Schwarz,
\[a_P(X)^2\leq\frac{1}{\|X\|^2}.\]
Outside the unit ball for $X$ the squared action is at most $1$. We thus, for $m\geq3$, have that
\begin{align*}
E[a_P(X)^2]&\leq K\int_{\|x\|\leq1}\frac{dx}{\|x\|^2}+1 <\infty,
\end{align*}
where the last inequality follows from \citet[p.\ 59]{steinshakarchi2005real}. If instead $m=1$ or $m=2$, positivity and continuity at zero give constants $c,\delta>0$ such that $f_\theta(x)\geq c$ whenever $\|x\|<\delta$. Thus,
\begin{align*}
E[a_P(X)^2]&\geq c\int_{\|x\|<\delta}\frac{(x'y)^2}{\|x\|^4}\,dx=c \frac{\|y\|^2}{m} \int_{\|x\|<\delta}\frac{dx}{\|x\|^2} = \infty,
\end{align*}
where the first equality follows from orthogonal invariance of Lebesgue measure, and the last equality follows again from \citet[p.\ 59]{steinshakarchi2005real} and $\|y\|>0$.

To go from here to risk, note that, for any scalar action $a$,
\[\frac12\|\theta\|^2a^2-1\leq\|y-\theta a\|^2\leq2+2\|\theta\|^2a^2.\]
Using the above bounds for $E[a_P(X)^2]$ it follows that risk is finite for every $\theta$ when $m\geq3$, and infinite for every $\theta\neq0$ when $m=1,2$.
\end{proof}

\subsection{Proof of \texorpdfstring{\cref{prop:cautious_risk}}{Proposition \ref{prop:cautious_risk}}(i)}
\label{app:prop:cautious_risk_i}

Write $Z \equiv X-\theta$, so that $E[Z]=0$ and $\var(\ve(Z))=\Sigma$ for every $\theta$. Let $z \equiv \nop{W^{1/2}ZC^{-1/2}}$ and $\zeta^2 \equiv E[z^2]$. Jensen's inequality gives $E\big[\sqrt{1+z^2}\big] \leq \sqrt{1+\zeta^2}$. Squaring the bound in \cref{lem:cautious_bound}(iii) in \cref{app:lemmas}, taking expectations, and applying Jensen, we get
\[R(\theta,a_C) \leq \tfrac14\left(2+\zeta^2+2\sqrt{1+\zeta^2}\right)\nW{y}^2 = \tfrac14\left(1+\sqrt{1+\zeta^2}\right)^2\nW{y}^2\]
for every $\theta \in \mathbb{R}^{m \times k}$. Since $\zeta^2 \leq k$ by \cref{lem:cautious_bound}, it follows that
\[\sup_{\theta \in \mathbb{R}^{m \times k}} R(\theta,a_C) \leq \tfrac14\left(1+\sqrt{1+k}\right)^2\nW{y}^2.\]
The proposition follows, since $R(\theta,0)=\nW{y}^2$. Note that the bound does not use normality of $Z$, only its first two moments. \qed

\subsection{Proof of \texorpdfstring{\cref{prop:cautious_risk}}{Proposition \ref{prop:cautious_risk}}(ii)}
\label{app:prop:cautious_risk_ii}

\emph{Step 1.} Write $Z \equiv X-\theta$, so that $Z$ and $-Z$ have the same distribution. By \cref{lem:normalization} in \cref{app:lemmas}, we may assume that $W=I_m$, $\|y\|=1$, and $C=\tr(\Sigma)=1$. The changes of variables in that lemma are linear in $Z$, so $Z$ and $-Z$ still have the same distribution. Then $E\|Z\|^2 = \tr(\Sigma) = 1$, and
\[a_C(X) = \frac{X'y}{1+\|X\|^2}, \quad R(\theta,0)=1.\]
Writing $\tau^2 \equiv \|\theta\|^2$, the loss of a scalar action $\alpha$ is
\begin{equation} \label{eqn:loss}
L(\theta,\alpha) = 1-2\alpha\,\theta'y+\alpha^2\tau^2.
\end{equation}

\emph{Step 2.} For vectors $\theta,Z \in \mathbb{R}^m$, write $a_\pm \equiv a_C(\theta \pm Z)$ and
\[\bar{L}(\theta,Z) \equiv \tfrac12\left[L(\theta,a_+)+L(\theta,a_-)\right].\]
Since $Z$ and $-Z$ have the same distribution,
\begin{equation} \label{eqn:pairing}
R(\theta,a_C)-1 = E\left[\bar{L}(\theta,Z)-1\right].
\end{equation}
By \eqref{eqn:pairing} and $E\|Z\|^2=1$, it now suffices to show that $\bar{L}(\theta,Z)-1 \leq (2+25\|Z\|^2)/160$ for all $\theta,Z \in \mathbb{R}^m$.

\emph{Step 3.} Fix $\theta,Z \in \mathbb{R}^m$, and write $Q \equiv \|Z\|^2$, $\eta \equiv \theta'Z$, and $\bar{S} \equiv 1+\tau^2+Q$, so that $1+\|\theta \pm Z\|^2 = \bar{S} \pm 2\eta$. The definition of the cautious rule then gives
\begin{equation} \label{eqn:pm}
(\bar{S} \pm 2\eta)a_\pm = (\theta \pm Z)'y.
\end{equation}
Define $\bar{a} \equiv (a_++a_-)/2$ and $\delta \equiv (a_+-a_-)/2$, so that $a_\pm = \bar{a} \pm \delta$. Averaging the ``plus'' and ``minus'' equations in \eqref{eqn:pm} yields
\begin{equation} \label{eqn:thetay}
\theta'y = \tfrac12\left[(\bar{S}+2\eta)(\bar{a}+\delta)+(\bar{S}-2\eta)(\bar{a}-\delta)\right] = \bar{S}\bar{a}+2\eta\delta.
\end{equation}

\emph{Step 4.} Since $\frac12(a_+^2+a_-^2) = \bar{a}^2+\delta^2$, \eqref{eqn:loss} and \eqref{eqn:thetay} give
\begin{align}
\bar{L}(\theta,Z)-1 &= \tau^2(\bar{a}^2+\delta^2)-2\bar{a}\,\theta'y = \tau^2\delta^2-(2\bar{S}-\tau^2)\bar{a}^2-4\eta\delta\bar{a} \nonumber \\
&\leq \sup_{\alpha \in \mathbb{R}} \left\lbrace \tau^2\delta^2-(2\bar{S}-\tau^2)\alpha^2-4\eta\delta\alpha \right\rbrace = \left(\tau^2+\frac{4\eta^2}{2\bar{S}-\tau^2}\right)\delta^2, \label{eqn:Lbar}
\end{align}
using $2\bar{S}-\tau^2 = 2+\tau^2+2Q>0$.

\emph{Step 5.} Since $\bar{S} \pm 2\eta \geq 1$ and hence $\bar{S}^2-4\eta^2 = (\bar{S}+2\eta)(\bar{S}-2\eta) \geq 1$, \eqref{eqn:pm} gives
\[\delta = \frac12\left(\frac{(\theta+Z)'y}{\bar{S}+2\eta}-\frac{(\theta-Z)'y}{\bar{S}-2\eta}\right) = \frac{(\bar{S}Z-2\eta\theta)'y}{\bar{S}^2-4\eta^2}.\]
Because $\bar{S}-\tau^2 = 1+Q$,
\[\|\bar{S}Z-2\eta\theta\|^2 = \bar{S}^2Q-4\eta^2(\bar{S}-\tau^2) = (\bar{S}^2-4\eta^2)Q-4\eta^2 = (\bar{S}^2-4\eta^2)(Q-\rho),\]
where $\rho \equiv 4\eta^2/(\bar{S}^2-4\eta^2) \geq 0$. By the Cauchy-Schwarz inequality and $\|y\|=1$,
\begin{equation} \label{eqn:delta}
\delta^2 \leq \frac{\|\bar{S}Z-2\eta\theta\|^2}{(\bar{S}^2-4\eta^2)^2} = \frac{Q-\rho}{\bar{S}^2-4\eta^2},
\end{equation}
and in particular $\rho \leq Q$.

\emph{Step 6.} Inserting \eqref{eqn:delta} into \eqref{eqn:Lbar} and using $4\eta^2/(\bar{S}^2-4\eta^2) = \rho$ and $\bar{S}^2/(\bar{S}^2-4\eta^2) = 1+\rho$,
\begin{equation} \label{eqn:shared}
\bar{L}(\theta,Z)-1 \leq (Q-\rho)\left(\frac{\tau^2}{\bar{S}^2-4\eta^2}+\frac{\rho}{2\bar{S}-\tau^2}\right) = (Q-\rho)\left[\frac{\tau^2}{\bar{S}^2}+\rho\left(\frac{\tau^2}{\bar{S}^2}+\frac{1}{2\bar{S}-\tau^2}\right)\right].
\end{equation}

\emph{Step 7.} Since $\bar{S} = \tau^2+(1+Q)$, we have $\bar{S}^2-4(1+Q)\tau^2 = (\tau^2-1-Q)^2 \geq 0$, so the first term in the square brackets in \eqref{eqn:shared} satisfies
\begin{equation} \label{eqn:first}
\frac{\tau^2}{\bar{S}^2} \leq \frac{1}{4(1+Q)}.
\end{equation}

\emph{Step 8.} We will now show that the coefficient of $\rho$ in the square brackets in \eqref{eqn:shared} is bounded by $5/[8(1+Q)]$. To that end, consider the expression
\[\frac{5}{8(1+Q)}-\frac{\tau^2}{\bar{S}^2}-\frac{1}{2\bar{S}-\tau^2} = \frac{5\bar{S}^3-11(1+Q)\bar{S}^2+8(1+Q)^3}{8(1+Q)\bar{S}^2\lbrace \bar{S}+(1+Q)\rbrace},\]
where the right-hand side substitutes $\tau^2 = \bar{S}-(1+Q)$ and $2\bar{S}-\tau^2 = \bar{S}+(1+Q)$. The numerator on the right-hand side above is positive because
\[5\bar{S}^3-11(1+Q)\bar{S}^2+8(1+Q)^3 = 5\left[\bar{S}-\tfrac{22}{15}(1+Q)\right]^2\left[\bar{S}+\tfrac{11}{15}(1+Q)\right]+\tfrac{76}{675}(1+Q)^3.\]
Hence
\begin{equation} \label{eqn:coef}
\frac{\tau^2}{\bar{S}^2}+\frac{1}{2\bar{S}-\tau^2} < \frac{5}{8(1+Q)}.
\end{equation}

\emph{Step 9.} Since $\rho \geq 0$ and $Q-\rho \geq 0$, inserting \eqref{eqn:first} and \eqref{eqn:coef} into \eqref{eqn:shared} gives
\[\bar{L}(\theta,Z)-1 \leq \frac{(Q-\rho)(2+5\rho)}{8(1+Q)} < \frac{2+25Q}{160},\]
where the second inequality holds because
\[(2+25Q)(1+Q)-20(Q-\rho)(2+5\rho) = 25\left(Q-2\rho-\tfrac{13}{50}\right)^2+\tfrac{31}{100}+14\rho>0.\]
This is the bound required in Step 2. \qed

\subsection{Proof of \texorpdfstring{\cref{prop:cautious_risk_best_case}}{Proposition \ref{prop:cautious_risk_best_case}}}
\label{app:prop:cautious_risk_best_case}

(i) By \cref{lem:normalization}(iii) in \cref{app:lemmas}, with $H=\delta^{-1}I_k$, we have $R_\Sigma(\theta_\delta,a_C) = R_{\Sigma/\delta^2}(M,a_C)$ for every $\delta>0$. As $\delta \to \infty$, $\Sigma/\delta^2 \to 0$, and $M$ has full column rank, so part (ii) gives $R_{\Sigma/\delta^2}(M,a_C) \to V(M)$. Finally, $V(M)=0$ because $y \in \col(M)$.

(ii) Fix $\theta$ with full column rank $k$, and consider an arbitrary sequence of symmetric positive definite matrices $\Sigma\to0$, holding $W$ and $y$ fixed; the matrices $C$, $X$, and $Z\equiv X-\theta$ vary along the sequence. Since $R_\Sigma(\theta,a_C)$ depends only on the distribution of $X$, we may write $\ve(Z)=\Sigma^{1/2}\xi$, where $\xi\sim N(0,I_{mk})$ does not depend on $\Sigma$.

By \cref{lem:cautious_bound}(iv) in \cref{app:lemmas}, $L(\theta,a_C(X))\leq\tfrac14(1+\sqrt{1+k\|\xi\|^2})^2\nW{y}^2$, a bound that does not depend on $\Sigma$ and is integrable since $E\|\xi\|^2=mk$.

For every fixed $\xi$, as $\Sigma\to0$ we have $\Sigma^{1/2}\to0$, so $Z\to0$ and $X\to\theta$; moreover, $C\to0$ since $C$ is linear in $\Sigma$. Thus $X'WX+C\to\theta'W\theta$, which is nonsingular because $\theta$ has full column rank. By continuity of matrix inversion, $a_C(X)=(X'WX+C)^{-1}X'Wy\to(\theta'W\theta)^{-1}\theta'Wy=a_\theta$, and so $L(\theta,a_C(X))\to L(\theta,a_\theta)=V(\theta)$. The dominated convergence theorem now yields $R_\Sigma(\theta,a_C)\to V(\theta)$. \qed

\subsection{Proof of \texorpdfstring{\cref{prop:risk_bias_m1}}{Proposition \ref{prop:risk_bias_m1}}}
\label{app:prop:risk_bias_m1}
For sufficiently small $\varepsilon>0$, the pair of parameter and bias values $(\theta, b)_1 \equiv (\varepsilon, -\varepsilon)$ and $(\theta, b)_2 \equiv (-\varepsilon,\varepsilon)$ lie in the respective parameter spaces; note that both configurations generate identical data $N(0, \sigma^2)$. We thus find that, for any rule $a(X)$,
\begin{equation*}
R((\theta,b)_1, a) + R((\theta,b)_2, a) = E_{0,0}\left [ (1 - \varepsilon a(X))^2 + (1 + \varepsilon a(X))^2 \right ] = 2 + 2\varepsilon^2 E_{0,0}\left [ a(X)^2 \right ],
\end{equation*}
and so $\max\lbrace R((\theta,b)_1, a), R((\theta,b)_2, a)\rbrace \geq 1 + \varepsilon^2 E_{0,0}\left [ a(X)^2 \right ]$, i.e., the worst-case risk strictly exceeds do-nothing unless $a(X) = 0$ almost everywhere. Finally, the regret statement follows from the same display: since $V(\theta) = 0$ for every $\theta \neq 0$ when $m = k = 1$, regret equals risk at both configurations, so $\max\lbrace r((\theta,b)_1, a), r((\theta,b)_2, a)\rbrace \geq 1 + \varepsilon^2 E_{0,0}[a(X)^2]$, whereas $r(\theta,b,0) = 1 - V(\theta) \leq 1$ for all $(\theta,b)$, with equality whenever $\theta \neq 0$. \qed

\subsection{Proof of \texorpdfstring{\cref{lem:nesting}}{Lemma \ref{lem:nesting}}}
\label{app:lem:nesting}

(i) (\ref{itm:parameter_space_iv}) $\Rightarrow$ (\ref{itm:parameter_space_iii}): If $\mathcal{B} \subset \Theta$ is an $\mathbb{R}^{m \times k}$-open neighborhood of $\theta_0 \in \mathcal{S}$ with $\rk(\theta_0)>0$, then $\theta_0 \neq 0$, and $\mathcal{B} \cap \mathcal{S} \subset \Theta$ is an $\mathcal{S}$-open neighborhood of $\theta_0$.

(\ref{itm:parameter_space_iii}) $\Rightarrow$ (\ref{itm:parameter_space_ii}): Let $\theta_0 \in \mathcal{S} \setminus \lbrace 0 \rbrace$. Fix a nonzero vector $w \in \col(\theta_0)$ and define, for small $\varepsilon>0$,
\begin{equation} \label{eqn:perturb}
\theta_j \equiv \theta_0+\varepsilon we_j', \quad j=1,\dots,k,
\end{equation}
where $e_j$ denotes the $j$-th standard unit vector of $\mathbb{R}^k$. By construction, $\col(\theta_j) \subset \col(\theta_0)$, implying $\theta_j \in \mathcal{S}$. Moreover, $\theta_0,\theta_1,\dots,\theta_k$ all lie in $\Theta \cap \mathcal{S}$ for sufficiently small $\varepsilon$. The stacked matrix $(\theta_0',\theta_1',\dots,\theta_k')'$ has full column rank $k$: if for some $\alpha \in \mathbb{R}^k$ we have $\theta_j\alpha = 0$ for all $j=0,1,\dots,k$, then in particular $\theta_0\alpha=0$, so for every $j \geq 1$, $0=\theta_j\alpha=\varepsilon(e_j'\alpha)w=\varepsilon \alpha_j w$, giving $\alpha_j=0$.

(\ref{itm:parameter_space_ii}) $\Rightarrow$ (\ref{itm:parameter_space_i}): Immediate.

(ii) If $m=1$, then $\mathcal{S}=\lbrace 0 \rbrace$, so (\ref{itm:parameter_space_ii}) and (\ref{itm:parameter_space_iii}) clearly fail; hence, (\ref{itm:parameter_space_iv}) fails by (i). If $k=1$, no $\theta_0$ satisfies $0<\rk(\theta_0)<1$, so (\ref{itm:parameter_space_iv}) fails.

(iii) Since $m \geq 2$, there exists $w \in \mathbb{R}^m \setminus \lbrace 0 \rbrace$ with $w'Wy=0$. Then $\theta_0 \equiv \varepsilon we_1' \in \mathcal{S} \setminus \lbrace 0 \rbrace$, and $\Theta$ contains an $\mathbb{R}^{m \times k}$-open neighborhood of $\theta_0$ for sufficiently small $\varepsilon>0$. The intersection of this neighborhood with $\mathcal{S}$ is an $\mathcal{S}$-open neighborhood of $\theta_0$ inside $\Theta$, so (\ref{itm:parameter_space_iii}) holds, and (\ref{itm:parameter_space_ii}) follows by part (i). If moreover $k \geq 2$, then $\theta_0$ satisfies $\rk(\theta_0)=1<k$, so (\ref{itm:parameter_space_iv}) holds. \qed

\subsection{Proof of \texorpdfstring{\cref{prop:minimax_general}}{Proposition \ref{prop:minimax_general}}}
\label{app:prop:minimax_general}

(i) Let $\theta_1,\dots,\theta_J \in \Theta \cap \mathcal{S}$ be as in \cref{asn:parameter_space}(\ref{itm:parameter_space_ii}). Let $a$ be minimax; by \cref{prop:minimax}(i), which applies by \cref{lem:nesting}(i), $R(\theta,a) \leq \nW{y}^2$ on $\Theta$. Since $\theta_j \in \mathcal{S}$, \cref{lem:loss_decomp} in \cref{app:lemmas} gives
\[\nW{y}^2 \geq R(\theta_j,a) = \nW{y}^2+E_{\theta_j}\left[\nW{\theta_j a(X)}^2\right],\]
so $\theta_j a(X)=0$ $P_{\theta_j}$-almost surely. Because $f_\theta$ is everywhere positive, ``$P_\theta$-a.s.''\ and ``Lebesgue-a.e.''\ are interchangeable for every $\theta$, and the $J$ exceptional null sets may be combined; hence $(\theta_1',\dots,\theta_J')' a(x) = 0$ a.e.\ $x$, implying $a=0$ a.e.\ by full rank.

(ii) If $R(\theta,a)\leq R(\theta,0)=\nW{y}^2$ for all $\theta\in\Theta$, then $a$ is minimax by \cref{prop:minimax}(i), hence $a=0$ a.e.\ by part (i), so the inequality is nowhere strict.

For parts (iii)--(iv), let $\theta_0$ be as in \cref{asn:parameter_space}(\ref{itm:parameter_space_iv}), so that $\theta_0 \in \mathcal{S}$, $0<\rk(\theta_0)<k$, and $\Theta$ contains an $\mathbb{R}^{m \times k}$-open neighborhood $\mathcal{B}$ of $\theta_0$. Construct $\theta_1,\dots,\theta_k$ as in \eqref{eqn:perturb} in \cref{app:lem:nesting} for small $\varepsilon>0$. As shown in the proof of \cref{lem:nesting}, $\theta_j \in \mathcal{S}$ and $\col(\theta_j) \subset \col(\theta_0)$, so $\rk(\theta_j)\leq\rk(\theta_0)<k$. Moreover, for sufficiently small $\varepsilon>0$, $\mathcal{B}$ contains a ball around each of $\theta_0,\theta_1,\dots,\theta_k$. For each $j=0,1,\dots,k$, choose a unit-length vector $c_j \in \ker(\theta_j)$ (possible due to reduced rank) and set $\theta_j^{(\delta)} \equiv \theta_j+\delta yc_j'$. For sufficiently small $\delta>0$, we then have
\begin{equation} \label{eqn:perturb2}
\theta_j^{(\delta)} \in \Theta, \quad \theta_j^{(\delta)}c_j = \delta y, \quad \text{so} \quad y \in \col(\theta_j^{(\delta)}) \quad \text{and thus} \quad V(\theta_j^{(\delta)})=0.
\end{equation}

(iii) By \cref{lem:loss_decomp} in \cref{app:lemmas}, $r(\theta,0)=\nW{y}^2-V(\theta)=\nW{\theta a_\theta}^2\leq\nW{y}^2$ for every $\theta$, with equality at $\theta=\theta_0^{(\delta)}$ by \eqref{eqn:perturb2}; hence, $\sup_{\theta\in\Theta}r(\theta,0)=\nW{y}^2$. Conversely, fix any rule $a$. By \eqref{eqn:perturb2}, $r(\theta_0^{(\delta)},a)=R(\theta_0^{(\delta)},a)$, so \cref{lem:loss_decomp,lem:lsc} give
\[\sup_{\theta\in\Theta}r(\theta,a) \geq \liminf_{\delta\downarrow0}R(\theta_0^{(\delta)},a) \geq R(\theta_0,a) \geq \nW{y}^2.\]

(iv) Let $a$ satisfy $r(\theta,a)\leq\nW{y}^2$ on $\Theta$. By \eqref{eqn:perturb2}, $R(\theta_j^{(\delta)},a)=r(\theta_j^{(\delta)},a)\leq\nW{y}^2$, so \cref{lem:lsc} yields $R(\theta_j,a)\leq\nW{y}^2$ upon letting $\delta\downarrow0$. Since $\theta_j \in \mathcal{S}$, \cref{lem:loss_decomp} in \cref{app:lemmas} then gives $E_{\theta_j}[\nW{\theta_ja(X)}^2]=0$. Since the proof of \cref{lem:nesting} in \cref{app:lem:nesting} shows that $(\theta_0',\dots,\theta_k')'$ has full column rank, the conclusion $a=0$ a.e.\ follows as in the proof of part (i). \qed

\begin{lem} \label{lem:lsc}
Under \cref{asn:density}, for every rule $a$, the map $\theta\mapsto R(\theta,a)$ is lower semicontinuous on $\Theta$.
\end{lem}
\begin{proof}
$R(\theta,a)=\int\nW{y-\theta a(x)}^2\,f_\theta(x)\,dx$. For each fixed $x$ the integrand is nonnegative and continuous in $\theta$, so Fatou's lemma yields $R(\theta,a)\leq\liminf_{n}R(\theta_n,a)$ whenever $\theta_n\to\theta$.
\end{proof}

\subsection{Proof of \texorpdfstring{\cref{prop:reduction}}{Proposition \ref{prop:reduction}}}
\label{app:prop:reduction}

Throughout, $W = I_m$, $\tr(\Sigma) = 1$, and $\|y\| = 1$ as in \cref{app:num_cautious}, so that $a_C(X) = X'y/(1+\|X\|^2)$.

(i) Define $\Pi \equiv I_m-yy'-\vartheta\vartheta'$. By Cauchy-Schwarz, $|\rho| \leq 1$. Because $\Sigma y = \sigma_1^2y+\rho\sigma_1\sigma_2\vartheta+\Pi \Sigma y$ is an orthogonal decomposition,
\begin{equation} \label{eqn:sperp}
\sigma_\perp^2 = \frac{\|\Pi \Sigma y\|^2}{\sigma_1^2} \geq 0.
\end{equation}
Write $Z \equiv X-\theta$, $a \equiv a_C(X)$, and $S \equiv 1+\|X\|^2$, and identify $\check{X}_1 = X'y$ and $\check{X}_2 = X'\vartheta$, which have the bivariate normal law of the proposition. Since $\Pi\theta = 0$,
\begin{equation} \label{eqn:decomp}
S = 1+\check{X}_1^2+\check{X}_2^2+\|\Pi Z\|^2 \geq \underline{S}.
\end{equation}
Since $a = \check{X}_1/S$ and $\|\theta\|^2 = \tau_1^2+\tau_2^2$, expanding $\|y-\theta a\|^2$ gives
\begin{align}
R(\theta,a_C)-1 &= -2\tau_1E[a]+\left(\tau_1^2+\tau_2^2\right)E\left[a^2\right] \nonumber \\
&=\left(\tau_1^2+\tau_2^2\right)E\left[\frac{\check{X}_1^2}{S^2}\right]-2E\left[\frac{(\tau_1\check{X}_1)^+}{S}\right]+2E\left[\frac{(\tau_1\check{X}_1)^-}{S}\right], \label{eqn:risk_expansion}
\end{align}
where $\tau_1a = (\tau_1\check{X}_1)^+/S-(\tau_1\check{X}_1)^-/S$. Conditionally on $\check{X}_1$, the vector $(Z'\vartheta,\Pi Z)$ is Gaussian with $E[Z'\vartheta \mid \check{X}_1] = \rho\sigma_2\zeta$, $\var(Z'\vartheta \mid \check{X}_1) = \sigma_2^2(1-\rho^2)$, $E[\Pi Z \mid \check{X}_1] = \Pi\Sigma y\,\zeta/\sigma_1$, and $\var[\Pi Z \mid \check{X}_1] = \Pi\Sigma\Pi-(\Pi\Sigma y)(\Pi\Sigma y)'/\sigma_1^2$. Hence $E[\check{X}_2 \mid \check{X}_1] = \check{X}_{2\mid 1}$, while $\tr(\Pi\Sigma\Pi) = \tr(\Sigma)-\sigma_1^2-\sigma_2^2 = 1-\sigma_1^2-\sigma_2^2$ and $E\left[\|\Pi Z\|^2 \mid \check{X}_1\right] = \sigma_\perp^2\left(\zeta^2-1\right)+\tr(\Pi\Sigma\Pi)$. Taking the conditional expectation of \eqref{eqn:decomp},
\[E[S \mid \check{X}_1] = 1+\check{X}_1^2+\check{X}_{2\mid 1}^2+\sigma_2^2\left(1-\rho^2\right)+\sigma_\perp^2 \zeta^2+1-\sigma_1^2-\sigma_2^2-\sigma_\perp^2 = \bar{S}.\]
The conditional covariance matrix of $\Pi Z$ is positive semidefinite, so its trace $1-\sigma_1^2-\sigma_2^2-\sigma_\perp^2$ is nonnegative.

Since $S \geq 1>0$, Jensen's inequality conditional on $\check{X}_1$ gives $E[1/S \mid \check{X}_1] \geq 1/\bar{S}$. Because $(\tau_1\check{X}_1)^+ \geq 0$ is a function of $\check{X}_1$, we then have $E[(\tau_1\check{X}_1)^+/S] \geq E[(\tau_1\check{X}_1)^+/\bar{S}]$. The other two terms in \eqref{eqn:risk_expansion} are bounded using \eqref{eqn:decomp}, yielding the bound $\Psi$ in \eqref{eqn:Psi}.

(ii) Since $\sigma_1^2+\sigma_2^2 = 1$, we have $\tr(\var(\Pi Z)) = \tr(\Pi\Sigma\Pi) = 0$, implying $\Pi Z=0$ almost surely; moreover, $\sigma_\perp=0$ since $\sigma_1^2+\sigma_2^2+\sigma_\perp^2\leq 1$. Since $|\rho|=|\corr(\check{X}_1,\check{X}_2)|=1$, we have $\check{X}_2 = \check{X}_{2\mid 1}$ almost surely. It follows from these results that $S = \underline{S} = \bar{S}$, so the bounds on the three terms in \eqref{eqn:risk_expansion} from the proof of part (i) are equalities, yielding $R(\theta,a_C)-1 = \Psi$.

Since $\Pi Z = 0$ almost surely, we have $\Pi \Sigma = \Sigma \Pi = 0$, implying $\Sigma = (I_m-\Pi)\Sigma(I_m-\Pi)$, or after substituting $I_m-\Pi = yy'+\vartheta\vartheta'$,
\[\Sigma = y\underbrace{(y'\Sigma y)}_{\sigma_1^2}y'+y\underbrace{(y'\Sigma\vartheta)}_{\rho\sigma_1\sigma_2}\vartheta'+\vartheta\underbrace{(\vartheta'\Sigma y)}_{\rho\sigma_1\sigma_2}y'+\vartheta\underbrace{(\vartheta'\Sigma\vartheta)}_{\sigma_2^2}\vartheta' = vv',\]
with $v = \sigma_1 y + \rho \sigma_2 \vartheta$, where the final equality in the display exploits $|\rho|=1$. \qed

\subsection{Proof of \texorpdfstring{\cref{lem:compact}}{Lemma \ref{lem:compact}}}
\label{app:lem:compact}

Put $\tau^2 \equiv \|\theta\|^2$, and write $Z \equiv X-\theta$, $a \equiv a_C(X)$, $\ell \equiv \|y-Xa\|$, and $D \equiv 2+\|X\|^2$. Since $W=I_m$, $\|y\|=1$, and $C=\tr(\Sigma)=1$, we have $(1+\|X\|^2)a = X'y$, so
\[\ell^2 = 1-2aX'y+a^2\|X\|^2 = 1-a^2\left(2+\|X\|^2\right), \quad\text{i.e.}\quad \ell^2+Da^2 = 1.\]
Because $y-\theta a = (y-Xa)+Za$ and $\|Za\| = |a|\,\|Z\|$, the triangle inequality followed by the Cauchy-Schwarz inequality in $\mathbb{R}^2$ gives
\begin{align*}
\|y-\theta a\| &\leq \ell+|a|\,\|Z\| = \ell \times 1+\left(\sqrt{D}|a|\right) \times \frac{\|Z\|}{\sqrt{D}} \\
&\leq \sqrt{\ell^2+Da^2} \sqrt{1+\frac{\|Z\|^2}{D}} = \sqrt{1+\frac{\|Z\|^2}{D}}.
\end{align*}
Squaring and taking expectations gives $R(\theta,a_C)-1 \leq E\left[\|Z\|^2/(2+\|X\|^2)\right]$. On $\lbrace \|Z\|^2 \leq \tau^2/4 \rbrace$ we have $\|X\| \geq \|\theta\|-\|Z\| \geq \|\theta\|/2$, so the integrand is at most $\|Z\|^2/(2+\tau^2/4)$; on the complement it is at most $\|Z\|^2/2 \leq \|Z\|^2-\tau^2/8$. Since $E\|Z\|^2 = \tr(\Sigma) = 1$,
\[R(\theta,a_C)-1 \leq \frac{1}{2+\tau^2/4}+E\left[\left(\|Z\|^2-\frac{\tau^2}{8}\right)^+\right].\]
Writing $\|Z\|^2 = \sum_i \lambda_i\xi_i^2$ with $\lambda_i \geq 0$, $\sum_i \lambda_i = 1$, and $\xi \sim N(0,I_m)$, the last expectation is convex in $\lambda$ and hence maximized at a vertex of the simplex, where $\|Z\|^2 \sim \chi_1^2$. Both terms decrease in $\tau^2$, and at $\tau^2 = 48$ their sum is $\frac{1}{14}+E\left[\left(\xi_1^2-6\right)^+\right] \approx 0.0972$. \qed

\subsection{Proof of \texorpdfstring{\cref{prop:cautious_risk_spherical}}{Proposition \ref{prop:cautious_risk_spherical}}}
\label{app:prop:cautious_risk_spherical}

Parameterize $\theta = \sigma\tau v$ with $\|v\| = 1$, so $\tau = \|\theta\|/\sigma$ is the signal-to-noise ratio, and let $\varsigma \equiv v'y/\|y\| \in [-1,1]$ be the cosine of the angle between $\theta$ and $y$. For any $\theta$, write the relative risk of the cautious rule vs.\ do-nothing as
\[\rho_m(\tau,\varsigma) \equiv \frac{R(\theta,a_C)}{\|y\|^2}.\]
We show that, for any $m \geq 1$,
\[\rho_m(\tau,\varsigma) = 1 \;+\; \tau^2A_m(\tau) \;+\; \varsigma^2\big(\tau^4D_m(\tau) - 2\tau^2B_m(\tau)\big)\]
where
\begin{gather*}
A_m(\tau) \equiv \int_0^\infty \frac{\nu}{1+2\nu}\,H_{m,\tau}(\nu)\,d\nu,\quad B_m(\tau) \equiv \int_0^\infty \frac{H_{m,\tau}(\nu)}{1+2\nu}\,d\nu,\\
D_m(\tau) \equiv \int_0^\infty \frac{\nu}{(1+2\nu)^2}\,H_{m,\tau}(\nu)\,d\nu,
\end{gather*}
with
\[H_{m,\tau}(\nu) \equiv e^{-m\nu}(1+2\nu)^{-m/2}\exp\Big\lbrace-\frac{\nu\,\tau^2}{1+2\nu}\Big\rbrace.\]
We furthermore show that the risk surface $\rho_m(\tau,\varsigma)$ is strictly decreasing in $\varsigma^2$ for $\tau > 0$, and so for $m \geq 2$ the worst case over $\theta$ is attained at $\varsigma = 0$, i.e., at $\theta \perp y$.

By \cref{lem:normalization} in \cref{app:lemmas}, we may set $\|y\|=\sigma = 1$ without loss of generality. Then $X = \tau v + Z$, $Z \sim N(0, I_m)$, and $C = \tr(I_m) = m$, so the denominator of the cautious rule is $X'X + C = \|X\|^2 + m \equiv S$, giving
\[a_C(X) = \frac{X'y}{S} = \frac{\check{X}_1}{S}\]
with $\check{X}_1 \equiv X'y$. The relative risk is therefore
\[\rho_m(\tau,\varsigma) = R(\theta,a_C) = 1 - 2E\left[y'\theta a_C\right] + \|\theta\|^2E\left[a_C^2\right] = 1 - 2\tau\varsigma\,E[\check{X}_1/S] + \tau^2E[\check{X}_1^2/S^2],\]
where we have substituted $\theta = \tau v$, so that $y'\theta = \tau\varsigma$ and $\|\theta\|^2 = \tau^2\|v\|^2 = \tau^2$. To evaluate these moments, we apply a standard Laplace transform trick \citep[e.g.,][]{Meng2005}. By an exponential tilting argument, $E[g(X)e^{-\nu S}] = H_{m,\tau}(\nu)E[g(\tilde{X})]$ for any integrable $g(\cdot)$, where $\tilde{X} \sim N(\tau v/(1+2\nu),I_m/(1+2\nu))$. Hence,
\[E[\check{X}_1e^{-\nu S}] = \frac{\tau\varsigma}{1+2\nu}H_{m,\tau}(\nu),\quad E[\check{X}_1^2e^{-\nu S}] = \Big(\frac{1}{1+2\nu} + \frac{\tau^2\varsigma^2}{(1+2\nu)^2}\Big)H_{m,\tau}(\nu).\]
Since $\int_0^\infty \nu^j e^{-\nu S} \, d\nu = S^{-j-1}$ for $j=0,1$, integrating over $\nu$ gives $E[\check{X}_1/S] = \tau\varsigma\,B_m(\tau)$ and $E[\check{X}_1^2/S^2] = A_m(\tau) + \tau^2\varsigma^2D_m(\tau)$, delivering the claimed expression for the risk surface.

Next we have that
\[A_m(\tau) - 2B_m(\tau) + \tau^2D_m(\tau) \;=\; \int_0^\infty H_{m,\tau}(\nu)\,\frac{-1 + (3-2m)\nu - 2m\nu^2}{1+2\nu}\,d\nu \;<\; 0.\]
To see this, note that $-\nu H_{m,\tau}'(\nu) = \nu H_{m,\tau}(\nu)\big[m + \tfrac{m}{1+2\nu} + \tfrac{\tau^2}{(1+2\nu)^2}\big]$, while integration by parts yields $\int_0^\infty(-\nu H')d\nu = \int_0^\infty H\,d\nu$. It follows from these two equations that $\tau^2D_m = \int \frac{\nu\tau^2}{(1+2\nu)^2}Hd\nu = \int [1-\nu(m + \frac{m}{1+2\nu})]H d\nu$. Substituting into $A_m - 2B_m + \tau^2D_m$ yields the above display. Finally note that the quadratic $-1 + (3-2m)\nu - 2m\nu^2$ is negative for all $\nu \geq 0$: for $m \geq 2$ every coefficient is negative, and for $m = 1$ it equals $-1 + \nu - 2\nu^2$, with discriminant $1 - 8 < 0$. We conclude that the coefficient of $\varsigma^2$ is $\tau^4D_m - 2\tau^2B_m = \tau^2(A_m - 2B_m + \tau^2D_m) - \tau^2A_m < 0$, since both terms are negative for $\tau > 0$.

Since $R(\theta,0) = \|y\|^2$, it follows that, for $m \geq 2$,
\[\sup_{\theta \in \mathbb{R}^m} \frac{R(\theta,a_C)}{R(\theta,0)} - 1 = \sup_{\tau \geq 0} \tau^2A_m(\tau). \qed\]

\subsection{Proof of \texorpdfstring{\cref{prop:adm}}{Proposition \ref{prop:adm}}}
\label{app:prop:adm}

Suppose a rule $a$ dominating $a_C$ exists. Both $a$ and $a_C$ have bounded risk by \cref{prop:cautious_risk}(i), so by \cref{lem:second_moment} their second moments are bounded uniformly in $\theta$, and hence their risk functions are continuous \citep[Example 5.7.10]{LehmannCasella1998}. Therefore $D(\theta) \equiv R(\theta,a_C)-R(\theta,a) \geq 0$ is continuous and positive somewhere, and $\mathcal{A} \equiv \lbrace \theta \colon D(\theta)>0 \rbrace$ has positive Lebesgue measure.

For a measure $\nu$ on $\mathbb{R}^{m \times k}$, write $R(\nu,a) \equiv \int R(\theta,a)\,\nu(d\theta)$. For $\tau>0$, let $\nu_\tau$ be the finite measure on $\mathbb{R}^{m \times k}$ with density
\[g_\tau(\theta) \equiv \exp\left\lbrace -\frac{1}{2\tau^2}\ve(\theta)'\Sigma^{-1}\ve(\theta)\right\rbrace,\]
and total mass
\begin{equation} \label{eqn:prior_mass}
c_\tau \equiv \nu_\tau(\mathbb{R}^{m \times k}) = (2\pi\tau^2)^{mk/2}|\Sigma|^{1/2}.
\end{equation}
Let $\pi_\tau \equiv \nu_\tau/c_\tau$, corresponding to the prior $\ve(\theta) \sim N(0,\tau^2\Sigma)$. For a probability measure $w$ on $[1,\infty)$ with compact support, let $\nu_w \equiv \int \nu_\tau\,w(d\tau)$, with density $g_w \equiv \int g_\tau\,w(d\tau)$ and total mass $c_w \equiv \int c_\tau\,w(d\tau)<\infty$, and let $\pi_w \equiv \nu_w/c_w$, corresponding to the hierarchical prior $\tau \sim c_\tau w(d\tau)/c_w$, $\ve(\theta) \mid \tau \sim N(0,\tau^2\Sigma)$. We will choose concrete mixtures $w$ below. All integrals below are finite, because $\sup_\theta R(\theta,a_C)<\infty$ by \cref{prop:cautious_risk}(i), $\nu_w$ is a finite measure, and $R(\nu_w,a) \leq R(\nu_w,a_C)$.

Since $g_1 > 0$ everywhere,
\[\eta \equiv \int_{\mathcal{A}} D(\theta)g_1(\theta)\,d\theta > 0 .\]
For every $\theta$, $g_\tau(\theta)$ is nondecreasing in $\tau$, so $g_w \geq g_1$ pointwise. Hence, since $D \geq 0$,
\begin{align*}
\eta &\leq \int D(\theta)g_w(\theta)\,d\theta = R(\nu_w,a_C)-R(\nu_w,a) \\
&\leq R(\nu_w,a_C)-\inf_{\tilde{a}}R(\nu_w,\tilde{a}) = c_w\left[R(\pi_w,a_C)-R(\pi_w,a_{\pi_w})\right],
\end{align*}
where the last equality uses \cref{lem:bayes} (which applies, due to \cref{lem:gauss}(i)), and $a_{\pi_w}$ is defined in \cref{lem:bayes}. By \cref{lem:gauss}(ii),
\begin{equation} \label{eqn:master}
\eta \leq c_w\,E\left[E[(1+\tau^2)^{-1} \mid X]^2\right]\nW{y}^2,
\end{equation}
where the expectations are under $\pi_w$.

If $mk \leq 3$, take $w = \delta_{\bar{\tau}}$ (a point mass at $\bar{\tau}$) for $\bar{\tau} \geq 1$: then \eqref{eqn:master} and \eqref{eqn:prior_mass} give
\[\eta \leq \frac{c_{\bar{\tau}}}{(1+\bar{\tau}^2)^2} \nW{y}^2 = (2\pi)^{mk/2}|\Sigma|^{1/2}\frac{\bar{\tau}^{mk}}{(1+\bar{\tau}^2)^2}\nW{y}^2 \to 0 \quad \text{as}\quad \bar{\tau} \to \infty.\]
If $mk = 4$, take $w = w_n$ with $n \geq 4$ as in \cref{lem:mix}: then \eqref{eqn:master} and \cref{lem:mix} give
\[\eta \leq \frac{16\pi^2|\Sigma|^{1/2}}{\log(n/2)}\left(2+\log\frac{\log(n/2)}{\log 2}\right)\nW{y}^2 \to 0 \quad \text{as} \quad n \to \infty.\]
Either way, this contradicts $\eta>0$. \qed

\begin{lem} \label{lem:bayes}
Let $\pi$ be a probability measure on $\mathbb{R}^{m \times k}$ with $\int \|\theta\|^2\,\pi(d\theta)<\infty$, and suppose $Q(X) \equiv E[\theta'W\theta \mid X]$ is positive definite for almost every $X$. Define $a_\pi(X) \equiv Q(X)^{-1}E[\theta \mid X]'Wy$. Then, for every decision rule $a$,
\[R(\pi,a) = R(\pi,a_\pi)+E\left[(a(X)-a_\pi(X))'Q(X)(a(X)-a_\pi(X))\right],\]
where $R(\pi,a_\pi) \leq \nW{y}^2<\infty$. In particular, $a_\pi$ minimizes $R(\pi,\cdot)$.
\end{lem}

\begin{proof}
For every $\alpha \in \mathbb{R}^k$,
\[E[L(\theta,\alpha) \mid X] = E[L(\theta,a_\pi(X)) \mid X]+(\alpha-a_\pi(X))'Q(X)(\alpha-a_\pi(X)),\]
by completing the square. Evaluate at $\alpha=a(X)$ and take expectations over $X$; both sides may equal $+\infty$. Finally, evaluating the risk equality at $a=0$, we get $R(\pi,a_\pi) \leq R(\pi,0)=\nW{y}^2$.
\end{proof}

\begin{lem} \label{lem:gauss}
Let $w$ be a probability measure on $[1,\infty)$ with compact support. Write $\Gamma \equiv X'WX$, $S \equiv \Gamma+C$, $b \equiv X'Wy$, $q(x) \equiv \ve(x)'\Sigma^{-1}\ve(x)$, $\rho(\tau) \equiv \frac{\tau^2}{1+\tau^2}$, and $h_\tau(x) \equiv \rho(\tau)^{mk/2}e^{-(1-\rho(\tau))q(x)/2}$. Under $\pi_w$:
\begin{enumerate}[(i)]
\item $X$ has density $c_w^{-1}\int h_\tau(\cdot)\,w(d\tau)$, the conditional distribution of $\tau$ given $X$ is
\[P(d\tau \mid X) = \frac{h_\tau(X)\,w(d\tau)}{\int h_{\tilde{\tau}}(X)\,w(d\tilde{\tau})},\]
and $\ve(\theta) \mid (X,\tau) \sim N(\rho(\tau)\ve(X),\rho(\tau)\Sigma)$. Hence, with $\bar\rho_j(X) \equiv E[\rho(\tau)^j \mid X]$ and $\kappa(X) \equiv \bar\rho_2(X)/\bar\rho_1(X)$,
\[Q(X) = \bar\rho_2 \Gamma+\bar\rho_1 C = \bar\rho_1\left(\kappa \Gamma+C\right) \succ 0, \quad a_{\pi_w}(X) = (\kappa \Gamma+C)^{-1}b ,\]
where the $X$ arguments are omitted from $\bar{\rho}_j$ and $\kappa$ for brevity.
\item $R(\pi_w,a_C)-R(\pi_w,a_{\pi_w}) \leq E\left[E[1-\rho(\tau) \mid X]^2\right]\nW{y}^2$.
\end{enumerate}
\end{lem}

\begin{proof}
(i) Given $\tau$, write $\rho \equiv \rho(\tau)$. The conditional posterior precision matrix of $\ve(\theta)$ is $\Sigma^{-1}+\tau^{-2}\Sigma^{-1} = \Sigma^{-1}/\rho$, so the conditional posterior covariance matrix is $\rho\Sigma$ and the conditional posterior mean is $\rho\Sigma\Sigma^{-1}\ve(X) = \rho\ve(X)$. Moreover, $\ve(X) \mid \tau \sim N(0,(1+\tau^2)\Sigma)$, whose density at $\ve(x)$ equals, by \eqref{eqn:prior_mass},
\[(2\pi)^{-mk/2}|(1+\tau^2)\Sigma|^{-1/2}e^{-q(x)/(2(1+\tau^2))} = h_\tau(x)/c_\tau ;\]
in particular, $\int h_\tau(x)\,dx = c_\tau$. Since $\tau$ has distribution $c_\tau w(d\tau)/c_w$ under $\pi_w$, the stated density of $X$ and conditional distribution of $\tau$ given $X$ follow. Hence
\[E[\theta'W\theta \mid X,\tau]_{ij} = (\rho X_{\bullet i})'W(\rho X_{\bullet j})+\tr(W \rho\Sigma^{ij}) = \rho^2 \Gamma_{ij}+\rho C_{ij}, \quad E[\theta \mid X,\tau]'Wy = \rho b,\]
and averaging over $P(d\tau \mid X)$ gives $Q(X) = \bar\rho_2 \Gamma+\bar\rho_1 C$ and $E[\theta \mid X]'Wy = \bar\rho_1 b$. Since $\rho(\tau) \geq 1/2$ for $\tau \geq 1$, we have $\bar\rho_1 \geq 1/2$, so $Q(X) \succeq \bar\rho_1 C \succ 0$ and $a_{\pi_w}(X) = Q(X)^{-1}E[\theta \mid X]'Wy = (\kappa \Gamma+C)^{-1}b$.

(ii) Since $\rho(\tau)^2 \leq \rho(\tau)$, we have $\kappa \leq 1$. Moreover, $\bar\rho_1^2 \leq \bar\rho_2$ by Cauchy-Schwarz, so
\begin{equation} \label{eqn:cs}
\frac{\bar\rho_1}{\kappa} = \frac{\bar\rho_1^2}{\bar\rho_2} \leq 1, \quad \text{and} \quad 0\leq 1-\kappa \leq 1-\bar\rho_1 = E[1-\rho(\tau) \mid X] .
\end{equation}
Algebra gives
\[a_C-a_{\pi_w} = \left[S^{-1}-(\kappa \Gamma+C)^{-1}\right]b = -(1-\kappa) M b,\]
where
\[M \equiv S^{-1}\Gamma(\kappa \Gamma+C)^{-1} = (\kappa \Gamma+C)^{-1}\Gamma S^{-1},\]
so $M$ is symmetric and, using the two expressions for $M$ on either side of $Q = \bar\rho_1(\kappa \Gamma+C)$,
\[MQ(X)M = \bar\rho_1S^{-1}\Gamma(\kappa \Gamma+C)^{-1}\Gamma S^{-1} .\]
By \cref{lem:bayes} and part (i),
\[R(\pi_w,a_C)-R(\pi_w,a_{\pi_w}) = E\left[(1-\kappa)^2\bar\rho_1\, b'S^{-1}\Gamma(\kappa \Gamma+C)^{-1}\Gamma S^{-1}b\right] .\]
It remains to bound the integrand by $E[1-\rho(\tau) \mid X]^2\nW{y}^2$ for every $X$. First, $\kappa \Gamma+C \succeq \kappa S$ because $(1-\kappa)C \succeq 0$, so $(\kappa \Gamma+C)^{-1} \preceq \kappa^{-1}S^{-1}$ and hence
\[b'S^{-1}\Gamma(\kappa \Gamma+C)^{-1}\Gamma S^{-1}b \leq \kappa^{-1}b'S^{-1}\Gamma S^{-1}\Gamma S^{-1}b .\]
Second, with $\check{\Gamma} \equiv S^{-1/2}\Gamma S^{-1/2}$ we have $0 \preceq \check{\Gamma} \preceq I_k$, so $\check{\Gamma}^2 \preceq I_k$ and
\[S^{-1}\Gamma S^{-1}\Gamma S^{-1} = S^{-1/2}\check{\Gamma}^2S^{-1/2} \preceq S^{-1} .\]
Third,
\[b'S^{-1}b = y'WXa_C(X) \leq \nW{y}\nW{Xa_C(X)} \leq \nW{y}^2,\]
by Cauchy-Schwarz and \cref{lem:cautious_bound}(i) in \cref{app:lemmas}. Combining the three displays and \eqref{eqn:cs} bounds the integrand as required.
\end{proof}

\begin{lem} \label{lem:mix}
Let $mk = 4$ and $n \geq 4$. Let $w_n$ be the distribution of $\tau = (1/\lambda-1)^{1/2}$ when $\lambda$ has density $(\Lambda\lambda)^{-1}$ on $[1/n,1/2]$, where $\Lambda \equiv \log(n/2)$. Then, with expectations under $\pi_{w_n}$,
\[c_{w_n}\,E\left[E[(1+\tau^2)^{-1} \mid X]^2\right] \leq \frac{16\pi^2|\Sigma|^{1/2}}{\Lambda}\left(2+\log\frac{\Lambda}{\log 2}\right).\]
\end{lem}

\begin{proof}
By \cref{lem:gauss}(i),
\[c_{w_n}\,E\left[E[(1+\tau^2)^{-1} \mid X]^2\right] = \int \frac{\left(\int (1+\tau^2)^{-1}h_\tau(x)\,w_n(d\tau)\right)^2}{\int h_\tau(x)\,w_n(d\tau)}\,dx .\]
By the definition of $w_n$, substituting $\lambda = (1+\tau^2)^{-1}$ replaces $w_n(d\tau)$ by $\mu_n(d\lambda) \equiv (\Lambda\lambda)^{-1}d\lambda$ on $[1/n,1/2]$, and $h_\tau(x)$ by $(1-\lambda)^2e^{-\lambda q(x)/2}$ since $mk=4$. Since $(1-\lambda)^2 \in [1/4,1]$ and $\lambda\,\mu_n(d\lambda) = \Lambda^{-1}d\lambda$, the numerator is at most $\omega(q(x))^2$ and the denominator at least $\Omega(q(x))/4$, where
\[\omega(q) \equiv \frac{1}{\Lambda}\int_{1/n}^{1/2}e^{-\lambda q/2}\,d\lambda = \frac{2}{\Lambda q}\Delta(q), \quad \Delta(q) \equiv e^{-q/(2n)}-e^{-q/4},\]
\[\Omega(q) \equiv \frac{1}{\Lambda}\int_{1/n}^{1/2}e^{-\lambda q/2}\,\frac{d\lambda}{\lambda}, \quad \Omega(0)=1 .\]
Changing variables to $\Sigma^{-1/2}\ve(x)$, with Jacobian $|\Sigma|^{1/2}$, and then to polar coordinates with radius $t$, so that $q(x)=t^2$ (the unit sphere in $\mathbb{R}^4$ has area $2\pi^2$),
\begin{equation} \label{eqn:polar}
c_{w_n}\,E\left[E[(1+\tau^2)^{-1} \mid X]^2\right] \leq 4|\Sigma|^{1/2}\cdot 2\pi^2\int_0^\infty \frac{\omega(t^2)^2}{\Omega(t^2)}\,t^3\,dt = 4\pi^2|\Sigma|^{1/2}\int_0^\infty q\,\frac{\omega(q)^2}{\Omega(q)}\,dq .
\end{equation}
Differentiating under the integral sign, $\Omega' = -\omega/2$, so $\omega/\Omega = 2\left(-\log \Omega\right)'$ and
\begin{equation} \label{eqn:ibp_setup}
\int_0^\infty q\,\frac{\omega(q)^2}{\Omega(q)}\,dq = \frac{4}{\Lambda}\int_0^\infty \Delta(q)\left(-\log\Omega\right)'(q)\,dq .
\end{equation}
Restricting the integral defining $\Omega$ to $\lambda \in [1/n,2/n]$ gives $\Omega(q) \geq e^{-q/n}\frac{1}{\Lambda}\int_{1/n}^{2/n}\frac{d\lambda}{\lambda} = \frac{\log 2}{\Lambda}e^{-q/n}$, that is,
\begin{equation} \label{eqn:logOmega}
0 \leq -\log\Omega(q) \leq \frac{q}{n}+\gamma, \quad \gamma \equiv \log\frac{\Lambda}{\log 2} .
\end{equation}
Since $\Delta(0) = -\log\Omega(0) = 0$ and $\Delta(q)\left(-\log\Omega(q)\right) \leq e^{-q/(2n)}\left(q/n+\gamma\right) \to 0$ as $q \to \infty$, integration by parts gives (all integrals being finite by \eqref{eqn:logOmega})
\[\int_0^\infty \Delta\left(-\log\Omega\right)'dq = \int_0^\infty \left(-\Delta'\right)\left(-\log\Omega\right)dq, \quad -\Delta'(q) = \frac{1}{2n}e^{-q/(2n)}-\frac{1}{4}e^{-q/4} .\]
Since $\frac{1}{4}e^{-q/4}\left(-\log\Omega\right) \geq 0$, dropping this term and applying \eqref{eqn:logOmega},
\[\int_0^\infty \left(-\Delta'\right)\left(-\log\Omega\right)dq \leq \frac{1}{2n}\int_0^\infty e^{-q/(2n)}\left(\frac{q}{n}+\gamma\right)dq = 2+\gamma .\]
Combining the last display with \eqref{eqn:polar} and \eqref{eqn:ibp_setup} yields the result.
\end{proof}

\begin{lem} \label{lem:second_moment}
Assume Gaussian data \eqref{eqn:X_gauss}. Every policy rule $a$ satisfies
\[\sup_{\theta \in \mathbb{R}^{m \times k}} E_\theta\|a(X)\|^2 \leq \frac{2^{mk/2+1}}{\lambda_{\min}(C)}\sup_{\theta \in \mathbb{R}^{m \times k}} R(\theta,a),\]
where $C$ is defined in \eqref{eqn:cautious}.
\end{lem}

\begin{proof}
Fix $\theta_0 \in \mathbb{R}^{m \times k}$. Let $Z_1$ and $Z_2$ be independent copies of $Z \equiv X-\theta$, let $\bar{X} \equiv \theta_0+Z_1+Z_2$, and let $E$ denote expectation over $(Z_1,Z_2)$. Then
\begin{align*}
\sup_\theta R(\theta,a) &\geq E\left[R(\theta_0+Z_2,a)\right] = \tfrac12 E\left[L\big(\theta_0+Z_1,a(\bar{X})\big)+L\big(\theta_0+Z_2,a(\bar{X})\big)\right] \\
&\geq \tfrac14 E\left[\nW{(Z_1-Z_2)a(\bar{X})}^2\right] = \tfrac12 E\left[a(\bar{X})'Ca(\bar{X})\right] \geq \tfrac12\lambda_{\min}(C)\,E\|a(\bar{X})\|^2 .
\end{align*}
The first equality uses exchangeability of $(Z_1,Z_2)$, the second inequality uses $\nW{v_1}^2+\nW{v_2}^2 \geq \frac{1}{2}\nW{v_1-v_2}^2$, and the second equality uses independence of $Z_1-Z_2$ and $Z_1+Z_2$ together with $E[(Z_1-Z_2)'W(Z_1-Z_2)]=2C$. Finally, if $f_{\theta;\Sigma}$ denotes the density of $N(\ve(\theta),\Sigma)$, then the density of $\ve(\bar{X})$ is $f_{\theta_0;2\Sigma}(x) = 2^{-mk/2}f_{0;\Sigma}\left((x-\ve(\theta_0))/\sqrt{2}\right) \geq 2^{-mk/2}f_{\theta_0;\Sigma}(x)$; hence, $E\|a(\bar{X})\|^2 \geq 2^{-mk/2}E_{\theta_0}\|a(X)\|^2$.
\end{proof}

\subsection{Proof of \texorpdfstring{\cref{prop:bias_costly}}{Proposition \ref{prop:bias_costly}}}
\label{app:prop:bias_costly}

The maximizer $\theta^*$ exists, and the worst-case excess risk $\gamma$ is strictly positive. To see this, first note that $R(\theta,0,a_C) - 1$ is continuous in $\theta$ (see the proof of \cref{prop:adm} in \cref{app:prop:adm}), and equals $\|\theta\|^2 E_\theta[a_C(X)^2] > 0$ at any $\theta \in \mathcal{S} \setminus \lbrace 0 \rbrace$ (non-empty since $m \geq 2$) by \cref{lem:loss_decomp} in \cref{app:lemmas}. Moreover, the decaying bound in \cref{app:lem:compact}, applied after the scale normalization $\tr(\Sigma) = 1$ of \cref{lem:normalization}(iii) in \cref{app:lemmas}, gives $\limsup_{\|\theta\| \to \infty} \big[R(\theta,0,a_C)-1\big] \leq 0$. Thus, the excess risk outside a sufficiently large closed ball is smaller than its value at a fixed $\theta_0 \in \mathcal{S} \setminus \lbrace 0 \rbrace$. By continuity, it attains its global maximum on that ball, and this maximum is strictly positive.

Consider any $b \in \mathcal{B}_K$ (so $bb' \preceq K^2 \Delta$), and note that $-b \in \mathcal{B}_K$. Since $\Theta = \mathbb{R}^m$, the parameter-bias configurations $(\theta^* - b, b)$ and $(\theta^* + b, -b)$ are both feasible, and both induce the same signal distribution $X \sim N(\theta^*, \Sigma)$. Expanding the risk,
\begin{equation*}
R(\theta^* \mp b, \pm b, a_C) = 1 + \gamma \pm 2 b' \left ( y \, E_{\theta^*}[a_C(X)] - \theta^* E_{\theta^*}[a_C(X)^2] \right ) + \|b\|^2 \, E_{\theta^*}[a_C(X)^2].
\end{equation*}
The worst-case risk is bounded below by the larger of the two, and so, for every $b \in \mathcal{B}_K$,
\begin{equation*}
\sup_{\theta \in \Theta}\sup_{\tilde{b} \in \mathcal{B}_K} R(\theta,\tilde{b},a_C) \; \geq \; 1 + \gamma + 2 \left | b' \left ( y \, E_{\theta^*}[a_C(X)] - \theta^* E_{\theta^*}[a_C(X)^2] \right ) \right | + \|b\|^2 \, E_{\theta^*}[a_C(X)^2].
\end{equation*}
We next verify that $\omega^* \neq 0$. Suppose that $y \, E_{\theta^*}[a_C(X)] - \theta^* E_{\theta^*}[a_C(X)^2] = 0$, i.e., $\theta^* = c \, y$ with $c \equiv E_{\theta^*}[a_C(X)]/E_{\theta^*}[a_C(X)^2]$. Then, evaluating the risk at the maximizer,
\begin{equation*}
1 + \gamma \; = \; R(\theta^*, 0, a_C) \; = \; 1 - 2 c \, E_{\theta^*}[a_C(X)] + c^2 \, E_{\theta^*}[a_C(X)^2] \; = \; 1 - \frac{E_{\theta^*}[a_C(X)]^2}{E_{\theta^*}[a_C(X)^2]} \; \leq \; 1,
\end{equation*}
contradicting $\gamma > 0$. Since $\Delta \succ 0$, it follows that $\omega^* \neq 0$.

Finally take $b = K \Delta^{1/2} v$ with $\|v\| = 1$, which lies in $\mathcal{B}_K$ by \eqref{eqn:bias_set}, and let $v$ be either the leading eigenvector of $\Delta$, so that $\|b\|^2 = K^2 \lambda_{\max}(\Delta)$ and we keep only the quadratic term, or the normalized direction $v = \omega^*/\|\omega^*\|$, so that $b' ( y \, E_{\theta^*}[a_C(X)] - \theta^* E_{\theta^*}[a_C(X)^2] ) = K \|\omega^*\|$ and we keep only the linear term. This delivers the stated maximum, and, since $\lambda_{\max}(\Delta) > 0$ and $\|\omega^*\| > 0$, both of its arguments are strictly positive, delivering the final strict inequality. \qed

\end{appendices}

\clearpage
\phantomsection
\addcontentsline{toc}{section}{References}
\bibliography{ref}

\end{document}

%% file: noisy_optpol_header.tex
\usepackage{amsmath, amsfonts, amssymb, amsthm, fullpage, color, bbm, enumerate, titlesec, graphicx, epstopdf, multirow, array, mathtools}
\usepackage[T1]{fontenc}
\usepackage{lmodern}
\usepackage{pdflscape}
\usepackage[title]{appendix}

\usepackage[onehalfspacing]{setspace}

\definecolor{darkred}{rgb}{0.5,0,0}

\titleformat*{\paragraph}{\sc}

\usepackage{hyperref}
\hypersetup{allbordercolors={1 1 1},allcolors=darkred,colorlinks=true}
\usepackage[capitalize,nameinlink,noabbrev]{cleveref}

\usepackage[font={small}]{caption}

\usepackage[round]{natbib}
\usepackage[runin]{abstract}

\theoremstyle{plain}
\newtheorem{asn}{Assumption}
\crefname{asn}{Assumption}{Assumptions}
\newtheorem{lem}{Lemma}
\crefname{lem}{Lemma}{Lemmas}
\newtheorem{prop}{Proposition}
\crefname{prop}{Proposition}{Propositions}

\crefname{cor}{Corollary}{Corollaries}
\newtheorem{num}{Numerical result}
\crefname{num}{Numerical result}{Numerical results}

\theoremstyle{definition}
\newtheorem{exm}{Example}
\crefname{exm}{Example}{Examples}

\newcommand\independent{\protect\mathpalette{\protect\independenT}{\perp}}
\def\independenT#1#2{\mathrel{\rlap{$#1#2$}\mkern2mu{#1#2}}}

\newcommand{\nW}[1]{\|#1\|_W}
\newcommand{\nop}[1]{\|#1\|_{\mathrm{op}}}

\DeclareMathOperator*{\argmin}{argmin}
\DeclareMathOperator*{\tr}{trace}
\DeclareMathOperator*{\ve}{vec}

\DeclareMathOperator*{\rk}{rank}
\DeclareMathOperator*{\diag}{diag}

\DeclareMathOperator{\var}{Var}
\DeclareMathOperator{\cov}{Cov}
\DeclareMathOperator{\corr}{Corr}

\DeclareMathOperator{\col}{col}

\allowdisplaybreaks


%% file: ref.bib
@incollection{Ramey2016,
	author = {Ramey, Valerie A.},
	booktitle = {Handbook of Macroeconomics},
	chapter = {2},
	doi = {10.1016/bs.hesmac.2016.03.003},
	editor = {Taylor, John B. and Uhlig, Harald},
	pages = {71--162},
	publisher = {Elsevier},
	title = {Macroeconomic Shocks and Their Propagation},
	url = {http://dx.doi.org/10.1016/bs.hesmac.2016.03.003},
	volume = {2A},
	year = {2016}}

@article{Eberly2020,
	author = {Eberly, Janice C. and Stock, James H. and Wright, Jonathan H.},
	journal = {International Journal of Central Banking},
	number = {1},
	pages = {5--71},
	title = {The {F}ederal {R}eserve's Current Framework for Monetary Policy: A Review and Assessment},
	volume = {16},
	year = {2020}}

@article{MontielOlea2026,
	author = {Montiel Olea, Jos{\'e} Luis and Qiu, Chen and Stoye, J{\"o}rg},
	journal = {Review of Economic Studies},
	note = {Forthcoming},
	publisher = {Oxford University Press UK},
	title = {Decision theory for treatment choice problems with partial identification},
	year = {2026}}

@article{Manski2004,
	author = {Manski, Charles F},
	journal = {Econometrica},
	number = {4},
	pages = {1221--1246},
	publisher = {Wiley Online Library},
	title = {Statistical treatment rules for heterogeneous populations},
	volume = {72},
	year = {2004}}

@article{Hendren2020,
	author = {Hendren, Nathaniel and Sprung-Keyser, Ben},
	journal = {Quarterly Journal of Economics},
	number = {3},
	pages = {1209--1318},
	publisher = {Oxford University Press},
	title = {A unified welfare analysis of government policies},
	volume = {135},
	year = {2020}}

@article{Andrews2026,
	author = {Andrews, Isaiah and Shapiro, Jesse M},
	journal = {Econometrica},
	number = {3},
	pages = {843--875},
	publisher = {Wiley Online Library},
	title = {Communicating scientific uncertainty via approximate posteriors},
	volume = {94},
	year = {2026}}

@article{Onatski2002,
	author = {Onatski, Alexei and Stock, James H},
	journal = {Macroeconomic Dynamics},
	number = {1},
	pages = {85--110},
	publisher = {Cambridge University Press},
	title = {Robust monetary policy under model uncertainty in a small model of the U.S. economy},
	volume = {6},
	year = {2002}}

@article{Barlevy2009,
	author = {Barlevy, Gadi},
	journal = {Economic Perspectives},
	number = {2},
	title = {Policymaking under uncertainty: Gradualism and robustness},
	volume = {33},
	year = {2009},
	pages = {38-55},
	note = {Federal Reserve Bank of Chicago}}

@article{Giannoni2002,
	author = {Giannoni, Marc P},
	journal = {Macroeconomic Dynamics},
	number = {1},
	pages = {111--144},
	publisher = {Cambridge University Press},
	title = {Does model uncertainty justify caution? Robust optimal monetary policy in a forward-looking model},
	volume = {6},
	year = {2002}}

@book{Hansen2008,
	author = {Hansen, Lars Peter and Sargent, Thomas J},
	publisher = {Princeton University Press},
	title = {Robustness},
	year = {2008},
	address = {Princeton, NJ}}

@techreport{Caravello2026,
	author = {Caravello, Tom{\'a}s E and McKay, Alisdair and Wolf, Christian K},
	title = {Evaluating Monetary Policy Counterfactuals: When Do We Need Structural Models?},
	institution = {National Bureau of Economic Research},
	type = {Working Paper},
	number = {32988},
	year = {2026}}

@article{Chetty2009,
	author = {Chetty, Raj},
	journal = {Annual Review of Economics},
	number = {1},
	pages = {451--488},
	publisher = {Annual Reviews},
	title = {Sufficient statistics for welfare analysis: A bridge between structural and reduced-form methods},
	volume = {1},
	year = {2009}}

@book{Woodford2003,
	author = {Woodford, Michael},
	publisher = {Princeton University Press},
	title = {{Interest and Prices: Foundations of a Theory of Monetary Policy}},
	year = {2003},
	address = {Princeton, NJ}}

@book{Gali2015,
	author = {Gal{\'\i}, Jordi},
	publisher = {Princeton University Press},
	title = {{Monetary Policy, Inflation, and the Business Cycle: An Introduction to the New Keynesian Framework and Its Applications}},
	year = {2015},
	edition = {2nd},
	address = {Princeton, NJ}}

@article{Lucas1976,
	author = {Lucas, Jr., Robert E.},
	journal = {Carnegie-Rochester Conference Series on Public Policy},
	pages = {19--46},
	title = {{Econometric policy evaluation: A critique}},
	volume = {1},
	year = {1976}}

@article{Sims1980,
	author = {Sims, Christopher A},
	journal = {Econometrica},
	number = 1,
	pages = {1--48},
	publisher = {JSTOR},
	title = {Macroeconomics and Reality},
	volume = 48,
	year = {1980}}

@article{Barnichon2023,
	author = {Barnichon, R{\'e}gis and Mesters, Geert},
	journal = {American Economic Review},
	number = {11},
	pages = {2809--2845},
	publisher = {American Economic Association 2014 Broadway, Suite 305, Nashville, TN 37203},
	title = {A sufficient statistics approach for macro policy},
	volume = {113},
	year = {2023}}

@article{McKay2023,
	author = {McKay, Alisdair and Wolf, Christian K},
	journal = {Econometrica},
	number = {5},
	pages = {1695--1725},
	publisher = {Wiley Online Library},
	title = {What can time-series regressions tell us about policy counterfactuals?},
	volume = {91},
	year = {2023}}

@article{Brainard1967,
	author = {William C. Brainard},
	issn = {00028282},
	journal = {American Economic Review},
	note = {Papers and Proceedings},
	number = {2},
	pages = {411--425},
	publisher = {American Economic Association},
	title = {Uncertainty and the Effectiveness of Policy},
	url = {http://www.jstor.org/stable/1821642},
	urldate = {2026-07-31},
	volume = {57},
	year = {1967}}

@book{Ferguson1967,
	address = {New York},
	author = {Ferguson, Thomas S.},
	number = {1},
	publisher = {Academic Press},
	series = {Probability and Mathematical Statistics},
	title = {Mathematical Statistics: A Decision Theoretic Approach},
	year = {1967}}

@incollection{JamesStein1961,
	address = {Berkeley},
	author = {James, W. and Stein, Charles},
	booktitle = {Proceedings of the Fourth Berkeley Symposium on Mathematical Statistics and Probability, Volume 1: Contributions to the Theory of Statistics},
	editor = {Neyman, Jerzy},
	pages = {361--379},
	publisher = {University of California Press},
	title = {Estimation with Quadratic Loss},
	year = {1961}}

@book{LehmannCasella1998,
	address = {New York},
	author = {Lehmann, Erich L. and Casella, George},
	edition = {2nd},
	publisher = {Springer},
	title = {Theory of Point Estimation},
	year = {1998}}

@book{Theil1958,
	address = {Amsterdam},
	author = {Theil, Henri},
	number = {15},
	publisher = {North-Holland},
	series = {Contributions to Economic Analysis},
	title = {Economic Forecasts and Policy},
	year = {1958}}

@book{Tinbergen1952,
	address = {Amsterdam},
	author = {Tinbergen, Jan},
	number = {1},
	publisher = {North-Holland},
	series = {Contributions to Economic Analysis},
	title = {On the Theory of Economic Policy},
	year = {1952}}

@article{BergerBerlinerZaman1982,
	author = {Berger, James O. and Berliner, L. Mark and Zaman, Asad},
	journal = {Annals of Statistics},
	number = {3},
	pages = {838--856},
	title = {General Admissibility and Inadmissibility Results for Estimation in a Control Problem},
	url = {https://www.jstor.org/stable/2240908},
	volume = {10},
	year = {1982}}

@article{Berliner1983,
	author = {Berliner, L. Mark},
	journal = {Annals of Statistics},
	number = {3},
	pages = {814--826},
	title = {Improving on Inadmissible Estimators in the Control Problem},
	volume = {11},
	year = {1983}}

@article{Zaman1981,
	author = {Zaman, Asad},
	journal = {Annals of Statistics},
	number = {4},
	pages = {812--821},
	title = {A Complete Class Theorem for the Control Problem and Further Results on Admissibility and Inadmissibility},
	volume = {9},
	year = {1981}}

@techreport{ZamanStein1981,
	address = {Pasadena, CA},
	author = {Zaman, Asad and Stein, Charles},
	institution = {California Institute of Technology},
	month = jun,
	number = {391},
	title = {Admissibility in Four but Not in Five Dimensions of the Diffuse Prior Bayes Rule for the Control Problem},
	type = {Social Science Working Paper},
	year = {1981}}

@article{Fisher1962,
	author = {Fisher, Walter D.},
	journal = {International Economic Review},
	number = {1},
	pages = {1--29},
	title = {Estimation in the Linear Decision Model},
	url = {https://www.jstor.org/stable/2525296},
	volume = {3},
	year = {1962}}

@article{Lindley1968,
	author = {Lindley, D. V.},
	journal = {Journal of the Royal Statistical Society, Series B},
	number = {1},
	pages = {31--53},
	title = {The Choice of Variables in Multiple Regression},
	url = {https://www.jstor.org/stable/2984458},
	volume = {30},
	year = {1968}}

@book{Zellner1971,
	address = {New York},
	author = {Zellner, Arnold},
	publisher = {John Wiley \& Sons},
	series = {Wiley Series in Probability and Mathematical Statistics},
	title = {An Introduction to Bayesian Inference in Econometrics},
	year = {1971}}

@misc{DLMF,
  author       = {{DLMF}},
  title        = {{NIST Digital Library of Mathematical Functions}},
  year         = {2026},
  howpublished = {Release 1.2.7 of 2026-06-15, \url{https://dlmf.nist.gov}},
  note         = {F.~W.~J. Olver, A.~B. {Olde Daalhuis}, D.~W. Lozier,
                  B.~I. Schneider, R.~F. Boisvert, C.~W. Clark, B.~R. Miller,
                  B.~V. Saunders, H.~S. Cohl, and M.~A. McClain, eds.}
}

@book{HansenSargent2013,
	address = {Princeton, NJ},
	author = {Hansen, Lars Peter and Sargent, Thomas J.},
	publisher = {Princeton University Press},
	series = {The Gorman Lectures in Economics},
	title = {Recursive Models of Dynamic Linear Economies},
	year = {2013}}

@article{Meng2005,
  author  = {Meng, Xiao-Li},
  title   = {From Unit Root to {Stein}'s Estimator to {Fisher}'s $k$ Statistics: If You Have a Moment, {I} Can Tell You More},
  journal = {Statistical Science},
  volume  = {20}, number = {2}, pages = {141--162}, year = {2005},
}

@article{Manski2021,
	author = {Manski, Charles F.},
	journal = {Econometrica},
	number = {6},
	pages = {2827--2853},
	title = {Econometrics for Decision Making: Building Foundations Sketched by {Haavelmo} and {Wald}},
	volume = {89},
	year = {2021}}

@article{GuKoenker2023,
	author = {Gu, Jiaying and Koenker, Roger},
	journal = {Econometrica},
	number = {1},
	pages = {1--41},
	title = {Invidious Comparisons: Ranking and Selection as Compound Decisions},
	volume = {91},
	year = {2023}}

@article{Christensen2026,
	author = {Christensen, Timothy and Moon, Hyungsik Roger and Schorfheide, Frank},
	doi = {10.1093/restud/rdag017},
	journal = {Review of Economic Studies},
	note = {Forthcoming},
	title = {Optimal Decision Rules When Payoffs are Partially Identified},
	year = {2026}}

@unpublished{Kitagawa2026,
title = {Evidence-Based Policy Choice in Macroeconomics},
author = {Toru Kitagawa and Simone Martinalli},
year = 2026,
note = {arXiv:2610.27405}
}

@article{HodgesLehmann1952,
	author = {Hodges, Jr., J. L. and Lehmann, E. L.},
	journal = {Annals of Mathematical Statistics},
	number = {3},
	pages = {396--407},
	title = {The Use of Previous Experience in Reaching Statistical Decisions},
	volume = {23},
	year = {1952}}

@techreport{Hurwicz1951,
	author = {Hurwicz, Leonid},
	institution = {Cowles Commission for Research in Economics},
	month = {February},
	number = {Statistics No.\ 356},
	title = {A Class of Criteria for Decision-Making under Ignorance},
	type = {Cowles Commission Discussion Paper},
	year = {1951}}

@article{Diamond1973,
	author = {Diamond, Peter A.},
	journal = {Bell Journal of Economics and Management Science},
	number = {2},
	pages = {526--538},
	title = {Consumption Externalities and Imperfect Corrective Pricing},
	volume = {4},
	year = {1973}}

@article{Sandmo1975,
	author = {Sandmo, Agnar},
	journal = {Swedish Journal of Economics},
	number = {1},
	pages = {86--98},
	title = {Optimal Taxation in the Presence of Externalities},
	volume = {77},
	year = {1975}}

@article{Harberger1964,
	author = {Harberger, Arnold C.},
	journal = {American Economic Review},
	note = {Papers and Proceedings},
	number = {3},
	pages = {58--76},
	title = {The Measurement of Waste},
	volume = {54},
	year = {1964}}

@article{Allcott2019,
	author = {Allcott, Hunt and Lockwood, Benjamin B. and Taubinsky, Dmitry},
	journal = {Quarterly Journal of Economics},
	number = {3},
	pages = {1557--1626},
	title = {Regressive Sin Taxes, with an Application to the Optimal Soda Tax},
	volume = {134},
	year = {2019}}

@article{Taubinsky2018,
	author = {Taubinsky, Dmitry and Rees-Jones, Alex},
	journal = {Review of Economic Studies},
	number = {4},
	pages = {2462--2496},
	title = {Attention Variation and Welfare: Theory and Evidence from a Tax Salience Experiment},
	volume = {85},
	year = {2018}}

@article{Li2024,
	author = {Li, Dake and Plagborg-M{\o}ller, Mikkel and Wolf, Christian K.},
	title = {Local Projections vs.\ {VARs}: Lessons From Thousands of {DGPs}},
	journal = {Journal of Econometrics},
	year = {2024},
	volume = {244},
	number = {2},
	pages = {105722},
	doi = {10.1016/j.jeconom.2024.105722}}

@article{MPQW2026,
	author = {Montiel Olea, Jos{\'e} Luis and Plagborg-M{\o}ller, Mikkel and Qian, Eric and Wolf, Christian K.},
	title = {Double Robustness of Local Projections and Some Unpleasant {VARithmetic}},
	journal = {Econometrica},
	year = {2026},
	volume = {94},
	number = {4},
	pages = {1313--1343},
	doi = {10.3982/ECTA23345}}

@article{ParmigianiTD,
author={Parmigiani, Giovanni},
title={Minimax, information and ultrapessimism},
journal={Theory and Decision},
year={1992},
volume={33},
number={3},
pages={241-252}}

@book{steinshakarchi2005real,
  author    = {Stein, Elias M. and Shakarchi, Rami},
  title     = {Real Analysis: Measure Theory, Integration, and Hilbert Spaces},
  series    = {Princeton Lectures in Analysis},
  volume    = {3},
  publisher = {Princeton University Press},
  address   = {Princeton, NJ},
  year      = {2005}
}
